\documentclass[apj,numberedappendix,twocolappendix,appendixfloats]{emulateapj}

\usepackage{epsfig}

\usepackage{textcomp}

\usepackage{float}

\usepackage{amssymb,amsmath}

\newcommand{\perbeam} {beam$^{-1}$}

\begin{document}

\title{Discovery of Megaparsec-Scale, Low Surface Brightness Nonthermal Emission in Merging Galaxy Clusters Using the Green Bank Telescope}


\author{
Damon Farnsworth\altaffilmark{1},
Lawrence Rudnick\altaffilmark{1},
Shea Brown\altaffilmark{2},
Gianfranco Brunetti\altaffilmark{3}
}

\altaffiltext{1}{Minnesota Institute for Astrophysics, University of Minnesota, 116 Church St. S.E., Minneapolis, MN 55455, USA}
\altaffiltext{2}{Department of Physics and Astronomy, University of Iowa, 203 Van Allen Hall, Iowa City, IA 52242, USA}
\altaffiltext{3}{INAF/Istituto di Radioastronomia, via Gobetti 101, I-40129 Bologna, Italy}

\begin{abstract}
We present results from a study of twelve X-ray bright clusters at 1.4~GHz with the 100-m Green Bank Telescope.  
After subtraction of point sources using existing interferometer data, we reach a median  (best) 1$\sigma$ rms sensitivity level of 0.01 (0.006) $\mu$Jy~arcsec$^{-2}$, and find a significant excess of diffuse, low surface brightness emission in eleven of twelve Abell clusters observed.
We also present initial results at 1.4~GHz of Abell 2319 from the Very Large Array.
In particular, we find:
(a) four new detections of diffuse structures tentatively classified as two halos (A2065, A2069) and two relics (A2067, A2073); 
(b) the first detection of the radio halo in A2061 at 1.4~GHz, which qualifies this as a possible ultra-steep spectrum halo source with a synchrotron spectral index of $\alpha\sim 1.8$ between 327~MHz and 1.4~GHz;
(c) a $\sim$2~Mpc radio halo in the sloshing, minor-merger cluster A2142; 
(d) a $>$2$\times$ increase of the giant radio halo extent and luminosity in the merging cluster A2319; 
(e) a $\sim$7$\times$ increase to the integrated radio flux and $>$4$\times$ increase to the observed extent of the peripheral radio relic in A1367 to $\sim$600~kpc, which we also observe to be polarized on a similar scale;
(f) significant excess emission of ambiguous nature in three clusters with embedded tailed radio galaxies (A119, A400, A3744).
Our radio halo detections agree with the well-known X-ray/radio luminosity correlation, but they are larger and fainter than current radio power correlation studies would predict.
The corresponding volume averaged synchrotron emissivities are 1-2 orders of magnitude below the characteristic value found in previous studies.
Some of the halo-like detections may be some type of previously unseen, low surface brightness radio halo or blend of unresolved shock structures and sub-Mpc scale turbulent regions associated with their respective cluster merging activity.
Four of the five tentative halos contain one or more X-ray cold fronts, suggesting a possible connection between gas sloshing and particle acceleration on large scales in some of these clusters.
Additionally, we see evidence for a possible inter-cluster filament between A2061 and A2067.
For our faintest detections, we note the possibility of residual contamination from faint radio galaxies not accounted for in our confusion subtraction procedure.
We also quantify the sensitivity of the NVSS to extended emission as a function of size.

\end{abstract}


\section{Introduction}
\label{sec:introduction}
Diffuse radio synchrotron emission, on scales of hundreds to thousands of kpc, is observed in some galaxy clusters, illuminating the presence of magnetic fields and relativistic (GeV) electrons.
The various morphologies of the observed structures are suggestive of multiple physical origins.
Those extended radio structures not directly associated with individual galaxies (e.g., jets and lobes of radio galaxies) are tied directly to the thermal intracluster medium (ICM), and are dubbed radio halos and radio relics (see \citealt{feretti2012} for a review).
Perhaps the best known example is the Coma cluster, which contains both a Mpc-scale halo and a peripheral relic (e.g., \citealt{willson1970}; \citealt{jaffe1979}; \citealt{giovannini1985}; \citealt{deiss1997}; \citealt{brown2011a}).
Halos and relics have relatively steep radio synchrotron spectral indices of $\alpha>1$, where we define the flux density $S_{\nu} \propto \nu^{-\alpha}$.

Radio halos are historically observed to be unpolarized, Mpc-scale smooth structures co-located with the X-ray ICM, observed in $\sim$1/3 of the most X-ray luminous clusters (\citealt{ferrari2008}), with a few tens of radio halos observed.
The central radio surface brightnesses of halos are fairly similar, with typical values ranging from $\sim$0.5 to a few $\mu$Jy~arcsec$^{-2}$ (e.g., \citealt{murgia2009}; \citealt{murgia2010}); this low characteristic surface brightness makes halo detection an observational challenge.
As the number of halo detections rises, so does the realization that radio halos are not all the same; their sizes range from hundreds to thousands of kpc, and their morphologies range from round to elongated and smooth to clumpy (e.g., \citealt{girardi2011}; \citealt{bonafede2012}; \citealt{boschin2012}, \citealt{venturi2013}).
The radio luminosity in halos is well correlated with halo extent (e.g., \citealt{cassano2007}; \citealt{murgia2009}) and X-ray properties such as luminosity and temperature (e.g., \citealt{cassano2006}).

Peripheral relics are typically elongated structures, up to and exceeding 1~Mpc in extent, found at the outskirts of a few tens of clusters.
In the NVSS\footnote{National Radio Astronomy Observatory (NRAO) VLA Sky Survey} sample, peripheral relics were found in $\sim$11\% of clusters with 0.1-2.4~keV X-ray luminosity $L_X > 5 \times 10^{44}$~erg~s$^{-1}$ (\citealt{giovannini2002}).
Often highly polarized (i.e, a few tens of percent), observations of these relics provide strong evidence of $\mu$G magnetic fields and cosmic rays (CRs) at cluster peripheries.
Predicted by cosmological simulations, peripheral relics are most likely tracers of merger or accretion processes, whereby the CRs are directly accelerated (or reaccelerated) by the resultant shocks, which also order and amplify the magnetic field (e.g., \citealt{hoeft2011}; \citealt{kang2012}).

There is much debate regarding the acceleration mechanism of the cosmic ray electrons (CRe) in halos, but it is now widely believed to occur in-situ due to the relatively short synchrotron lifetimes of the CRe compared to the diffusion timescales required to fill the cluster volume ($\sim$10~Mpc$^3$).
The two most likely origins of CRe in halos are re-acceleration of mildly relativistic electrons by merger-induced turbulence $-$ also known as the turbulent re-acceleration model $-$ and as decay products of collisions between the thermal ICM protons and long-lived cosmic ray protons (CRp) $-$ also known as the hadronic model of secondary CRe.  See \cite{brunetti2011a} for a recent review.

Clues to the dominant mechanism of CRe production in halos, which may be a function of the cluster's mass or merger state (e.g., \citealt{brunetti2011b}), can be gleaned from observables such as number counts and integrated spectral indices of halos.
There is an observed bimodality in the presence of radio halos for clusters with X-ray luminosity $>$5$\times$10$^{44}$~erg~s$^{-1}$, i.e., these clusters either have a halo with radio luminosity that correlates well with X-ray luminosity, or no radio halo at all at levels below $\sim$10\% of their expected radio luminosity (\citealt{brunetti2009}).
This is interpreted as evidence for an evolutionary cycle whereby clusters transition between radio halo ``on'' and ``off'' states driven by merger-induced turbulence (e.g., \citealt{brunetti2009}; \citealt{ensslin2011}; \citealt{donnert2013}).
Additionally, a correlation has been recently observed between the radio luminosity and integrated SZ signal for clusters, further strengthening the thermal-nonthermal relationship in clusters (\citealt{basu2012}; \citealt{cassano2013}).

In addition, turbulent re-acceleration models alone predict the presence of a population of ultra-steep spectrum (USS) halos, with $\alpha \gtrsim 1.5$, in clusters undergoing less energetic mergers (e.g., \citealt{cassano2006}; \citealt{brunetti2008}); for this reason, many radio halos are expected to be discovered at sub-GHz frequencies (e.g., \citealt{cassano2010c}).
To date, only about five USS halos are known (e.g., \citealt{feretti2012}); future sensitive surveys at $\sim$100~MHz, e.g., by the Low Frequency Array\footnote{www.lofar.org} (LOFAR), are expected to make great contributions to this issue.

\subsection{Detection Bias At Low $z$ And The Case For Single Dish Observations}
\label{sec:lowzbias}
It is well known that radio interferometers suffer from the so-called ``missing spacings problem,'' whereby the lack of sampling at short baselines results in decreased sensitivity to emission that is smooth on large spatial scales -- e.g., Mpc-scale radio halos.
Separate from the point source sensitivity of an interferometer $-$ which is dependent only on the total collecting area and receiver properties $-$ this problem is often times understated and ill-quantified, so we conducted a simple experiment to quantify the sensitivity of the NVSS to emission on various spatial scales.

\begin{figure}[t]
   \centering
   \epsfig{file=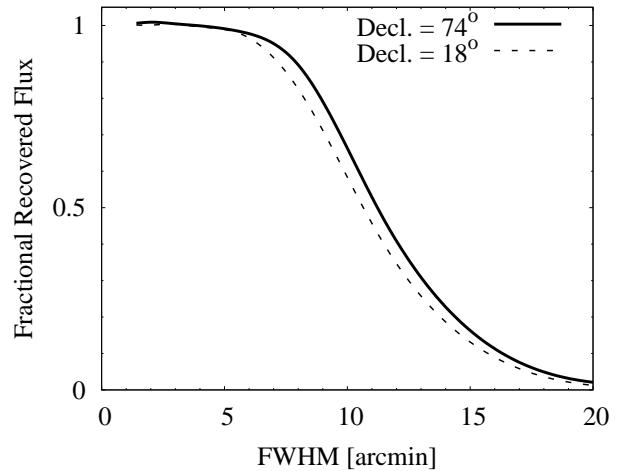,angle=0,width=0.49\textwidth}
   \caption{Fractional recovery of extended emission in the NVSS for two values of Declination (see text for details).  For emission on scales $>$11$'$, less than 50\% of the total flux is recovered.}
   \label{fig:NVSSsens}
 \end{figure}

The NVSS, which has been used as a finding survey for halos and relics (e.g.\citealt{giovannini1999}), was conducted using short ``snapshots'' which resulted in only modest $u$-$v$ coverage at short spacings.
As a result, the NVSS is increasingly insensitive to larger scale features, and an upper limit of $\sim$15$'$ is often quoted in the literature.
To estimate this effect more quantitatively, and to separate it from other issues such as signal:noise, we inserted very bright two-dimensional Gaussian components into two representative sets of NVSS $u$-$v$ data at declinations of 74\textdegree~ and 18\textdegree, respectively. 
We then constructed and cleaned the images, and measured the total flux in boxes drawn manually around the source as seen in the clean image.
The results of this experiment are shown in Figure~\ref{fig:NVSSsens}.
We find that the recovered flux falls to half of its true value at $\sim$10-11$'$ for snapshot observations, which will depend weakly on declination and the exact $u$-$v$ coverage, field overlaps, true source shape, etc.
At size scales of 15$'$, only $\sim$10\% of the flux is recovered.
This experiment was conducted using a 1000~Jy Gaussian; clean biases and other difficulties in detecting extended sources would occur at brightness levels near the detection limit.

This is important at low redshift because at $z = 0.1$~(0.05), a 1~Mpc halo would subtend roughly 9~(17) arcminutes.
One would then expect that surveys such as NVSS and WENSS\footnote{The Westerbork Northern Sky Survey}, which seek to maximize sky coverage by reducing integration times, might miss a significant number of radio halos at low redshift.
Furthermore, halo detections will be biased towards those with high surface brightness or spatially concentrated emission -- possibly excluding clusters of certain dynamical qualities.
The loss of flux on large scales by interferometers can be mitigated somewhat by improving the $u$-$v$ coverage for short baselines, either using longer integration times or compact arrays $-$ thus, ``filling'' the aperture more completely $-$ although increasing the integration time does not reduce the minimum baseline length. 
It is thus likely that the sizes and luminosities of halos at low redshift are underestimated.
The lack of short spacings for interferometers can be alleviated by combining single dish data (effectively zero baseline length) with the interferometer data (e.g., \citealt{stanimirovic2002}; \citealt{fletcher2011}).
Single-dish observations are the only way to recover the total flux of such highly extended sources, but suffer more from confusion from point source background and diffuse Galactic foreground.
For example, at 1.4~GHz the extragalactic point source confusion has an rms\footnote{Estimated using Equation 3E6 in Essential Radio Astronomy by Condon, J. \& Ransom, S.; http://www.cv.nrao.edu/course/astr534/ERA.shtml} of $\sim$90~$\mu$Jy \perbeam~ for the Karl G. Jansky Very Large Array (VLA) in D-configuration (45$''$ beam), but $\sim$13~mJy \perbeam~ for the Robert C. Byrd Green Bank Telescope (GBT; 9$'$ beam, nominal).
We will demonstrate that much of this confusion can be successfully removed, allowing these extended, low surface brightness halos to be observed where interferometers have failed.

Throughout this paper, unless otherwise stated, we assume $H_0 = 70$~km~s$^{-1}$~Mpc$^{-1}$, $\Omega_{\Lambda} = 0.7$, $\Omega_{m} = 0.3$.
We use $L_X$ to denote the 0.1-2.4~keV X-ray luminosity unless otherwise noted.
All linear (angular) sizes reported are deconvolved (observed) quantities unless otherwise noted.

\section{Observations and Data Reduction}
\label{sec:observations}

\subsection{GBT Observations}

We observed twelve Abell galaxy clusters with the 100-m GBT between June and September 2009 (see Table~\ref{tab:tableOfObservations}).  
The clusters were selected due to the possibility of extended radio polarization features present in reprocessed NVSS data (\citealt{rudnick2009a}), and restricted to $z\lesssim0.1$ in order to (at least slightly) resolve structure on Mpc scales; we also included two clusters with $z>0.1$ because they were serendipitously located in the same field of view as some of our sample members.
A brief summary of relevant parameters for each cluster is listed in Table~\ref{tab:literatureClusterParameters}.
The observations were taken with the GBT's Spectrometer in full polarization mode with a 50~MHz bandpass centered on 1.41~GHz.  
To create an image of each field we first employed on-the-fly mapping to create 1-D stripes of constant Declination.  
Each stripe was made from three successive ``back and forth'' scans, where Declination was held constant as the GBT was driven at a rate of 0.1\textdegree~s$^{-1}$ in Right Ascension, sampling at 2.4$'$ intervals.  
In anticipation of necessary baseline removal due to low spatial frequency gain drifts and foreground (atmospheric and Galactic) emission on scales of $\gtrsim$0.5\textdegree, our scans subtended at least a few degrees in Right Ascension (see Table~\ref{tab:tableOfObservations}).  
In order to adequately sample the GBT beam, the constant Declination stripes were separated by $3.3'$.
Similar observations were made each night for flux and polarization calibrators.

We give an overview here of the reduction procedure; for a more detailed discussion, see \cite{brown2011a}.
Using an internal correlated calibrator signal ($\sim$19~K) we determined the relative $X$ and $Y$ dipole gains and internal $X$-$Y$ phase offset.  
From each triplet of back and forth scans, the pair whose difference yielded the lowest rms value were averaged together to construct a stripe -- thus mitigating instabilities due to receiver or atmospheric fluctuations.  
This yielded fully calibrated stripes with sky position and Stokes $I, Q, U, V$ in units of surface brightness (i.e., Kelvin).  
Nominal $I$, $Q$, $U$ Gaussian beam dimensions and scalings of K/(Jy \perbeam) were determined using the calibrator source 3C286.
We note the dimensions of our beam ($\approx 9.5'$) are larger than the $9' \pm 0.1'$ listed in the GBT Proposal Guide\footnote{https://science.nrao.edu/facilities/gbt/proposing}.
This is likely due to a combination of factors such as: 1) the on-the-fly mapping technique, which smears the beam in the direction of telescope motion; 2) scan separation of 3.3$'$ in Declination; 3) interpolation onto a regular grid to create each map.
Parallactic angle correction was performed to transform the Stokes $Q$ and $U$ amplitudes from telescope to sky values.  
Due to insufficient parallactic angle coverage of a polarized calibrator, the full Mueller matrix was not computed, resulting in $\sim$1\% deviations of our $Q/I$, $U/I$ amplitudes with respect to the VLA calibrator manual values for 3C286.
Since we were only sensitive to very high fractional polarizations ($\gtrsim$30\%) for our low surface brightness features, no correction for the instrumental polarization was made.
We verified that the $V$ (circular polarization) stripes were consistent with zero to ensure that no significant residual polarization leakage remained.

\subsubsection{Stripe Baseline Removal}
Linear baseline subtraction in $I$, $Q$, $U$ for each stripe was performed to remove the effects of receiver drifts and smooth atmospheric and Galactic foregrounds, which vary on scales $\gtrsim$30$'$ (larger than the clusters observed).  
The ``sky'' for each GBT stripe was isolated by subtracting a preliminary point source model stripe, constructed from the 45$''$ resolution NVSS survey.  
To create the NVSS stripe for each Stokes parameter, we convolved the NVSS image to the nominal GBT beam measured from our 3C286 images, then interpolated to the GBT stripe sample positions.  
The desired linear baseline was then fit to the residuals and removed from the original stripe. 
Because Galactic emission is a source of confusion in some of the fields, we experimented with higher order polynomial (e.g., cubic) baseline fitting.  
We found that the baseline could not be well fit with higher order (i.e., $>$1) polynomial functions, which introduced large scale artefacts in many cases, and so we used linear baseline fitting only.  
This preliminary subtraction was meant only to help set the background ``zero'' level for each stripe and allow for image construction from the set of stripes; because the GBT gains, beam dimensions, and position correction varied slightly between fields, we later performed a more careful point source subtraction using the constructed GBT images after optimization of these parameters for each individual field.

\begin{table*}[t]
	\centering
	\small
	\caption{Galaxy Clusters Observed}  
	\begin{tabular}{ l *{6}{c} }
 \hline\hline
 Source & z$^a$ & Scale & R.A. & Decl. & Effective Beam & Local $\sigma_{map}^b$ \\
 & & (kpc/$''$) & (J2000) & (J2000) & $\theta_{Bmaj}$ $\times$ $\theta_{Bmin}, \phi_{Bpa}$ & (mJy \perbeam) \\
    \hline
 A119  & 0.0442 & 0.87 & 00:56:21.4 & -01:15:47 & $9.5' \times 9.3'$, 80\textdegree & 6.0 \\
 A400  & 0.0244 & 0.49 & 02:57:38.6 & +06:02:00 & $9.5' \times 9.3'$, 100\textdegree & 7.6 \\
 A1367 & 0.0220 & 0.44 & 11:44:29.5 & +19:50:21 & $9.5' \times 9.3'$, 80\textdegree & 4.5 \\
 A2056 & 0.0846 & 1.59 & 15:19:12.3 & +28:16:10 & $9.5' \times 9.3'$, 100\textdegree & 3.4 \\
 A2061 & 0.0784 & 1.48 & 15:21:15.3 & +30:39:17 & $9.5' \times 9.3'$, 100\textdegree & 2.3 \\
 A2065 & 0.0726 & 1.38 & 15:22:42.6 & +27:43:21 & $9.5' \times 9.3'$, 100\textdegree & 3.8 \\
 A2067 & 0.0739 & 1.40 & 15:23:14.8 & +30:54:23 & $9.5' \times 9.3'$, 100\textdegree & 2.3 \\
 A2069 & 0.1160 & 2.10 & 15:23:57.9 & +29:53:26 & $9.5' \times 9.3'$, 100\textdegree & 3.0 \\
 A2073 & 0.1717 & 2.92 & 15:25:41.5 & +28:24:32 & $9.5' \times 9.3'$, 80\textdegree & 3.3 \\
 A2142 & 0.0909 & 1.69 & 15:58:16.1 & +27:13:29 & $9.5' \times 9.3'$, 80\textdegree & 2.3 \\
 A2319 & 0.0557 & 1.08 & 19:20:45.3 & +43:57:43 & $9.7' \times 9.5'$, 110\textdegree & 6.0 \\
 A3744 & 0.0381 & 0.76 & 21:07:13.8 & -25:28:54 & $9.5' \times 9.3'$, 100\textdegree & 7.3 \\
 \hline
	\end{tabular}

	$^a$ Redshifts taken from the NASA/IPAC Extragalactic Database (NED) \\
	$^b$ After baseline removal and point source subtraction \\
  \label{tab:tableOfObservations}
\end{table*}

\begin{table*}[t]
  \centering
	\small
  \caption{Cluster Parameters From Literature$^a$}
  \begin{tabular}{ l *{7}{c} }
    \hline\hline
 Source & ${L_X}$ & $kT_X$ & $S_{1.4}$ & ${P_{1.4}}$ & LLS & Classification & References \\
 & (10$^{44}$ erg s$^{-1}$) & (keV) & (mJy) & (10$^{24}$ W Hz$^{-1}$) & (kpc) & & \\
    \hline
 A119  & 1.648 & 5.1 & -- & -- & -- & -- & 1 \\
 A400  & 0.204 & 2.1 & -- & -- & -- & -- & 2 \\
 A1367 & 0.816 & 3.5 & 35 & 0.038 & 130 & Relic & 2, 10, 11 \\
 A2056 & 0.116 & -- & -- & -- & -- & -- & 4 \\ 
 A2061H & 2.015 & 5.6 & -- & -- & N/A$^b$ & Halo & 2, 7 \\
 A2061R & 2.015 & 5.6 & 27.6 & 0.42 & 680 & Relic & 2, 6 \\
 A2065 & 2.520 & 8.4 & -- & -- & -- & -- & 2 \\
 A2067 & 0.439 & 3.1 & -- & -- & -- & -- & 3 \\
 A2069 & 4.551 & 7.9 & -- & -- & -- & -- & 2 \\
 A2073 & 1.908 & 5.6 & -- & -- & -- & -- & 3 \\
 A2142 & 10.58 & 11.0 & 18.3 & 0.38 & 200 & Mini-halo & 2, 9 \\
 A2319 & 6.995 & 9.9 & 153 & 1.13 & 1030 & Halo & 1, 8 \\
 A3744 & 0.180 & -- & -- & -- & -- & -- & 5 \\ 
		\hline
  \end{tabular}

	$^a$ All parameters are corrected for cosmology, where applicable. \\
	$^b$ The radio halo in A2061 was detected at 327~MHz by \cite{rudnick2009b} but no size was reported \\
	\textbf{Columns:} (1) Cluster; (2) X-ray luminosity, 0.1-2.4~keV; (3) X-ray temperature; (4) Flux density, 1.4~GHz; (5) Radio luminosity, 1.4~GHz; (6) Largest linear scale of radio halo or relic type emission; (7) Classification of diffuse radio emission not associated with radio galaxies. \\
	\textbf{References:} \textit{X-ray:} (1) \cite{ebeling1996}; (2) \cite{ebeling1998}; (3) \cite{ebeling2000}; (4) \cite{ledlow2003}; (5) \cite{bohringer2004}; \textit{Radio:} (6) \cite{vanweeren2011}; (7) \cite{rudnick2009b}; (8) \cite{feretti1997}; (9) \cite{giovannini2000}; (10) \cite{gavazzi1983}; (11) \cite{gavazzi1987} \\

  \label{tab:literatureClusterParameters}
\end{table*}

\subsubsection{GBT Imaging and Subtraction}
\label{sec:imagingAndSubtraction}
The calibrated and baseline subtracted stripes were then used to create $I$, $Q$, $U$ images for each field via interpolation with a square pixel scale of $2'$.  
Once the GBT images were constructed, we attempted to remove the contribution of point sources more carefully, enhancing the procedure described above; this was necessary because of small pointing errors and because the GBT beam properties vary slightly from field to field.
Since we are not limited by extragalactic point source confusion in polarization, image subtraction was performed only for the total intensity data.

We now describe the subtraction method for the total intensity images.
To mitigate the contribution of the NVSS image noise ($\sigma_{rms} = 0.45$~mJy~\perbeam, 45$''$ resolution), which is $\approx$5.5~mJy~\perbeam~ at the nominal 9$'$ GBT resolution, we first clipped each full resolution NVSS image at $3\sigma_{rms} = 1.35$~mJy~\perbeam.
The clipping procedure results in a small loss of flux for compact sources; the residual errors due to this are much less than the other sources of error in the final images.
We then optimized a set of six parameters to yield the lowest GBT-NVSS rms residuals around several moderately bright ($\sim$100 mJy \perbeam) sources in the vicinity of each target.
The parameters fit were: beam major and minor FWHM dimensions, $\theta_{Bmaj}$ and $\theta_{Bmin}$, and position angle, $\phi_{Bpa}$; x and y image shift; and flux scaling (Kelvin to Jy \perbeam).
In general, the shifts applied to each GBT image were $\lesssim$0.5 pixels in Right Ascension and zero pixels in Declination.  
The optimized scaling from Kelvin to Jy \perbeam~ was applied to each GBT image; the flux scaling varied by $\lesssim$2\% from field to field.

\begin{figure*}[t]
  \centering
  \epsfig{file=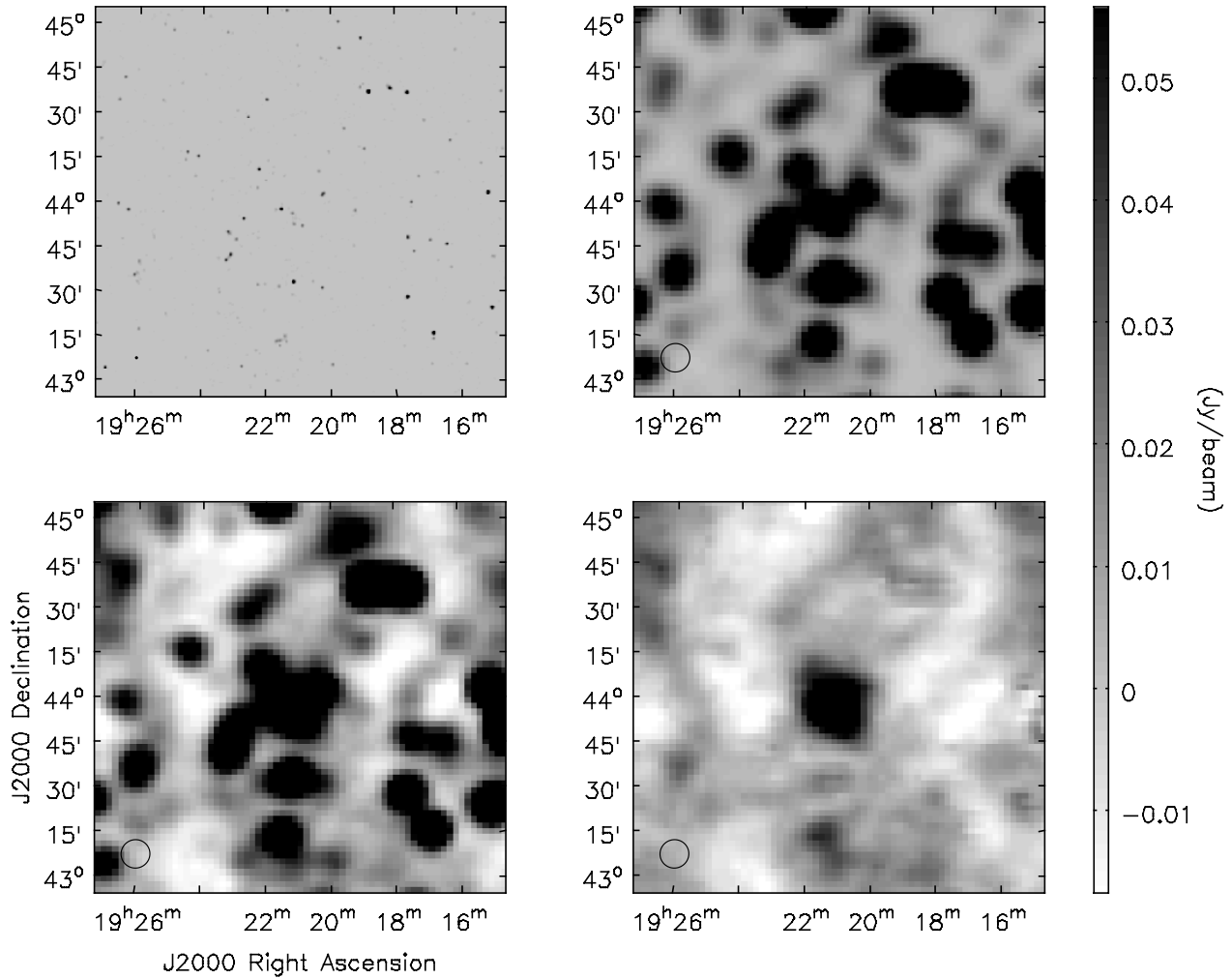,angle=0,width=0.95\textwidth}
  \caption{Illustration of the subtraction procedure. Top left: clipped NVSS image at $45''$ resolution.  Top right: clipped NVSS image convolved to the GBT beam.  Bottom left: GBT image.  Bottom right: GBT-NVSS residual image showing the radio halo in A2319.  The corresponding beam is shown in the lower left of each image.  The NVSS image has been clipped at $3\sigma = 1.35$~mJy~($45''$ beam)$^{-1}$.}
  \label{fig:subtrComparison}
\end{figure*}

Figure~\ref{fig:subtrComparison} illustrates the point source removal process for the A2319 field, displaying the input NVSS and GBT images, and the resulting GBT-NVSS residual image which shows a diffuse halo-like structure along with patchy Galactic emission and typical subtraction artefacts.

A final zero-level subtraction was performed on each residual image to minimize the contribution from the local Galactic foreground, which could contaminate the measurement of the diffuse emission associated with the cluster (e.g., halo or relic).
We first calculated and subtracted an initial estimate of the mean background level within an aperture of area $>$10 beams placed around the cluster with inner radius $\sim$1~Mpc.
Potential detections were then identified by examining contour maps of the images for coherent structures above the 2$\sigma$ level.
All fields except for A2056 exhibited a potential detection; the residual images for A400 and A3744 contained possible extended emission associated with each cluster, but suffered from strong subtraction artefacts due to the presence of embedded radio galaxies.
For the images with potential detections, the following iterative method was employed: 
define a ``background'' aperture located just outside the 2$\sigma_{rms}$ contour of the potential detection, made as large as possible (but $\gtrsim$10$\times$ the area of the cluster aperture) while excluding residual artefacts, e.g., of a bright radio galaxy $>$30$'$ away); calculate the mean and $\sigma_{rms}$ of the background aperture; subtract the mean of the background aperture from the entire image and adopt the $\sigma_{rms}$ for the next iteration.
For fields where no potential detection was apparent, the same iterative process was employed, but the background aperture was defined with an inner radius of 0.5~Mpc and an outer radius of $\sim$1.5~Mpc at the cluster redshift (again excluding artefacts).
After each iteration, the field was reassessed for the presence of a potential detection.
This process was repeated until the absolute value of the mean background level was $\lesssim$0.1~mJy~\perbeam~ and changed by $\lesssim$10\% between iterations; typically less than five iterations were required per field.
The background level removed was generally less than a few mJy \perbeam~ for each field.

For fields where significant patchy Galactic emission was present near the cluster, a final, more careful background subtraction was performed; we describe this process in the discussion of each relevant field in Section~\ref{sec:results}.
The adopted value of $\sigma_{map}$ (reported in Table~\ref{tab:tableOfObservations}) was then calculated within a background aperture on the final image.

The NVSS is partially sensitive to extended emission, and it is inevitable that some of our desired diffuse flux will be subtracted in some sources; we address this point for particular clusters in Section~\ref{sec:results}.

Residual artefacts due to non-Gaussianity of the GBT beam, imperfect image alignment, scaling, scan baseline removal, etc., remained after subtraction.  
This is most noticeable for bright ($>$100 mJy \perbeam) or particularly extended radio sources where the amplitudes of these artefacts are observed to be roughly 2-5\% of the original peak flux.  
This mainly posed a problem for clusters with an embedded RG, of which there are two in our sample.
We address these issues in Section~\ref{sec:results}.

For the polarization imaging, we used our GBT observations of the polarized calibrator source 3C286 to quantify the beam properties in $Q$ ($610'' \times 510'', 180$\textdegree) and $U$ ($590'' \times 530'', 45$\textdegree).
Each $Q$, $U$ image was then convolved to a common beam size ($10.5' \times 10.0', 0$\textdegree), which we adopted as our ``effective'' GBT polarization beam.
Images for $P$ (linear polarization intensity) and $\chi$ (linear polarization angle) were then constructed in the usual way, i.e.:
\begin{equation}
  P = \sqrt{Q^2 + U^2}
\end{equation}
\begin{equation}
  \chi = \frac{1}{2} \arctan \frac{U}{Q}.
\end{equation}

No correction was made for the polarized noise bias since our only detection, A1367, was at a high signal to noise ratio.

\subsection{VLA Observations of A2319}
Abell 2319 was observed on the recently upgraded VLA in its D-configuration at 1.4~GHz for three 7-hour sessions on March 27, March 28, and April 5, 2010 with 20 newly-configured antennas. 
Two pointings were used, centered at 19h21m15s, 43\textdegree52$'$ and 19h20m45s, 44\textdegree03$'$, with the northwest field added because of residuals seen from a beta-function fit to the Rosat PSPC X-ray data (\citealt{feretti1997}).

Data were taken in two bands of 128 MHz bandwidth each, centered at 1328 and 1860 MHz.  
Due to interference and receiver problems only the lower frequency band is presented here, and approximately 30\% of those data were also flagged as problematic.  
Calibration and self-cal using both amplitude and phase were performed in CASA\footnote{The \textit{Common Astronomy Software Applications} package; http://casa.nrao.edu}.  
Imaging and deconvolution were also performed in CASA resulting in two maps with resolutions of 43$''$$\times$39$''$ at -66\textdegree.
The maps were exported into AIPS\footnote{Astronomical Image Processing System; http://www.aips.nrao.edu} to be primary beam corrected and interpolated onto a single grid centered at the first pointing above.  
Compact sources totaling 1.1~Jy within 1000$''$ of the map center were removed from the map using the multiscale filtering technique described by \cite{rudnick2002}, with a box size of 207$''$.  
The residual image was then convolved to a 240$''$ circular beam to increase the signal:noise, resulting in a sensitivity of 3~mJy~beam$^{-1}$.

\section{Results}
\label{sec:results}
From the GBT residual images we identified detections of diffuse radio emission, requiring the 3$\sigma$ contour to be  extended in at least one dimension with respect to the GBT beam for all detections.
This identification was done separately in total intensity and polarization.  

After point source subtraction, nine of the twelve clusters exhibited one or more detections of excess diffuse emission, including one in polarization as well.
For the detections we measured the angular dimensions and radio fluxes within the 3$\sigma$ contours.
For the halo dimensions we adopt the following notation:  $\theta_{maj}$ and $\theta_{min}$ are the major and minor axis angular widths as measured from the 3$\sigma$ contours.
The 3$\sigma$ sizes were used to compare with those in the literature, where available, as listed in Table~\ref{tab:literatureClusterParameters}.
$I_{peak}$ and $S_{1.4}$ are the peak and integrated 1.4~GHz Stokes $I$ fluxes, respectively, and $P_{peak}$ and $P_{int}$ are the peak and integrated polarized intensities, respectively.
We also noted the location of the diffuse radio structure relative to the X-ray emission (e.g., centrally or peripherally located) in order to classify the nature of the emission (e.g., halo or relic).

If we assume elliptical Gaussianity to the diffuse intrinsic flux profile and a circular Gaussian beam (valid to $<$2\%) of effective FWHM
\begin{equation}
 \theta_{Beff} = \sqrt{\theta_{Bmaj} \times \theta_{Bmin}},
\end{equation}
we can estimate the deconvolved major and minor angular dimensions of the 3$\sigma$ contours, $\theta'_{maj}$ and $\theta'_{min}$, as:
\begin{equation}
 \theta'_{maj,min} \approx \sqrt{ \theta_{maj,min}^2 - \Delta_{Beff}^2 },
\label{eqn:general_deconv}
\end{equation}
where we define $\Delta_{Beff}$ as the full width of the effective circular Gaussian beam at the normalized amplitude of the 3$\sigma$ contour, i.e., 3$\sigma$/$I_{peak}$.  
Thus, when $I_{peak} = 6\sigma$, $\Delta_{Beff} = \theta_{Beff}$, and so on.
The deconvolved angular dimensions can then be converted to linear size.
$L_{maj}$ and $L_{min}$ are the deconvolved major and minor 3$\sigma$ dimensions of our detections in kpc, and we define the largest linear scale as $\textrm{LLS} = L_{maj}$.
For sources unresolved along the minor axis, we adopt $1/2$ the beam width as an upper limit to the deconvolved linear size.

The 3$\sigma$ sizes, measured angular and deconvolved linear, and fluxes for the total intensity detections are reported in Table~\ref{tab:observedClusterParameters}.
The integrated fluxes within the 3$\sigma$ contours should be considered ``biased'' estimates due to the likelihood that considerable emission lies below the 3$\sigma$ noise level.
Errors for the angular dimensions reflect the sizes of the 2$\sigma$ and 4$\sigma$ contours.
We report the sizes, measured and deconvolved, and fluxes for the polarization detection in Table~\ref{tab:polarizationDetectionParameters}.
Errors to the integrated flux within 3$\sigma$ contours, reported in Tables \ref{tab:observedClusterParameters} and \ref{tab:polarizationDetectionParameters}, are calculated as $\sigma_{map} \times \sqrt{N_{beams}}$, where $N_{beams}$ is the area of the 3$\sigma$ contour in beams.

\begin{table*}[t]
	\centering
	\small
  \caption{Total Intensity Results From 3$\sigma$ Contours}
  \begin{tabular}{ l | c c c c }
   \hline\hline
 Source & $P_{1.4}$ & $S_{1.4}$ & $\theta_{maj} \times \theta_{min}$ & $L_{maj} \times L_{min}$ \\
 & (10$^{24}$~W~Hz$^{-1}$) & (mJy) & (arcmin) & (kpc) \\
    \hline
A119 & 1.1 & 243$\pm$16 & $ 27.2_{-1.5}^{+1.8} \times 15.5_{-0.9}^{+0.5} $ & $ 1100_{-55}^{+61} \times <$250$^a$ \\
A400 & $<$0.48 & $<$352$^b$ & -- & -- \\
A1367H & $<$0.16 & $<$148$^b$ & -- & -- \\
A1367R$^c$ & 0.25 & 232$\pm$28 & $ 27.5_{-2.1}^{+2.1} \times 27.2_{-1.7}^{+1.8} $ & $ 610_{-50}^{+47} \times 600_{-37}^{+37} $ \\
A2056 & $<$0.28 & $<$15.7$^d$ & -- & -- \\
A2061H & 0.26 & 16.9$\pm$4.2 & $ 21.0_{-3.8}^{+2.8} \times 10.3_{-2.0}^{+2.0} $ & $ 1700_{-280}^{+170} \times 410_{-70}^{+70} $ \\
A2061R & 0.38 & 25.3$\pm$5.6 & $ 15.8_{-3.1}^{+1.5} \times 13.7_{-1.5}^{+2.3} $ & $ 760_{-350}^{+350} \times 290_{-30}^{+170} $ \\
A2065 & 0.42 & 32.9$\pm$11 & $ 16.7_{-3.7}^{+4.3} \times 16.3_{-4.3}^{+5.7} $ & $ 1100_{-270}^{+310} \times 1000_{-350}^{+460} $ \\
A2067 & 0.18 & 12.4$\pm$4.3 & $ 14.2_{-3.4}^{+8.8} \times 10.5_{-2.2}^{+3.2} $ & $ 840_{-240}^{+790} \times 240_{-79}^{+250} $ \\
A2069 & 1.0 & 28.8$\pm$7.2 & $ 23.7_{-3.0}^{+2.8} \times 12.0_{-2.3}^{+1.8} $ & $ 2800_{-290}^{+240} \times 1000_{-110}^{+170} $ \\
A2073 & 1.8 & 21.7$\pm$6.2 & $ 17.5_{-5.0}^{+6.2} \times 11.8_{-2.5}^{+3.0} $ & $ 2500_{-820}^{+1100} \times 1100_{-250}^{+340} $ \\
A2142 & 1.3 & 64.0$\pm$6.1 & $ 26.2_{-1.7}^{+1.5} \times 15.0_{-1.3}^{+1.3} $ & $ 2200_{-120}^{+68} \times <$480$^a$ \\
A2319 & 2.4 & 328$\pm$28 & $ 34.7_{-2.8}^{+3.3} \times 27.3_{-2.3}^{+2.2} $ & $ 2000_{-160}^{+190} \times 1400_{-130}^{+100} $ \\
A3744 & $<$0.16 & $<$1140$^b$ & -- & -- \\
	\hline
  \end{tabular}

 	$^a$ Upper limit for unresolved minor dimension (deconvolved) is set to one half the beam width \\
	$^b$ Upper limit within 500~kpc radius of cluster center; may contain residual contamination from strong radio galaxies \\
	$^c$ Measurements taken from the 4$\sigma$ contours.  Size errors reflect the dimensions of the 3$\sigma$ and 5$\sigma$ contours \\
	$^d$ Assuming 1~Mpc Gaussian halo \\
	\textbf{Columns:} (1) Cluster; (2) Radio luminosity at 1.4~Ghz; (3) Integrated 1.4~GHz total intensity flux; (4) Observed major and minor angular dimensions of the 3$\sigma$ contour.  Errors reflect the dimensions of the 2$\sigma$ and 4$\sigma$ contours (except A1367R); (5) Deconvolved major and minor linear dimensions of the 3$\sigma$ contour.    Errors reflect the dimensions of the 2$\sigma$ and 4$\sigma$ contours (except A1367R). \\

  \label{tab:observedClusterParameters}
\end{table*}

\begin{table}[t]
	\centering
	\small
  \caption{Polarization Detection From 3$\sigma$ Contours}
  \begin{tabular}{ l | c c c }
    \hline\hline
 Source & $P_{int}^a$ & $\theta_{maj} \times \theta_{min}$ & $L_{maj} \times L_{min}$ \\
 & (mJy) & (arcmin) & (kpc) \\
	\hline
	A1367R & 25.4$\pm$4.5 & $27.0 \times 17.0$ & $590 \times 210$ \\
	\hline
  \end{tabular}

	$^a$ Integrated 1.4~GHz polarized flux within 3$\sigma$ contours\\
  \label{tab:polarizationDetectionParameters}
\end{table}

\subsection{Individual Sources}

We now discuss the individual fields, including those with likely contamination, and add information from the literature.
For each field we illustrate our GBT detections $-$ for total intensity we overlay the NVSS raw image (greyscale) with the 1.4~GHz GBT-NVSS contours (red) and X-ray contours (blue).
The X-ray images are from the Rosat All Sky Survey\footnote{http://heasarc.gsfc.nasa.gov/docs/rosat/rass.html} (RASS) unless otherwise noted.

\subsubsection{Abell 119}
A119 ($z=0.0444$) is a merging cluster, as suggested by optical analyses (e.g., \citealt{way1997}; \citealt{tian2012}), which show multiple substructures and are suggestive of a dynamically young system with merging along the line of sight.
\cite{markevitch1998} found evidence for a mild merger shock in ASCA X-ray data.
Preliminary X-ray observations with $XMM-Newton$ (\citealt{whitaker2003}) show a long, cool filament ($\sim$4~keV) stretching to the N/NE of the core, further demonstrating that this system is likely undergoing a merger.
Short $Chandra$ observations suggest the presence of a cold front and possible merger shock to the north of the core, indicative of a state of near core passage in the merger (\citealt{sarazin2006}).

There is no previously known detection of halo or relic type radio emission.
\cite{giovannini2000} found no diffuse flux at 0.3~GHz with the VLA ($60'' \times 55''$ HPBW and $\sigma_{noise} = 5.0$~mJy \perbeam).
There is no evidence of halo emission in the 74~MHz VLA Low-Frequency Sky Survey redux (VLSSr; \citealt{lane2012}) image ($80''$ resolution).

\begin{figure}[t]
  \centering
  \epsfig{file=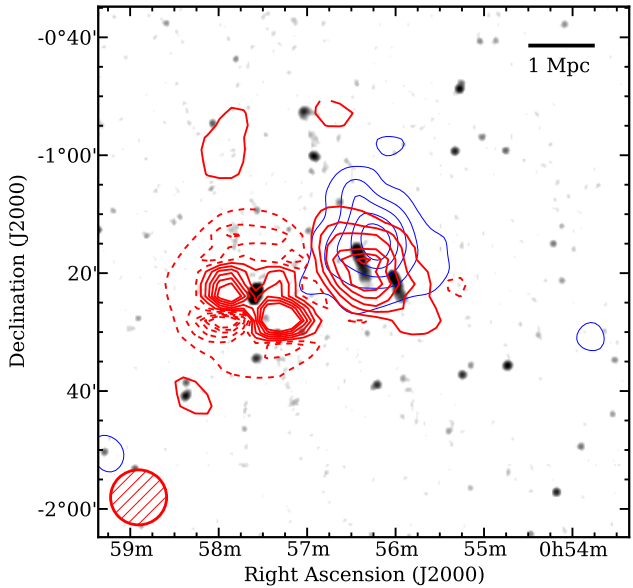,angle=0,width=0.47\textwidth}
  \caption{\textbf{A119}.  NVSS image (greyscale), clipped at 1.35 mJy (45$''$ beam)$^{-1}$ with overlaid GBT-NVSS 1.4 GHz residuals (red contours) and RASS X-ray image (smoothed with a 5$'$ Gaussian kernel, blue contours).  Radio contours are at $\pm$(3,9,15,21,27,33)$\times$$\sigma_{map}$ (negative contours dashed, if present).  Strong artefacts from the bright RG 3C29 to the SE of the cluster remain after subtraction, but the integrated residual flux for 3C29 is $<$1\% of the original flux. The GBT beam is shown in the lower left of the image.}
  \label{fig:a119}
\end{figure}

In the GBT-NVSS residual image we found an extended structure with largest linear scale (LLS) $\sim$1100~kpc and $S_{1.4} = 243$~mJy as measured from the 3$\sigma$ contours (Figure~\ref{fig:a119}).
3C29 is located at the cluster periphery, so the residual artefacts resulting from subtraction of this bright source do not appear to significantly contaminate the cluster diffuse emission.
There are multiple possibilities for the physical nature of the diffuse emission; the location of the radio structure, coincident with the X-ray emission, suggests a possible halo origin.
However, much or all of this diffuse flux is likely from the two tailed radio galaxies (TRGs) clearly visible in the NVSS image; it is probable that the extended radio tails have a low surface brightness component that NVSS has missed and that the GBT is picking up this missing flux.
We can not rule out halo or relic type emission, and therefore consider this as ``unclassified'' detection, worthy of deep followup interferometric observations.

\subsubsection{A400}
Abell 400 (z=0.0244), which contains the bright double TRG system 3C75 ($z=0.0232$), is believed to be undergoing a merger of two subgroups (\citealt{eilek2002} and references therein).
No diffuse ICM emission is apparent in NVSS or VLSSr images.

Due to the presence of the extremely bright 3C75 ($>$5~Jy \perbeam~ peak in the NVSS image convolved to the GBT beam), strong subtraction artefacts ($\sim$13\% of the peak in the residual image) prevented direct measurement of diffuse halo-type emission.  
To place a limit upon the diffuse radio emission, we integrated the fluxes inside an aperture of radius 500~kpc in both the convolved NVSS image and our GBT image.  
Comparison shows an excess of 352~mJy in the GBT image, which is $\sim$6\% of the integrated NVSS flux of 6.04~Jy.  
It is difficult to determine how much, if any, of this is ICM emission because there is likely a contribution from the extended radio tails of 3C75.  
We therefore report this as an upper limit to diffuse halo emission.

\subsubsection{Abell 1367}
A1367 ($z=0.022$), part of a filamentary supercluster structure with the Coma cluster, is a merging cluster which hosts a well known peripheral relic to the northwest.
A temperature structure analysis using X-ray observations was performed by \cite{donnelly1998}, which suggested a merger of two subclusters along a SE-NW line, with the lower luminosity component to the NW.
This interpretation is supported by the optical analysis of \cite{cortese2004} which suggests the NW component is in an early merging phase with the SE component; additional merging subclusters are also present, demonstrating the active dynamical state of this system.
\cite{ghizzardi2010} detected an X-ray brightness discontinuity -- which they label a ``merging'' cold front -- $\sim$350$''$ south of the X-ray center ($\sim$70$''$ from the peak) using \textit{XMM-Newton}.

Diffuse radio emission was first detected at 1.4~GHz by \cite{gavazzi1983} and observed again by \cite{gavazzi1987}, each time labeled a radio halo.
\cite{ensslin1998} re-labeled the radio structure a relic, noting the peripheral location and irregular shape of the radio structure; they predict a small polarization of $\sim$4\% due to the low expected viewing angle of the relic.
No known halo-type radio emission associated with the cluster has been detected.

In both the total intensity (GBT-NVSS residual) and polarization images, the relic is clearly visible but adjacent to a ridge of Galactic emission that runs through the cluster.
This confusing feature makes quantitative analysis difficult, and so it was removed by subtracting a large-scale Gaussian component from the total intensity and polarization images, similar to the method for A2319 (described in Section~\ref{sec:a2319}); see Figure~\ref{fig:galSubCompar}.

\begin{figure*}[t]
  \centering
	\begin{tabular}{cc}
  \epsfig{file=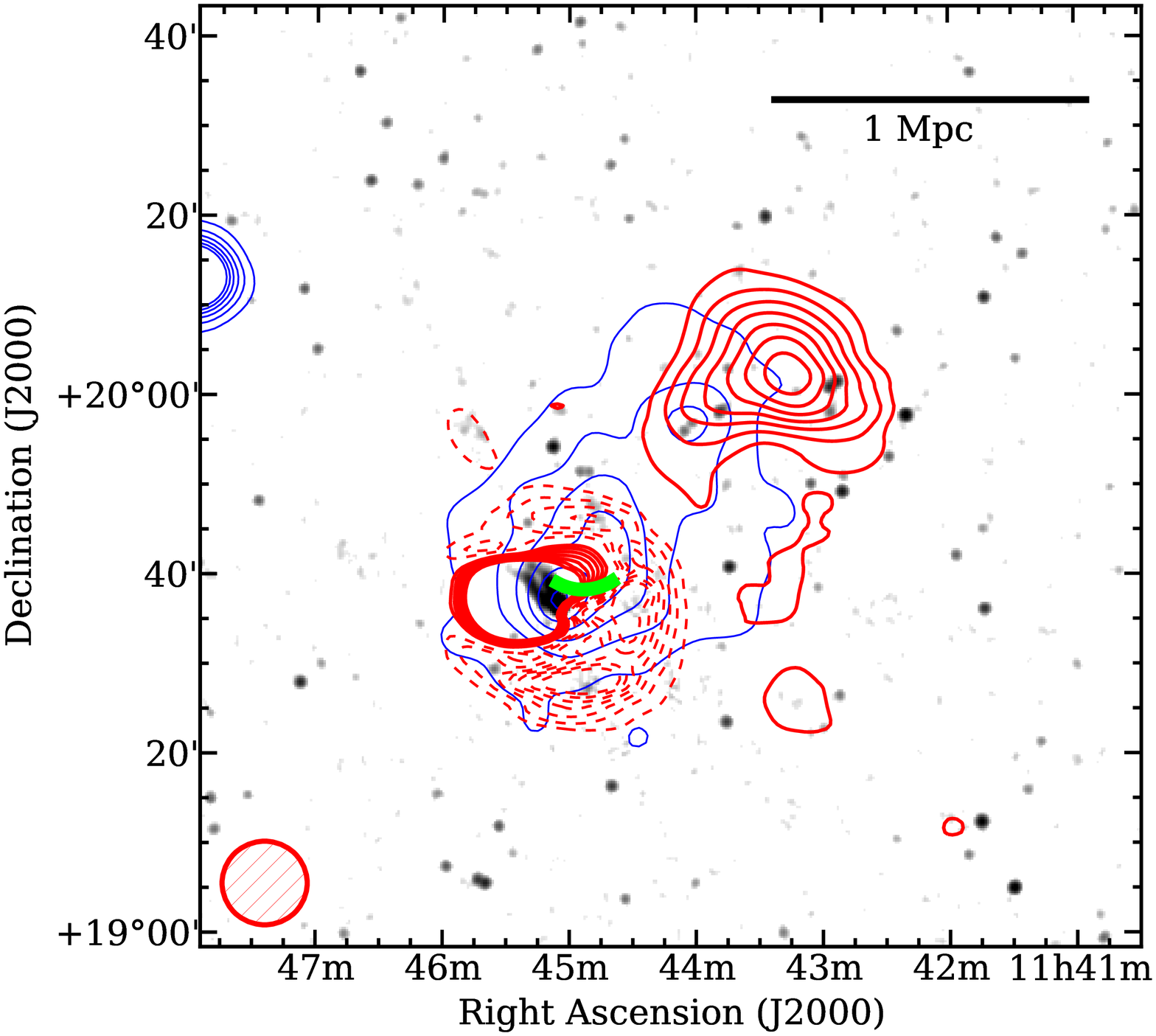,angle=0,width=0.47\textwidth} &
  \epsfig{file=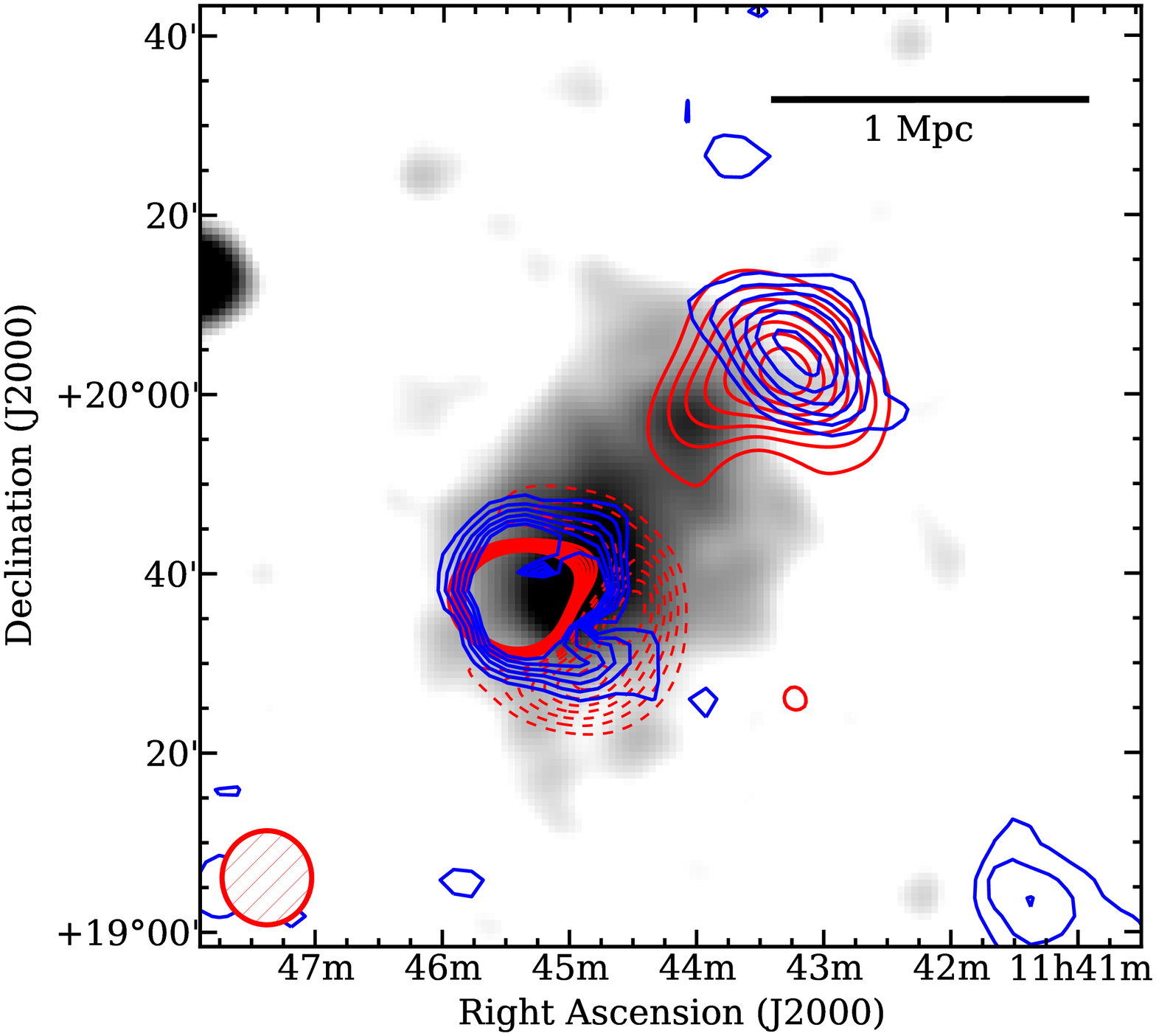,angle=0,width=0.47\textwidth}
	\end{tabular}
	\caption{\textbf{A1367}.  \textit{Left:} NVSS image (greyscale), clipped at 1.35 mJy (45$''$ beam)$^{-1}$ with overlaid GBT-NVSS 1.4 GHz total intensity residuals (red contours) and RASS X-ray image (smoothed with a 5$'$ Gaussian kernel, blue contours).  Radio contours are at $\pm$(4,7,10,13,16,19,22)$\times$$\sigma_{map}$ (negative contours dashed, if present).  Strong artefacts from the bright NAT 3C264 to the SE of the cluster remain after subtraction, preventing us from unambiguously detecting halo emission.  The X-ray cold front is shown as a green arc.  \textit{Right:} RASS X-ray image (smoothed with a 5$'$ Gaussian kernel, greyscale) with overlaid GBT 1.4 GHz polarized intensity (blue contours) and GBT-NVSS total intensity (red contours).  Polarized intensity contours are at (3,5,7,9,11,13)$\times$1.3~mJy~\perbeam; total intensity contours are at $\pm$(5,8,11,14,17,20,23)$\times$4.6~mJy~\perbeam~ (negative contours dashed, if present); $P$ and $I$ are at the resolution of the ``effective'' GBT $P$ beam ($10.5' \times 10'$), shown in the bottom left of the image.}
  \label{fig:a1367}
\end{figure*}

In total intensity we find a diffuse structure at the location of the known relic with LLS~$\sim$~600~kpc and $S_{1.4} = 232$~mJy as measured from the 4$\sigma$ contours, which we use as a conservative estimate of the relic boundary due to the presence of a peculiar residual feature which does not appear to be directly associated with the relic itself; see the left panel of Figure~\ref{fig:a1367}.
This arc-shaped residual feature, which is no longer a coherent structure above the 4$\sigma$ level and invisible above the 5$\sigma$ level, could be: a) Galactic emission not removed by our large-scale subtraction effort; b) a GBT beam side-lobe effect of 3C264; c) an NVSS imaging artefact, a common feature around many bright sources; d) cluster halo-type emission.
There is tentative evidence for a bridge of emission from the relic extending SE towards the main cluster.

The relic is also detected in linear polarization, with an integrated flux of 25.4~mJy (17.5~mJy~\perbeam~ peak; $\sigma_P = 1.3$~mJy \perbeam).
Because the signal to noise ratio was sufficiently high, correction for polarization bias was unnecessary.
We find a fractional polarization of roughly 18\% (15\%) at the location of peak polarized (total) intensity -- much higher than that predicted by \cite{ensslin1998}.
In the right panel of Figure~\ref{fig:a1367} we overlay the RASS X-ray image with the GBT-NVSS total intensity and GBT polarized intensity contours.

Due to the presence of the extended radio galaxy 3C264, strong subtraction artefacts prevented direct measurement of diffuse halo-type emission.  
To place a limit upon the diffuse radio emission, we integrated the flux inside an aperture of radius 500~kpc in the GBT-NVSS residual image, measuring 148~mJy.
It is difficult to determine how much, if any, of this is halo-type emission because there is likely a large contribution from the extended radio tails of 3C264.
We therefore report this as an upper limit to diffuse halo emission.

\subsubsection{A2056}
A2056 ($z=0.0846$) is a poorly studied member of the Corona-Borealis Supercluster (CrB-SC; e.g., \citealt{small1998}).  
With a relatively low X-ray luminosity of $L_X \approx 1.2 \times 10^{44}$~erg~s$^{-1}$, it is expected to host a very faint -- if any -- radio halo ($P_{1.4} \sim 5 \times 10^{20}$~W~Hz$^{-1}$) by the $P_{1.4}$ - $L_X$ correlation.  
It exhibits no diffuse emission in the NVSS image, and does not display any significant evidence of diffuse radio emission in the GBT residual image.  
We classify this as a ``clean'' non-detection due to the lack of residual subtraction artefacts from strong RGs present in A400 and A3744.  

\begin{figure*}[t]
	\centering
  \begin{tabular}{c c}
		\epsfig{file=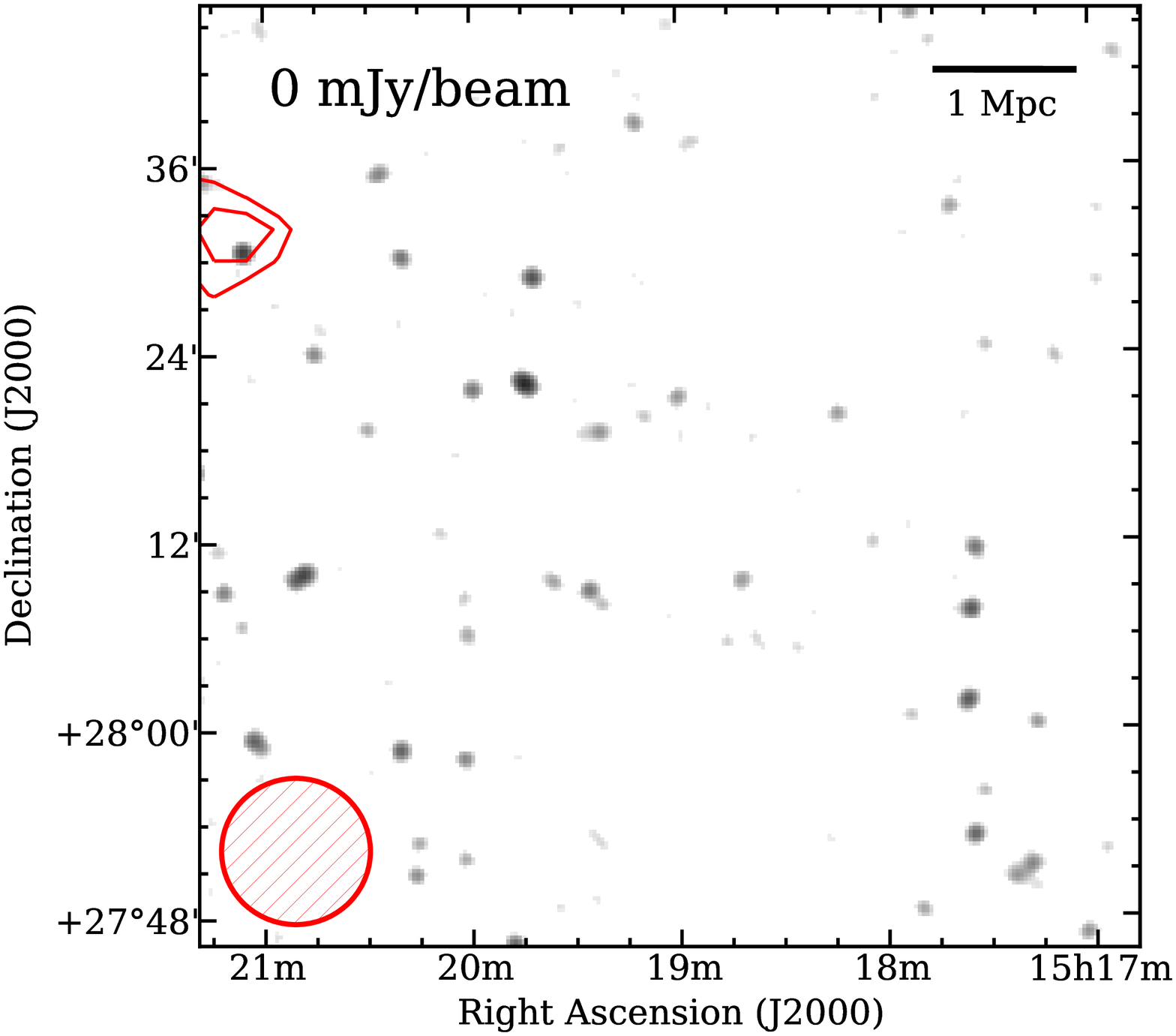,angle=0,width=0.4\textwidth} &
		\epsfig{file=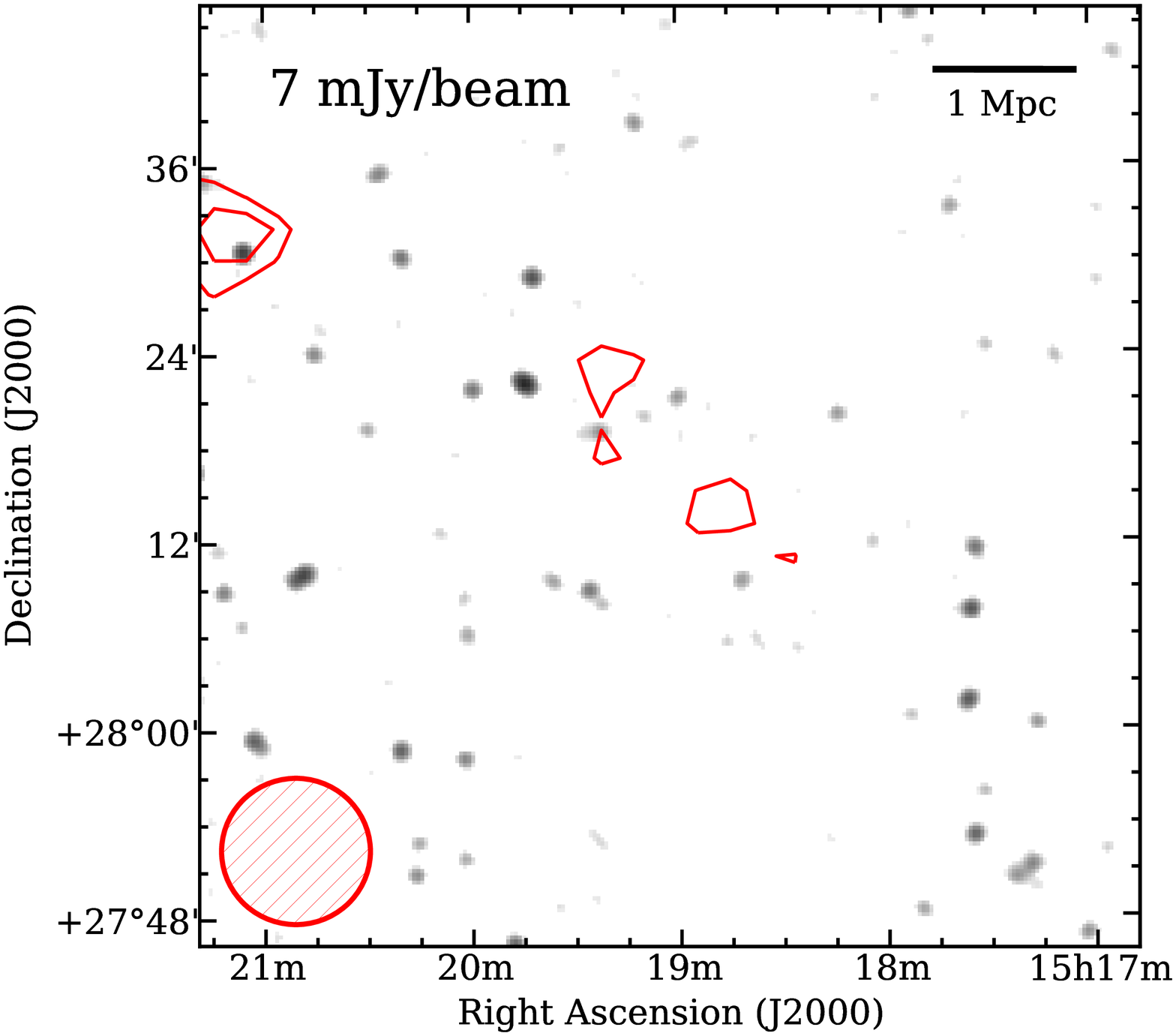,angle=0,width=0.4\textwidth} \\
		\epsfig{file=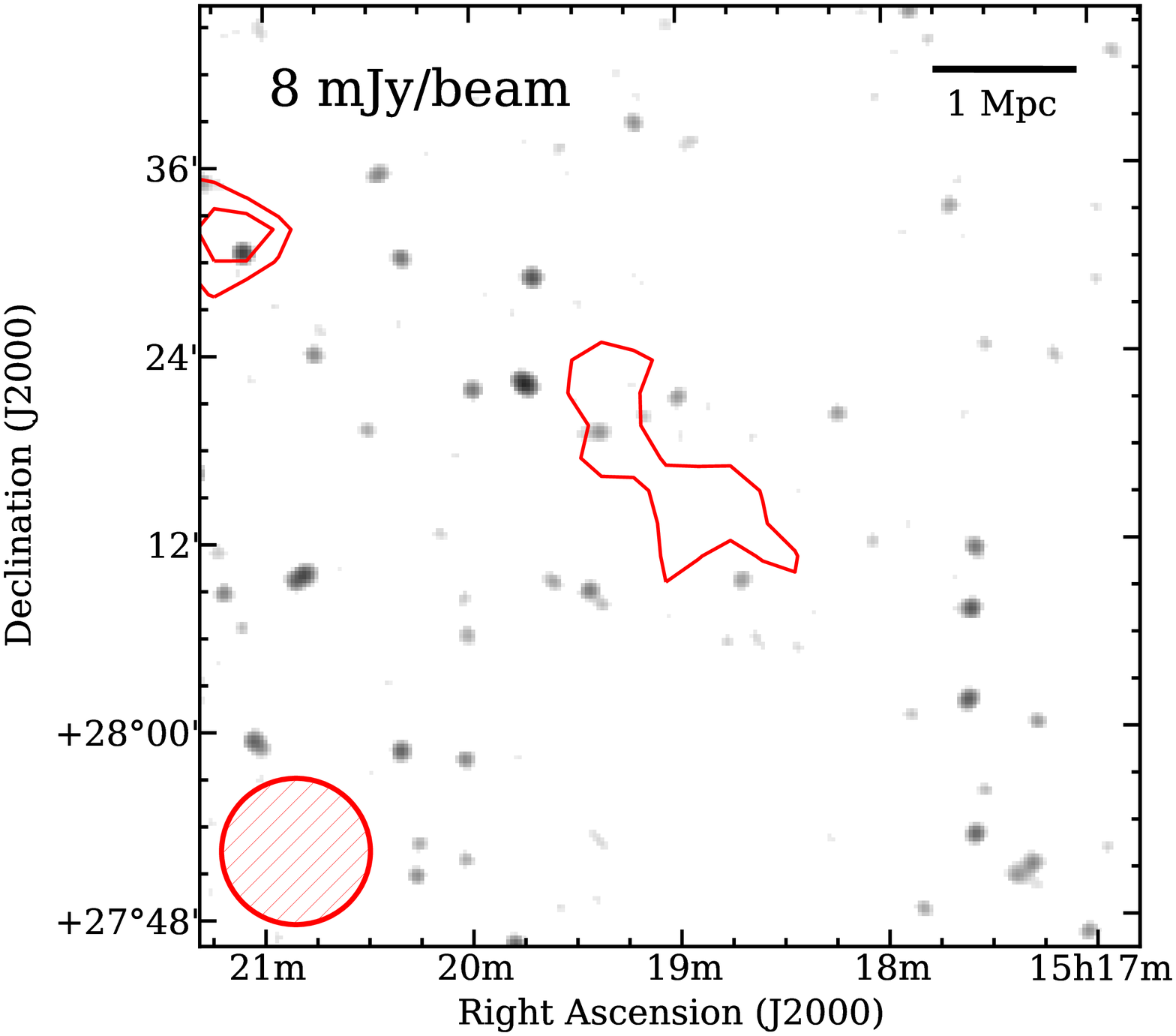,angle=0,width=0.4\textwidth} &
		\epsfig{file=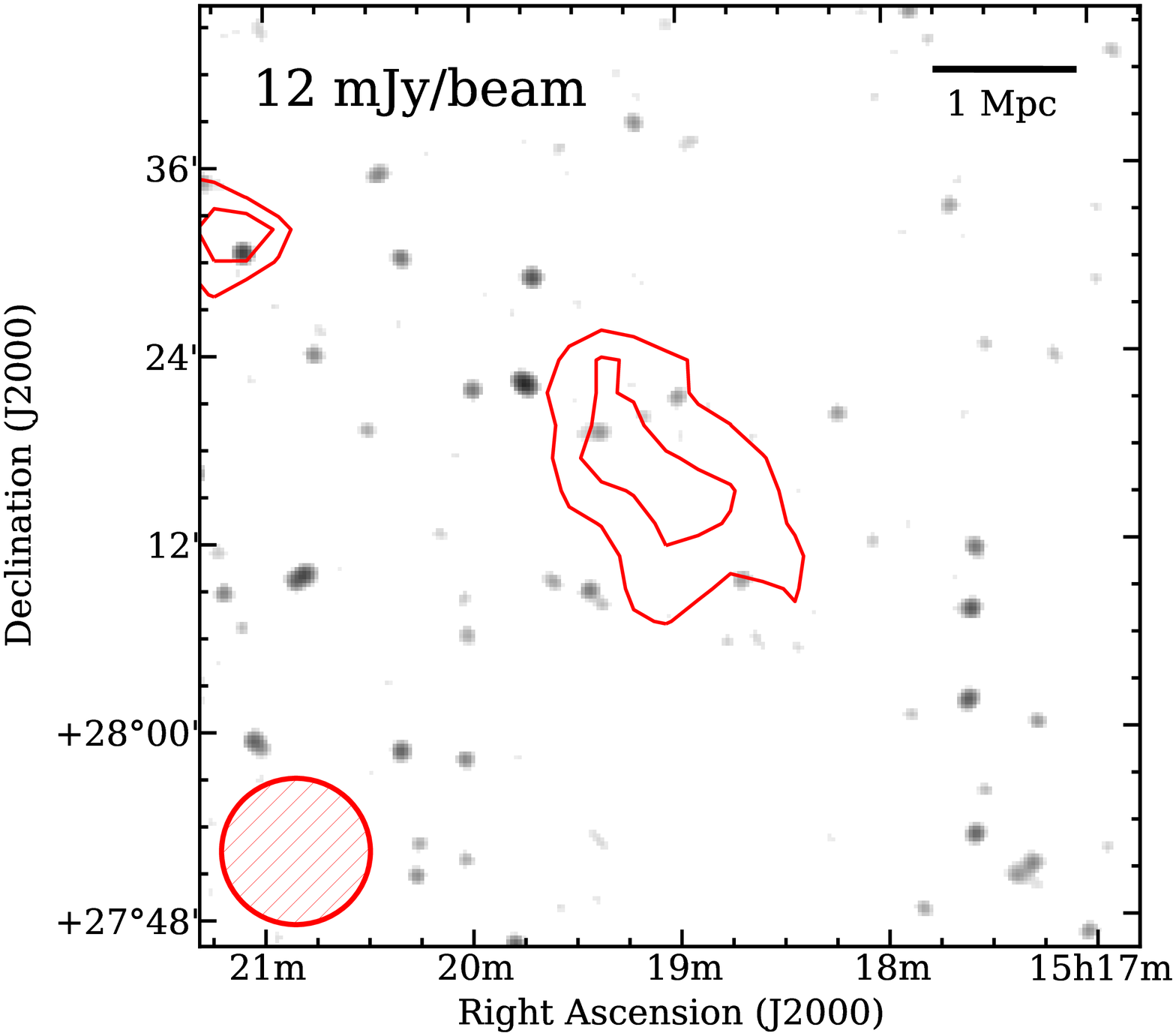,angle=0,width=0.4\textwidth}
  \end{tabular}
	\caption{\textbf{A2056.} Determination of an upper limit to the radio halo flux of A2056 using injection of a synthetic Gaussian halo with FWHM = 1~Mpc (see text for details).  Each image displays the NVSS image (greyscale), clipped at 1.35 mJy (45$''$ beam)$^{-1}$ with overlaid GBT-NVSS residuals plus injected halo (red contours).  For each frame, the contour levels are at $\pm$(3,4,5,...)$\times$$\sigma_{map}$ (negative contours are dashed, if present), the peak level of the synthetic halo is stated in the upper left corner, and the GBT beam is shown in the lower left corner.  The synthetic halo would be classified as a detection when injected with peak of 8~mJy~\perbeam, so we adopt 7~mJy~\perbeam~ as the peak flux of a non-detection.  The feature displaying residual flux in the upper left is an unrelated radio galaxy.}
	\label{fig:a2056synthGauss}
\end{figure*}

To estimate an upper limit to its radio halo flux we have injected synthetic Gaussian halos (e.g., \citealt{brunetti2007}; \citealt{venturi2008}; \citealt{rudnick2009b}).
We simulate a radio halo by constructing a circular Gaussian halo with a FWHM = 1~Mpc, convolved with the GBT beam for the A2056 field.
We then add scaled versions of this model to the original residual image until a detection (i.e., a coherent structure resolved in at least one dimension at the 3$\sigma$ contour) is observed.
Our halo simulation method is illustrated in Figure~\ref{fig:a2056synthGauss}.  
The synthetic halo is unambiguously detected when a Gaussian of peak flux 8~mJy~\perbeam~ is injected, so we adopt a peak value of 7~mJy~\perbeam~ as the upper limit for a non-detection.
To estimate a conservative upper limit for the 1.4~GHz radio luminosity, we use the total integrated flux of the model Gaussian with 7~mJy~\perbeam~ peak flux (i.e., integrate the model to infinity with zero noise).
The integrated flux of 15.7~mJy corresponds to an upper limit for the radio halo power of $P_{1.4} = 2.79 \times 10^{23}$~W~Hz$^{-1}$ $-$ more than two orders of magnitude above the value predicted by the observed X-ray correlation.

\subsubsection{Abell 2061}
A2061 ($z=0.0784$) is a CrB-SC member believed to be in a pre-merger state with the nearby cluster A2067; at their mean redshift of $z=0.0762$, the projected distance between A2061 and A2067 is 2.5~Mpc and \cite{rines2006} note they are separated by only $\sim$600~km~s$^{-1}$ in redshift.
\cite{marini2004} suggest the presence of an internal shock $\sim$3$'$ between the A2061 cluster core and a galaxy group infalling from the NE of the X-ray peak, as evidenced by an X-ray temperature jump in their \textit{BeppoSAX} data; its presence is further supported by preliminary \textit{Chandra} observations (\citealt{hogge2013}).

A radio relic $\sim$17$'$ to the SW of the X-ray peak was discovered in the WENSS and NVSS images by \cite{kempner2001}, with reported fluxes of 104~mJy and 19~mJy at 327~MHz and 1400~MHz, respectively; these measurements yield a spectral index of $\alpha$ = 1.17 $\pm$ 0.23.
Using the WSRT, \cite{vanweeren2011} measured the 1382 and 1714 MHz fluxes for the relic to be 27.6 and 21.2 mJy, respectively; combining their measurements with those of the literature they derived a spectral index of $\alpha$ = $1.03 \pm 0.09$ for the relic.
The projected dimensions of the relic are estimated to be $675\times320$~kpc (\citealt{vanweeren2011}).

We detect the SW relic in the GBT total intensity image; it is not detected in our polarization image.
It is marginally resolved, and the deconvolved 3$\sigma$ dimensions in Table~\ref{tab:observedClusterParameters} are thus unreliable.
However, it already has a measured size of 675~kpc $\times$ 320~kpc (\citealt{vanweeren2011}).
We measure 25.3~mJy of integrated flux within the 3$\sigma$ contours of the GBT residual image; this value includes $\sim$6~mJy of integrated relic flux present in the NVSS image above the clipping value of 1.35~mJy~\perbeam~ which we have added back in to the residual image.
Our integrated flux value for the relic is consistent, within errors, with the 27.6~mJy at 1382~MHz measured by \cite{vanweeren2011}.

A radio halo was discovered in reprocessed WENSS data by \cite{rudnick2009b}, who measured the 327 MHz fluxes for the halo and relic to be 270~mJy and 120~mJy, respectively, within separate apertures of radius 500~kpc.
They noted that their total fluxes are suspect ``because of extensive nearby emission'' present in the image.

\begin{figure}[t]
  \centering
	\epsfig{file=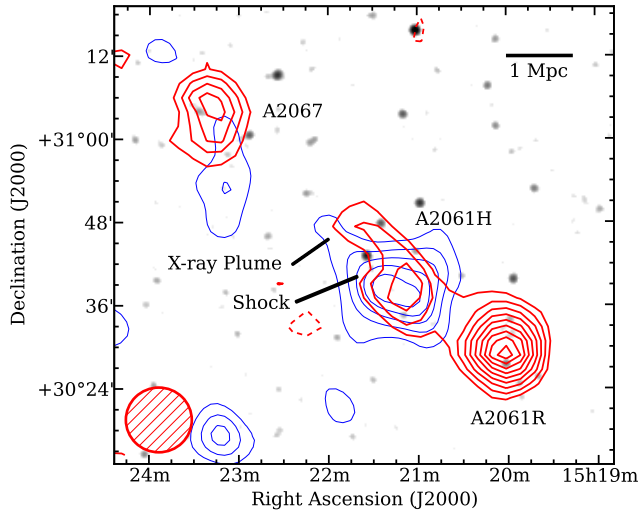,angle=0,width=0.47\textwidth}
  \caption{\textbf{A2061-A2067}.  NVSS image (greyscale), clipped at 1.35~mJy (45$''$ beam)$^{-1}$ with overlaid GBT-NVSS 1.4~GHz residuals (red contours) and RASS X-ray image (smoothed with a 5$'$ Gaussian kernel, blue contours).  The X-ray shock and plume are labeled.  Radio contours are at $\pm$(3,4,5,...)$\times$$\sigma_{map}$ (negative contours dashed, if present).  The GBT beam is shown in the lower left of the image.}
  \label{fig:a2061-a2067}
\end{figure}

Using the GBT, we have now made the first detection at 1.4~GHz of the radio halo, measuring 16.9~mJy of integrated flux within the 3$\sigma$ contours; see Figure~\ref{fig:a2061-a2067}.
The elongated halo morphology, with a largest linear extent of $\sim$1700~kpc, displays the fingerlike extension towards the NE, also seen at 327~MHz by \cite{rudnick2009b} which may be associated with the NE X-ray plume originally seen in Rosat-PSPC images (\citealt{marini2004}).
The halo classification is tentative, since relics, internal shocks, small scale turbulent regions, etc., associated with the extension to the NE may be unresolved by the GBT, appearing to be a coherent, Mpc-scale structure.
There appears to be a blending of the SW relic with the halo in the GBT image or perhaps a bridge of emission joining the two structures, as evident in Figure~\ref{fig:a2061-a2067}.

No evidence of a radio halo was found by \cite{vanweeren2011} at 1382~MHz with the WSRT, although a few small patches of diffuse radio emission are apparent in their WSRT image at the location of our GBT halo detection.
Their 1$\sigma$ rms sensitivity of 22~$\mu$Jy~(32$''$$\times$16$''$~beam)$^{-1}$ equates to a 3$\sigma$ detection limit of $\approx$41~mJy~\perbeam~ at the GBT resolution (570$''$$\times$560$''$), well above the peak flux of 13.5~mJy~\perbeam~ observed in our GBT image.
Thus, it is not surprising that the radio halo was not detected by \cite{vanweeren2011}.

\begin{figure}[t]
  \centering
	\epsfig{file=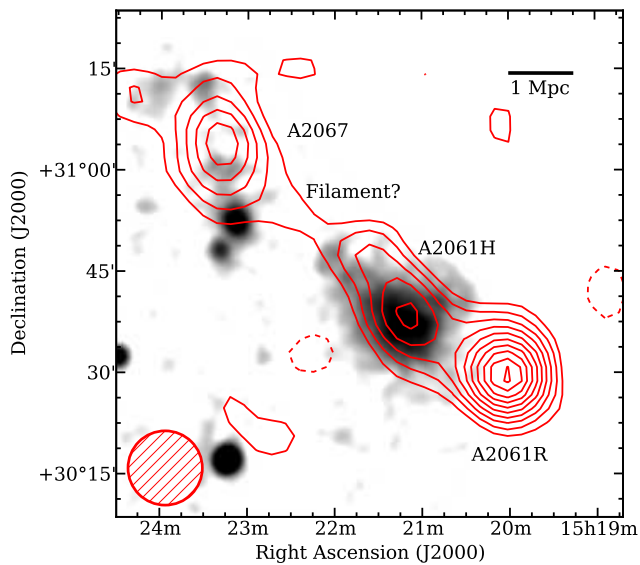,angle=0,width=0.47\textwidth}
  \caption{\textbf{Possible inter-cluster filament in A2061-A2067}.  Rosat PSPC X-rays (smoothed with a 2$'$ Gaussian kernel, greyscale) with overlaid GBT-NVSS 1.4~GHz residuals (red contours).  Radio contours are at $\pm$(2,3,4,...)$\times$2.4~mJy~(11$'$ beam)$^{-1}$ (negative contours dashed, if present).  The GBT beam is shown in the lower left of the image.}
  \label{fig:a2061-a2067_2sigFilament}
\end{figure}

To explore the possibility of an inter-cluster filament between A2061 and A2067, we show the GBT view of the A2061-A2067 system at 11$'$ resolution in Figure~\ref{fig:a2061-a2067_2sigFilament}.
At the 2$\sigma$ level we see an apparent bridge of emission between the clusters.

We also reprocessed the WENSS data to mitigate residual contamination from point sources; once convolved to the GBT resolution, this yielded a more reliable estimate of the 327~MHz halo flux within our GBT halo boundary (defined by the 3$\sigma$ contour) to be 250$\pm$42~mJy.
By combining the new 327~MHz halo flux with our 1.4~GHz measurement, we derive an integrated flux spectral index of $\alpha_{0.3}^{1.4} \approx 1.8 \pm 0.3$ for the halo.
This results in a tentative classification of the halo in A2061 as an ultra steep spectrum source, a class of objects whose spectral behavior has serious implications for the nature of halo generation -- we will return to this topic in Section \ref{sec:USS}.
We note that the WENSS image shows the halo to extend several arcminutes northeast beyond the GBT halo boundary, and reiterate that our spectral index estimate is for the emission within the boundary defined by our GBT 3$\sigma$ contour.

\subsubsection{Abell 2065}
A2065 ($z=0.0726$) is also part of CrB-SC.
\cite{markevitch1999} suggest A2065 is in the late stage of an ongoing merger, as evidenced by ROSAT and ASCA X-ray observations which show a double gas density peak about two central galaxies that appear to have survived a merger shock passage.
Similarly, \cite{belsole2005} used \textit{XMM-Newton} X-ray observations to label A2065 an ongoing merger of two subclusters in a compact phase, the evolutionary state of mergers where resultant strong shocks in the ICM are most easily detected.
They also state that the relatively cool main core suggests that the colliding object was probably of smaller mass, but that the quality of their observations are not good enough to be certain.
\cite{chatzikos2006} used $Chandra$ to detect a discontinuity they identify as a probable cold front and two ``cold'' cores coinciding with the cluster cD galaxies.
They suggest that evidence of shocks to the SE of the merger appear in the temperature maps, and that the deprojected density distribution in that region indicates the presence of a supersonic flow with $M \approx 1.7$; the quality of the data do not allow them to definitively distinguish between the shock wave and cold front interpretation for the discontinuity.
\cite{bourdin2008} detected a bow-like feature with \textit{XMM-Newton} that corresponds to the discontinuity detected by \cite{chatzikos2006}.
The X-ray brightness discontinuity is also detected $\sim$100$''$ SE of the X-ray center by \cite{ghizzardi2010} in an \textit{XMM-Newton} study, where it is classified as a cold front.
There is no known detection of diffuse radio emission in the literature for this cluster.

\begin{figure}[t]
  \centering
  \epsfig{file=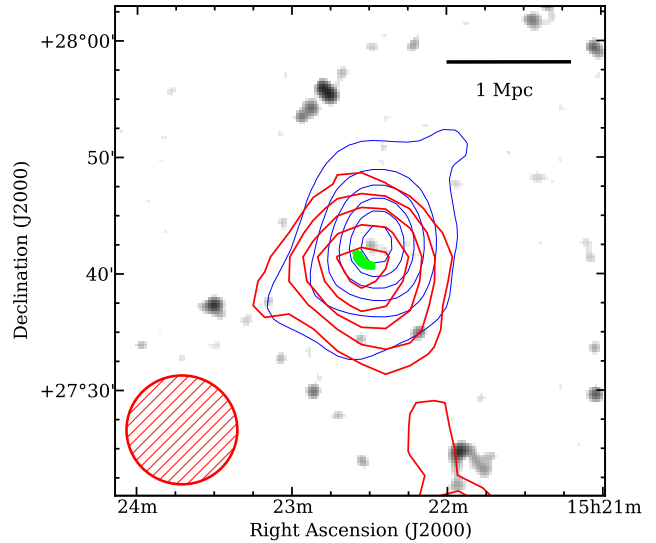,angle=0,width=0.47\textwidth}
  \caption{\textbf{A2065.}  NVSS image (greyscale), clipped at 1.35 mJy (45$''$ beam)$^{-1}$ with overlaid GBT-NVSS 1.4 GHz residuals (red contours) and RASS X-ray (smoothed with a 5$'$ Gaussian kernel, blue contours).  The X-ray cold front is shown as a green arc.  Radio contours are at $\pm$(3,4,5,...)$\times$$\sigma_{map}$ (negative contours dashed, if present).  The GBT beam is shown in the lower left of the image.}
  \label{fig:a2065}
\end{figure}

We detect a smooth diffuse structure of $S_{1.4} = 32.9$~mJy, roughly 1~Mpc in extent, within the 3$\sigma$ contours; see Figure~\ref{fig:a2065}.
Due to its $\sim$1~Mpc size and peak location, which is roughly coincident with the X-ray peak, we classify this structure as a possible giant radio halo.
The location of our radio centroid is  $\sim$3.5$'$ (290~kpc) SE of the X-ray peak in the RASS image, which corresponds roughly to the location of the $Chandra$ surface brightness discontinuity observed by \cite{chatzikos2006}.

\subsubsection{Abell 2067}
A2067 ($z=0.073858$), a relatively low X-ray luminosity cluster ($L_X \sim 4 \times 10^{43}$~erg~s$^{-1}$ in the 0.1-2.4~keV band), displays an elongated morphology in the RASS image.
\cite{marini2004} used \textit{BeppoSAX} X-ray observations to estimate the ICM gas temperature to be $kT \sim 1.5$~keV.  
Likely in a pre-merger state with A2061, there appear to be internal dynamics as well; this is suggested by the presence of two dominant galaxies -- one near the X-ray peak and the other $\sim$4.3$'$ to the south along an X-ray extension illuminating a galaxy overdensity (\citealt{marini2004}).
There is no previous detection of diffuse radio emission in the literature

We detect a marginally resolved feature in total intensity $\sim$12$'$ to the north of the X-ray peak; see Figure~\ref{fig:a2061-a2067}.
Within the 3$\sigma$ contours, we measure a flux of 12.4~mJy and LLS of $\sim$800~kpc.
We classify our detection as a possible radio relic based on its peripheral location with respect to the X-ray ICM emission; there is no evidence for a giant radio halo.

\subsubsection{Abell 2069}
A2069 ($z=0.116$) is undergoing a merger between two main X-ray components separated by $\sim$9$'$; \cite{owers2009} detected a cold front in the smaller component, A2069B ($z=0.1178$).
They suggest the cold front has arisen due to gas sloshing after an encounter with A2069A, with the subcluster motion primarily in the plane of the sky.

\begin{figure}[t]
  \centering
	\begin{tabular}{c}
  \epsfig{file=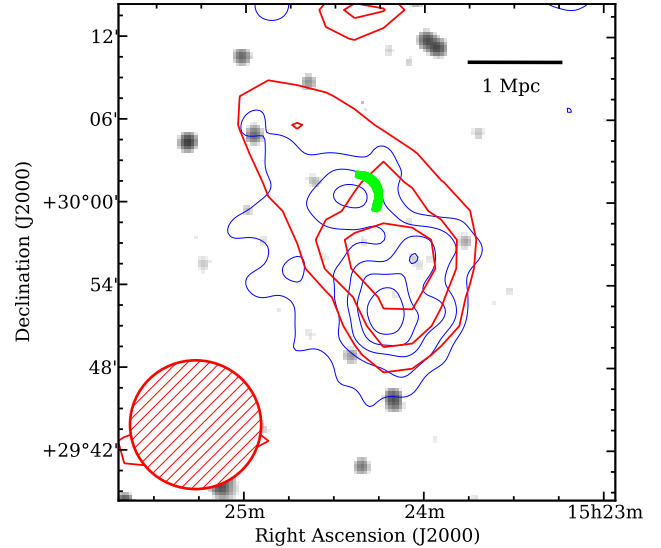,angle=0,width=0.47\textwidth}
	\end{tabular}
  \caption{\textbf{A2069.} NVSS image (greyscale), clipped at 1.35 mJy (45$''$ beam)$^{-1}$ with overlaid GBT-NVSS 1.4 GHz residuals (red contours) and Rosat PSPC X-ray image (smoothed with a 2$'$ Gaussian kernel, blue contours).  The X-ray cold front is shown as a green arc.  Radio contours are at $\pm$(3,4,5,...)$\times$$\sigma_{map}$ (negative contours dashed, if present).  The GBT beam is shown in the lower right of the image.}
  \label{fig:a2069}
\end{figure}

We have made the first detection of Mpc-scale radio emission in A2069, and classify this as a possible radio halo due to its coincidence with the diffuse X-rays; see Figure~\ref{fig:a2069}.
The radio emission, $\sim$2.8~Mpc long with integrated flux of 28.8~mJy, coincides roughly with the elongated X-ray emission connecting and surrounding the two main cluster components.
The radio peak is offset to the north of the X-ray peak (and location of A2069A) by $\sim$4$'$ ($\sim$700~kpc in projection), in the direction of A2069 and its cold front.
The radio halo axial ratio, as measured by the 3$\sigma$ contour, is $L_{maj}/L_{min} \approx 3$, which is at the high end for typical halos.
Another possible scenario is the existence of one or more merger-induced radio relics or small scale halo-like structures.
\cite{rudnick2009b} provide an upper limit at 327~MHz of 70~mJy for the diffuse flux within a 500~kpc radius of the X-ray centroid.
Using an aperture of 500~kpc radius centered at the location of the X-ray centroid we measure a GBT 1.4~GHz flux density of 10.7~mJy; this yields an upper limit to the spectral index from 327~MHz to 1.4~GHz of $\alpha^{1.4}_{0.3} \lesssim 1.3$ in this region.

\cite{giovannini1999} detected a diffuse radio structure $\sim$6~Mpc SE of the cluster center in the NVSS survey, but classify it as ``uncertain'' in type, although they speculate that it could be a relic; estimates of the flux and dimensions are absent.
By inspection of their radio image, this feature appears to be $\sim$5$'$ in extent, and corresponds to an unresolved radio feature in our GBT-NVSS residual image (not shown) that is $\sim$30$'$ ($\sim$4~Mpc in projection) SE of our halo-like detection.
There is no significant X-ray emission at this location.
It could be a peripheral relic, as \cite{owers2009} mentions that the X-ray structure of A2069A is elongated in this direction (SE-NW).  
Inspection of the NVSS full resolution image reveals several radio galaxies in the vicinity (between this feature and the cluster core, about 15-20$'$ SE of the cluster core) and a corresponding feature in our residual image, so perhaps this is some diffuse emission from one or more TRGs.  

\subsubsection{Abell 2073}
A2073 ($z=0.1717$) is a poorly studied cluster in the direction of CrB-SC but likely too distant to be dynamically associated.
\cite{flin2006} find evidence of substructure in the optical galaxy distribution, and suggest this may be a dynamically young system.
This cluster has not been studied in detail in X-rays, although it does have an integrated X-ray flux listed in \cite{ebeling2000}, with a corresponding 0.1-2.4~keV X-ray luminosity of $1.9 \times 10^{44}$~erg~s$^{-1}$.
There is no known detection of diffuse radio emission in this cluster.

\begin{figure}[t]
  \centering
  \epsfig{file=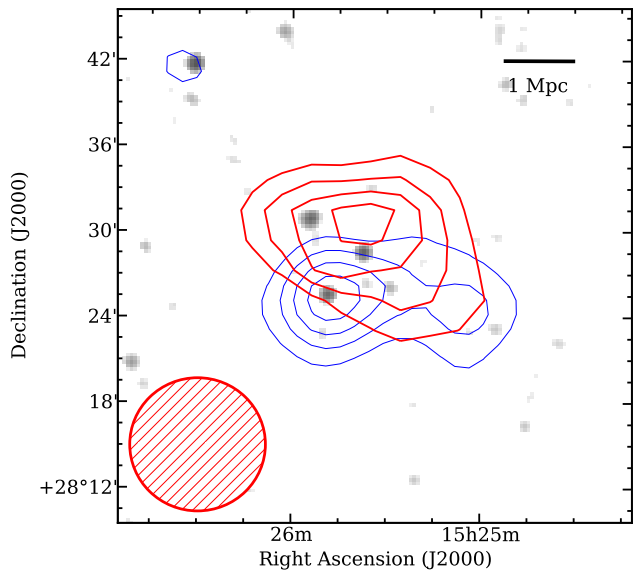,angle=0,width=0.47\textwidth}
  \caption{\textbf{A2073}.  NVSS image (greyscale), clipped at 1.35~mJy~($45''$ beam)$^{-1}$ with overlaid GBT-NVSS 1.4~GHz residuals (red contours) and RASS X-ray image (smoothed with a 5$'$ Gaussian kernel, blue contours).  Radio contours are at $\pm$(3,4,5,...)$\times$$\sigma_{map}$ (negative contours dashed, if present).  The GBT beam is shown in the lower left of the image.}
  \label{fig:a2073}
\end{figure}

We detect an extended radio feature of 21.7~mJy within the 3$\sigma$ contours, with a LLS of 2.5~Mpc; see Figure~\ref{fig:a2073}.
The peak of the radio flux is to the north of the X-ray peak by $\sim$5.5$'$ ($\sim$700~kpc in projection).
Given the peripheral location of the radio structure with respect to the X-ray ICM emission, we classify this as a possible relic source.

\subsubsection{Abell 2142}
A2142 ($z=0.0894$), whose diffuse X-ray emission is elongated SE-NW, was the first cluster where X-ray cold fronts were discovered with \textit{Chandra} (\citealt{markevitch2000}).
The origin(s) of the two \textit{Chandra} cold fronts near the X-ray core has long been debated (e.g., \citealt{markevitch2000}; \citealt{markevitch2007}); it is now widely accepted that gas sloshing in the core is at least partly responsible.
Optical analysis by \cite{owers2011} shows evidence of minor merging activity which they suggest may also play a role in the formation of the cold fronts.  
This merging activity could be inducing turbulence and creating an extended, low surface brightness radio halo.
Recent \textit{XMM-Newton} observations have revealed a third X-ray cold front nearly 1~Mpc SE of the core (\citealt{rossetti2013}), the largest CF to cluster center distance known to date.

A2142 was originally suggested to host a radio halo by \cite{harris1977}.
\cite{giovannini1999} detected diffuse emission in the NVSS with a size of $\sim$350~kpc but reported no flux measurement.
VLA observations by \cite{giovannini2000} yielded a 1.4~GHz flux of 18.3~mJy and an extent of 270~kpc (roughly 3$'$$\times$4$'$) to the diffuse synchrotron emission, located in the core just north of the southern central cold front detected by \textit{Chandra}.
The sub-Mpc extent of the diffuse emission led to its classification as a mini-halo (MH), although A2142 lacks other qualities typically observed in MH systems -- such as a relaxed X-ray ICM morphology and central AGN (e.g., \cite{govoni2009}).

We reduced confusion from nearby large-scale Galactic emission by subtracting a large-scale Gaussian component from the total intensity image, similar to the method for A2319 (described in Section~\ref{sec:a2319}); see Figure~\ref{fig:galSubCompar}.

\begin{figure}[t]
  \centering
  \epsfig{file=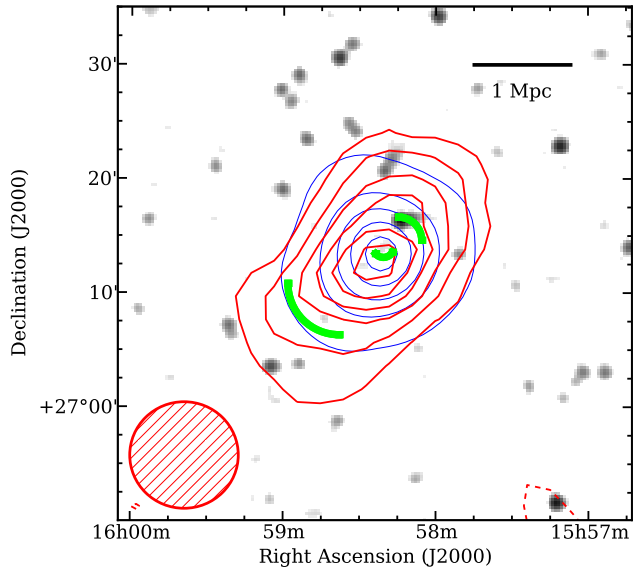,angle=0,width=0.47\textwidth}
  \caption{\textbf{A2142}.  NVSS image (greyscale), clipped at 1.35 mJy (45$''$ beam)$^{-1}$ with overlaid GBT-NVSS (plus reconstructed NVSS halo; see text) 1.4 GHz residuals (red contours) and RASS X-ray image (smoothed with a 5$'$ Gaussian kernel, blue contours) in the A2142 region.  Radio contours are at $\pm$(3,6,9,...)$\times$$\sigma_{map}$ (negative contours dashed, if present).}
  \label{fig:a2142}
\end{figure}

In order to represent the total diffuse emission, we added back in the $\sim$4~mJy of diffuse emission from the NVSS which had been subtracted.
Within the 3$\sigma$ contours we measure a structure elongated in the same SE-NW orientation as the diffuse X-rays, with  LLS of $\sim$2.2~Mpc and $S_{1.4} = 64.0$~mJy; see Figure~\ref{fig:a2142}.
This halo-like structure extends beyond the \textit{XMM-Newton} cold front to the SE, hinting at a possible connection between the two phenomena. 

\subsubsection{Abell 2319}
\label{sec:a2319}
A2319 ($z=0.0559$) is a merging cluster which hosts a well known GRH.
\cite{ghizzardi2010} detected an X-ray brightness discontinuity, which they label a ``merging'' cold front, $\sim$150$''$ SE of the X-ray center using \textit{XMM-Newton}.

\cite{feretti1997} studied A2319 in detail at 20~cm and 90~cm with the VLA and WSRT, providing halo flux and size measurements at multiple frequencies.  
Noting that they likely miss low surface brightness emission due to missing short baselines, they report an integrated flux at 1.4~GHz of 153 mJy and an average surface brightness of $\sim$0.45~$\mu$Jy~arcsec$^{-2}$ after subtraction of discrete sources.  
They report a largest linear extent of 1320~kpc, which yields 1030~kpc in our cosmology.
\cite{giovannini1999} reported a flux of 23~mJy and an extent of 420~kpc (310~kpc in our cosmology), measured from the NVSS image.  
We estimate that roughly 12~mJy of diffuse flux remained in the raw resolution NVSS image after 3$\sigma$ clipping, which would then be absent from our GBT-NVSS residuals.  
This is only $\sim$3\% of the integrated flux in our residual image, and so we have chosen not to attempt a recovery of this lost flux as we did for A2142.

We attempted to mitigate the large scale Galactic emission, which is patchy on $\sim$1\textdegree~ scales.
To do this, we modeled the Galactic emission in the vicinity of A2319 as a very large vertical Gaussian ($3600''\times580''$), convolved with the GBT beam.
This was subtracted from the original GBT-NVSS residual image, and the background level of the residual image was then rezeroed in the cluster vicinity.
The Galactic structure evident in Figures~\ref{fig:subtrComparison} was greatly reduced, and the effective map rms lowered by nearly a factor of two; see also Figure~\ref{fig:galSubCompar}.

\begin{figure*}[t]
  \centering	
	\begin{tabular}{cc}
  \epsfig{file=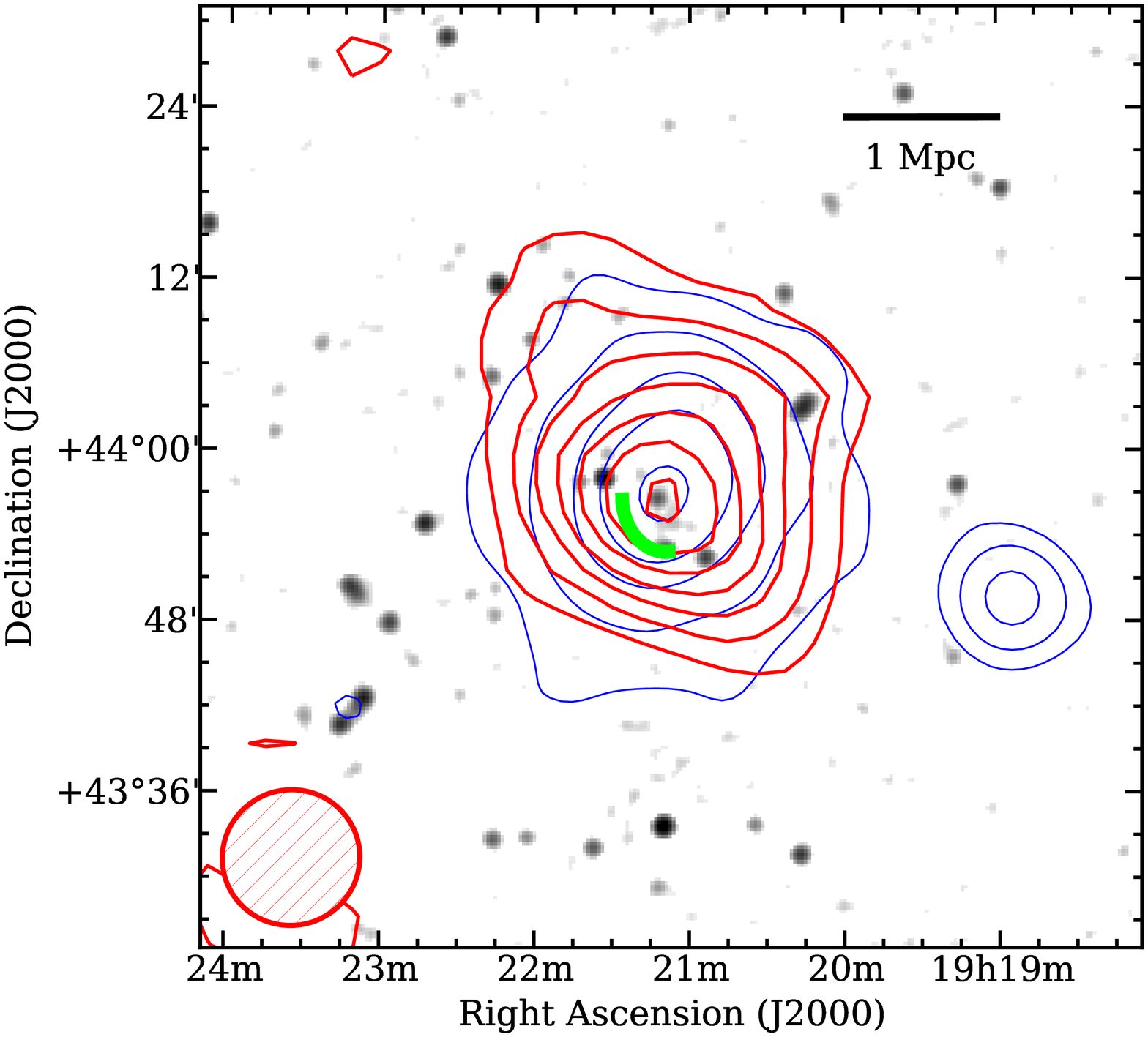,angle=0,width=0.47\textwidth} &
	\epsfig{file=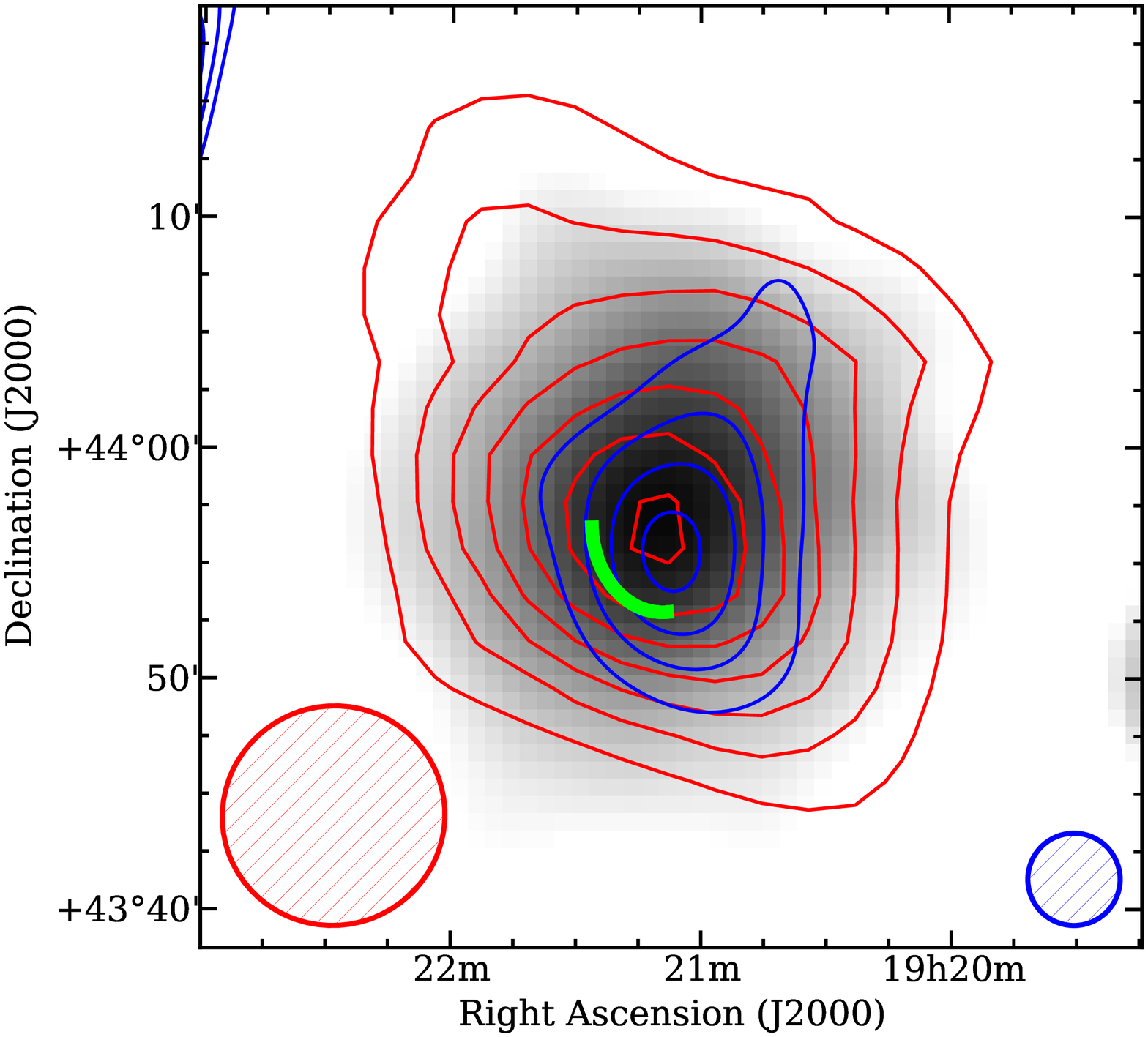,angle=0,width=0.47\textwidth}
	\end{tabular}
  \caption{\textbf{A2319}.  \textit{Left:} NVSS image (greyscale), clipped at 1.35 mJy (45$''$ beam)$^{-1}$ with overlaid GBT-NVSS 1.4 GHz residuals (red contours) and RASS X-ray (smoothed with a 5$'$ Gaussian kernel, blue contours).  Radio contours are at $\pm$(3,6,9,...)$\times$$\sigma_{map}$ (negative contours dashed, if present).  The GBT beam is shown in the lower left of the image.  \textit{Right:} RASS X-ray image (greyscale, convolved with a 5$'$ Gaussian kernel) with overlaid VLA diffuse flux (blue contours) and GBT-NVSS residuals (red contours).  The VLA contours are at (3,5,7,...)$\times$3~mJy~(240$''$ beam)$^{-1}$; the GBT-NVSS contours are the same as in the left panel.}
  \label{fig:a2319}
\end{figure*}

Within the 3$\sigma$ contours we measure a LLS of 2~Mpc and an integrated flux of 328~mJy (see Figure~\ref{fig:a2319}, left panel), more than twice the integrated flux found by \cite{feretti1997}.
We used the 0.4~GHz and 0.6~GHz flux measurements reported in \cite{feretti1997} and our 1.4~GHz flux measurement to derive a single spectral index from 0.4~GHz to 1.4~GHz of $\alpha_{0.4}^{1.4} = 1.2$ with only a slight steepening at 0.6~GHz, contrary to the report of \cite{feretti1997}.

The halo is also detected in our VLA image (Figure~\ref{fig:a2319}, right panel), which shows an extension to the NW in the direction of the X-ray excess found by \cite{feretti1997}.
The residual, diffuse flux within 1000$''$ of the center is 270$\pm$25~mJy, although the noise value is only a rough estimate.  
Approximately 50~mJy of diffuse emission was also removed by the filtering process, yielding a total diffuse flux of approximately 320~mJy, virtually the same as determined by our GBT measurements.
A more precise measurement of the diffuse structure and total flux will be made combining these measurements with higher resolution (C-configuration) data in a future publication. 

\subsubsection{A3744}
A3744 (z=0.0381) is a poorly studied cluster whose diffuse X-ray emission, concentrated near the cluster center, has 0.1-2.4~keV luminosity $L_X \approx 1.8 \times 10^{43}$~erg~s$^{-1}$ (\citealt{bohringer2004}).
The cluster contains two bright tailed radio galaxies, NGC~7016 and NGC~7018, near the cluster center.
\cite{rudnick2009a}, after a reprocessing of the NVSS polarization data, found large scale polarization features coincident with the tailed radio galaxies.
In addition they found a polarized structure $\sim$1.8~Mpc east of the cluster center, $\sim$1.4~Mpc in extent, which seems to have no total intensity counterpart.
They suggest that this could be a peripheral relic structure, a rarity in clusters with such low $L_X$.

Due to the strong subtraction artefacts ($\sim$10\% of the peak value in the residual image) from the bright radio galaxies, we estimated an upper limit to diffuse cluster emission by measuring the integrated fluxes inside an aperture of radius 500~kpc in both the NVSS image (convolved to the GBT beam) and our GBT image (no NVSS subtraction).
Comparison revealed an excess of 1.14~Jy in the GBT image, which is $\sim$11\% of the integrated NVSS flux of 10.5~Jy.
As we are unable to separate the radio galaxy flux from the GBT measurement, we report this as an upper limit to the diffuse flux.

Although we detect the (unresolved) central radio galaxies in polarization, we do not detect -- in either total intensity nor polarization -- a structure corresponding to the peripheral Mpc-scale polarization feature reported in \cite{rudnick2009a}.

\subsection{Residual Contamination From Faint Radio Galaxies}
It is likely that some of the residual flux in our GBT detections comes from faint cluster radio galaxies that are below the NVSS detection limit and, hence, not subtracted in our procedure.
We evaluate this contribution for each halo detection by using cluster radio luminosity functions (RLFs) and, for three of the clusters, deep interferometric data.

The published luminosity functions (\citealt{ledlow1996}; \citealt{miller2003}; \citealt{branchesi2006}) use data from low and high redshift clusters with completeness down to a minimum 1.4~GHz luminosity limit of $\sim$8$\times$10$^{21}$~W~Hz$^{-1}$ (in a $\Lambda$CDM cosmology) and some (incomplete) information down to $\sim$2.5$\times$10$^{21}$~W~Hz$^{-1}$.
To then derive estimates for each GBT cluster, we focused on data from A2255 and 19 other nearby clusters to compute scalings between galaxy numbers, $N_{gal}$, and X-ray luminosities, $L_X$ (e.g., \citealt{bahcall1977}; \citealt{abramopoulos1983}).
We then integrated the radio galaxy luminosity function between 2.5$\times$10$^{21}$~W~Hz$^{-1}$ and the corresponding luminosity limit of the NVSS for that cluster (determined by NVSS 1.35~mJy flux clipping level).
This integrated luminosity $-$ which we boosted by a scaling of $L_X^{0.5}$ $-$ was then converted to 1.4~GHz flux at the cluster distance, yielding an estimate of the residual contamination not accounted for in our NVSS subtraction; we list these values in Table \ref{tab:RGcontamination}.
Due to variations in RLF determination, $N_{gal} - L_X$ scaling, etc., our estimates of residual contamination vary by a factor of $\sim$1.5 for each field (reflected in the table values).
It should be noted that these estimates are biased low because the luminosity
functions that we use are cut off at 2.5$\times$10$^{21}$~W~Hz$^{-1}$.
Although the contribution of elliptical galaxies is found to drop off at these levels (e.g., \citealt{ledlow1996}), we do not know the contribution from starburst galaxies at such low luminosities.
The cases most likely to be problematic are those where the residual flux is only about twice as large as the estimated contamination level: A2061, A2065, and A2069.

To compare these estimates of residual contamination with direct observations, we analyzed deep interferometric images of three of the five clusters with halo detections, A2061, A2065, and A2142.
For A2061 we used a 1.3~GHz GMRT image graciously provided by T. Venturi and S. Bardelli (unpublished), with a characteristic 1$\sigma$ rms sensitivity of $\sim$25~$\mu$Jy beam$^{-1}$ (3$''$$\times$2.5$''$ resolution); for A2065 we reduced and imaged archival 1.4~GHz VLA data (Program AD0375, L-band, C-configuration) with a characteristic 1$\sigma$ rms sensitivity of $\sim$75~$\mu$Jy beam$^{-1}$ (15$''$$\times$14$''$ resolution); for A2142 we used snapshot 1.5~GHz VLA data (Program VLA11B-156; L-band, C-configuration) with a characteristic 1$\sigma$ rms sensitivity of 90~$\mu$Jy beam$^{-1}$ (11$''$ resolution).
These images are nearly an order of magnitude deeper than the NVSS, and are near or below the luminosity limit of the published radio luminosity functions used above.
We used the AIPS task SAD to extract radio galaxy locations and fluxes, and established cluster membership based on published redshifts in the SDSS catalog.
Where available, spectroscopic redshifts were used; for a few sources the photometric redshift errors made their cluster membership uncertain, so we estimate the residual contamination with and without these sources.
To summarize:
\begin{itemize}
 \item For A2061 we find that the NVSS subtraction missed roughly 1.4~mJy of faint RG flux, much lower than the 9-13~mJy estimated by extrapolation of the published RLFs.
This lowers our estimate of the halo flux in A2061 from 16.9~mJy to about 15.4~mJy.

 \item For A2065 we find that the NVSS subtraction missed roughly 5-9~mJy\footnote{The uncertainty comes from the inclusion/exclusion of two RGs whose large photometric redshifts (with large errors) make their cluster membership uncertain.} of faint RG flux.
This is in reasonable agreement with the 11-15~mJy estimated by extrapolation of the published RLFs.
This lowers our estimate of the halo flux in A2065 from 32.9~mJy to 23-28~mJy.

 \item For A2142 we find that the NVSS subtraction missed roughly 8.3~mJy of faint RG flux.
 This is roughly a factor of two lower than the estimation of residual contamination using published RLFs, and lowers our estimate of the halo flux in A2142 from 64.0~mJy to 55.7~mJy.
\end{itemize}

Thus we conclude that the level of contamination for each cluster that is expected from the published RLFs is likely an overestimate, although the unknown contribution from starforming galaxies is an issue that needs further investigation.
This is most problematic for the low surface brightness detections in A2061 and A2069.

\begin{table*}[t]
	\centering
	\small
	\caption{Potential Residual Contamination From Faint Radio Galaxies Below NVSS Limit}
		\begin{tabular}{ l c c c }
		\hline\hline
 Source & NVSS Limit & Calculated Residual$^a$ & Observed Residual$^b$ \\
				& $P_{1.4}$ (W Hz$^{-1}$) & (mJy) & (mJy) \\
		\hline
 A2061 & 2.0$\times$10$^{22}$ & 9-13  & 1.4 \\
 A2065 & 1.7$\times$10$^{22}$ & 11-15 & 5.0-9.0$^c$ \\
 A2069 & 4.7$\times$10$^{22}$ & 8-13  & - \\
 A2142 & 2.8$\times$10$^{22}$ & 14-25 & 8.3 \\
 A2319 & 1.0$\times$10$^{22}$ & 12-17 & - \\
		\hline
		\end{tabular}

	$^a$ Calculated by integrating the RLF between 2.5$\times$10$^{21}$~W~Hz$^{-1}$ and the NVSS 3$\sigma$ clipping level (1.35~mJy), whose equivalent luminosity is given in Column~2. \\
	$^b$ Faint RG flux between 75 (225, 270) $\mu$Jy and 1.35 mJy observed for A2061 (A2065, A2142) at the GMRT (VLA, VLA); equivalent limiting luminosity is 1.1$\times$10$^{21}$ (2.9$\times$10$^{21}$, 5.6$\times$10$^{21}$) W~Hz$^{-1}$. \\
	$^c$ Uncertainty due to two sources with large photometric redshift uncertainties \\
	\label{tab:RGcontamination}
\end{table*}

\subsection{Tentative Classification Summary}
We now summarize the classifications of our total intensity detections.  
Because we are resolution limited these classifications are considered tentative except where already classified by interferometric observations:
\begin{itemize}
  \item \textit{Radio Halos} $-$ A2319 hosts a well known ``classical'' GRH for which we have increased the observed size and luminosity, as well as observed the NW extension.  A2142 has now been detected as a 2~Mpc radio halo structure, in addition to the smaller, possible MH previously seen.  We have made three new 1.4~GHz detections of radio halos: A2065 and A2069 (both entirely new radio halo detections), and A2061, whose radio halo has previously been detected only at 327~MHz.  The diffuse detections in A2061 and A2069 may be multi-structure (e.g., halo+relic) systems.
  
  \item \textit{Radio Relics} $-$ A1367 and A2061 harbor known relics which we have detected in this study.  We have tentatively classified our detections in A2067 and A2073 as relics due to their peripheral location relative to the X-ray emission, although interferometer observations are necessary for proper diagnosis.
  
  \item \textit{Unclassified} $-$ A119, A400, and A3744 each contain an excess of extended emission, some of which is very likely associated with the diffuse tails of the cluster TRGs missed by interferometers.  Due to inadequate resolution, however, halo or relic type emission associated with the ICM of these merging clusters can not be ruled out.  Deep interferometric observations are desired.
\end{itemize}

\section{Analysis and Scaling Relations}
\label{sec:scalingRelations}

In order to bring our detections into context we looked at various aspects of the diffuse radio structures: measured quantities such as physical size and surface brightness, and derived quantities such as radio luminosity and volume-averaged synchrotron emissivity.

\begin{figure*}[t]
\centering
  \begin{tabular}{c c c}
		\epsfig{file=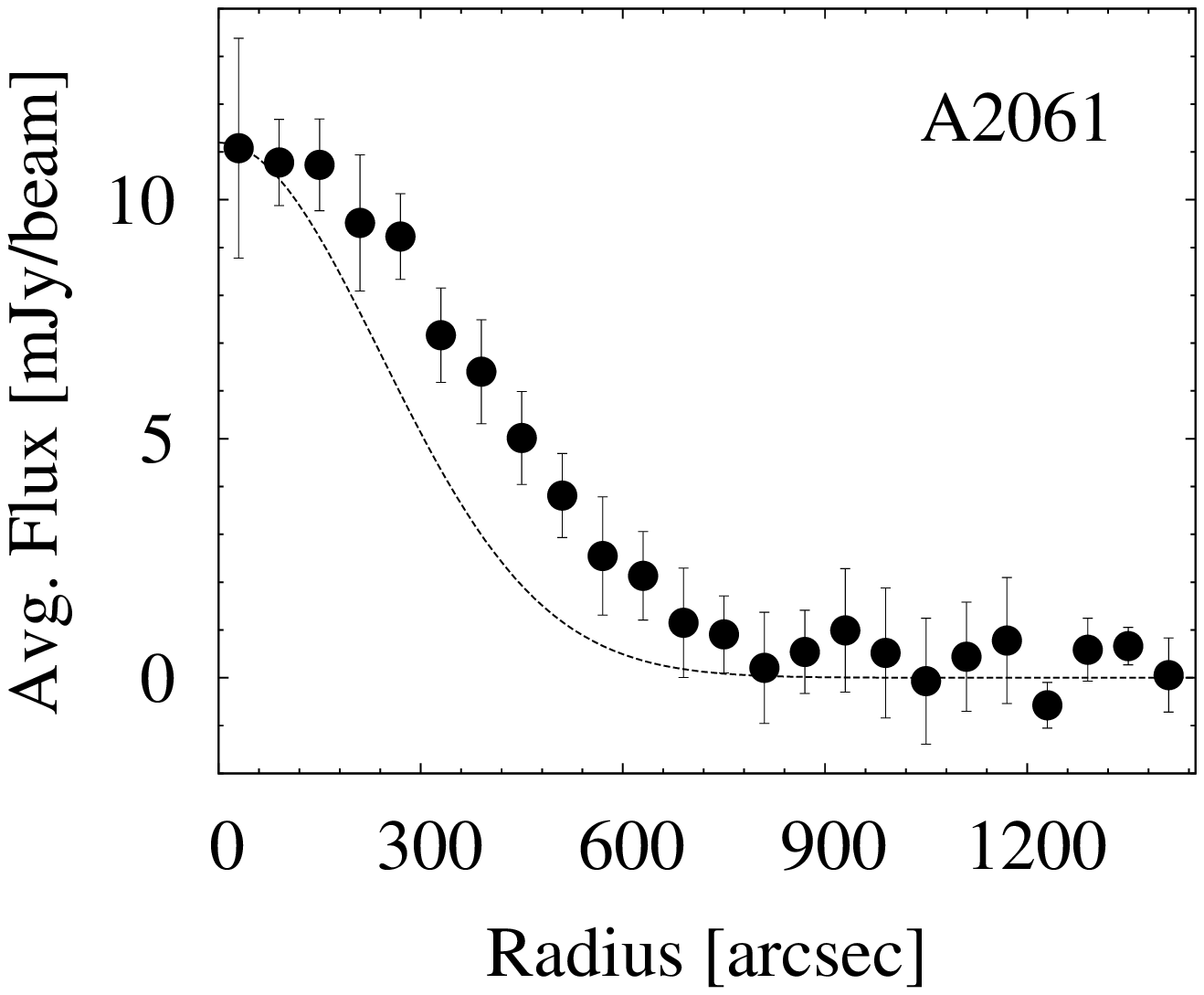,angle=0,width=0.32\textwidth} &
		\epsfig{file=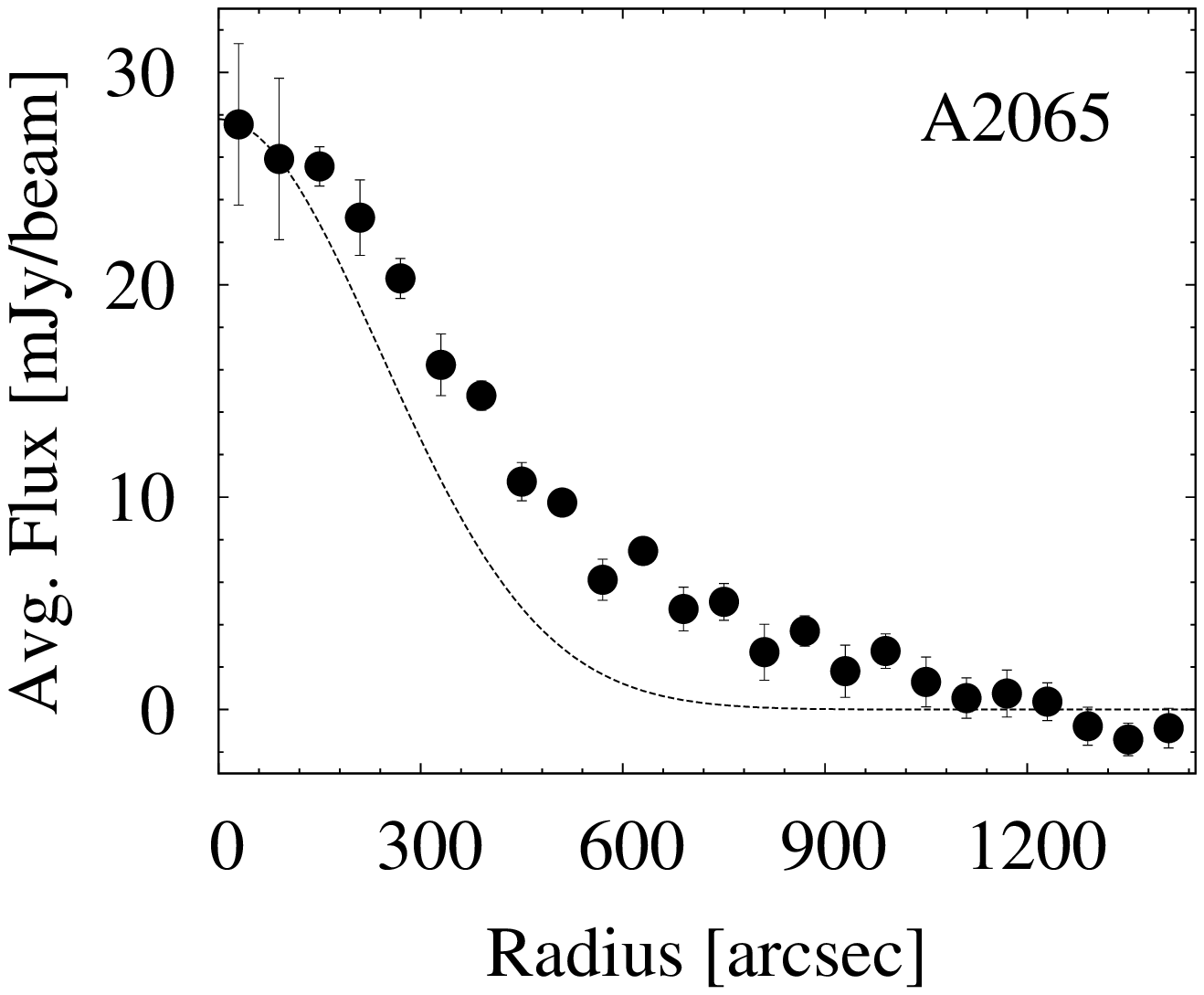,angle=0,width=0.32\textwidth} &
		\epsfig{file=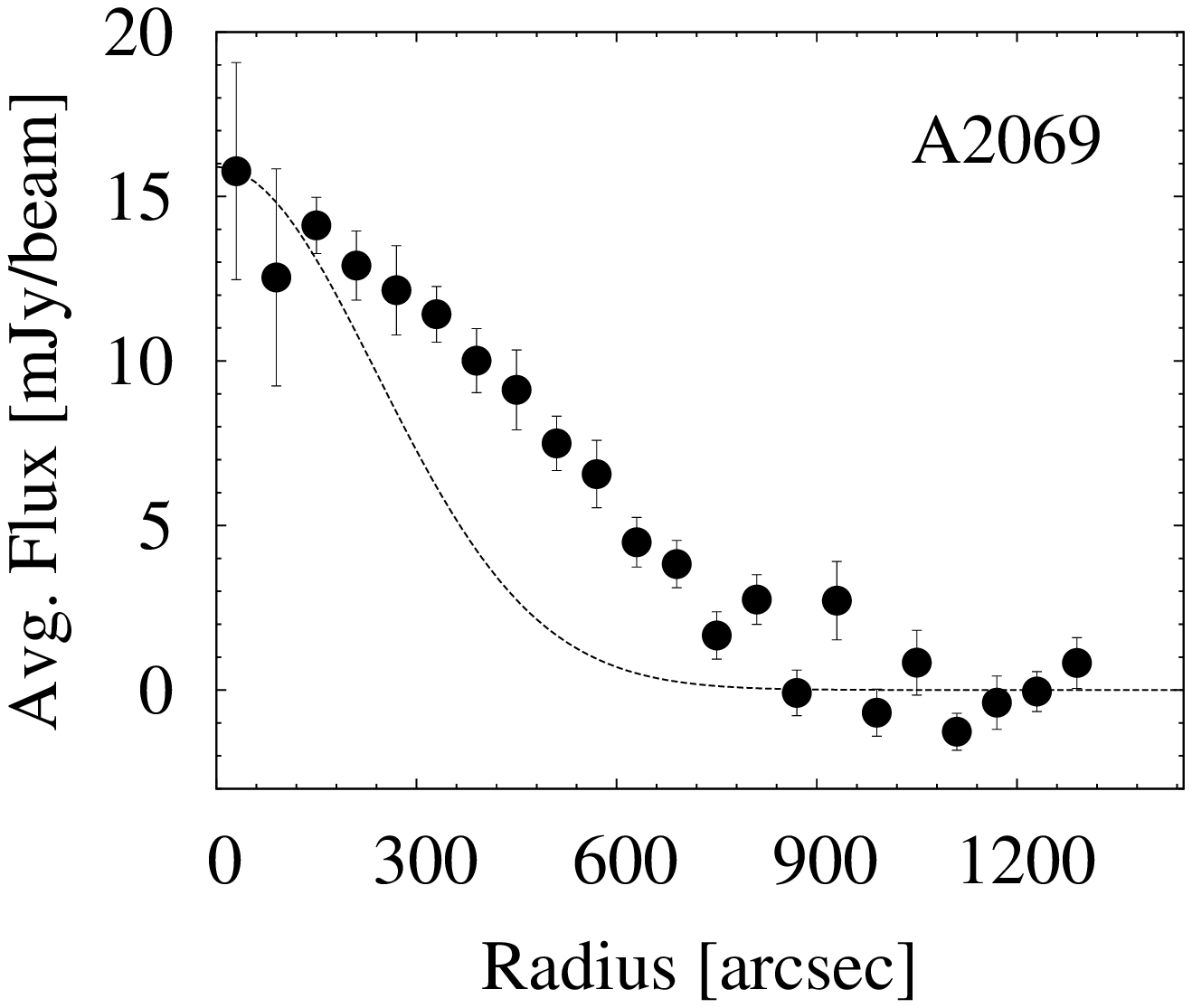,angle=0,width=0.32\textwidth} \\
		\epsfig{file=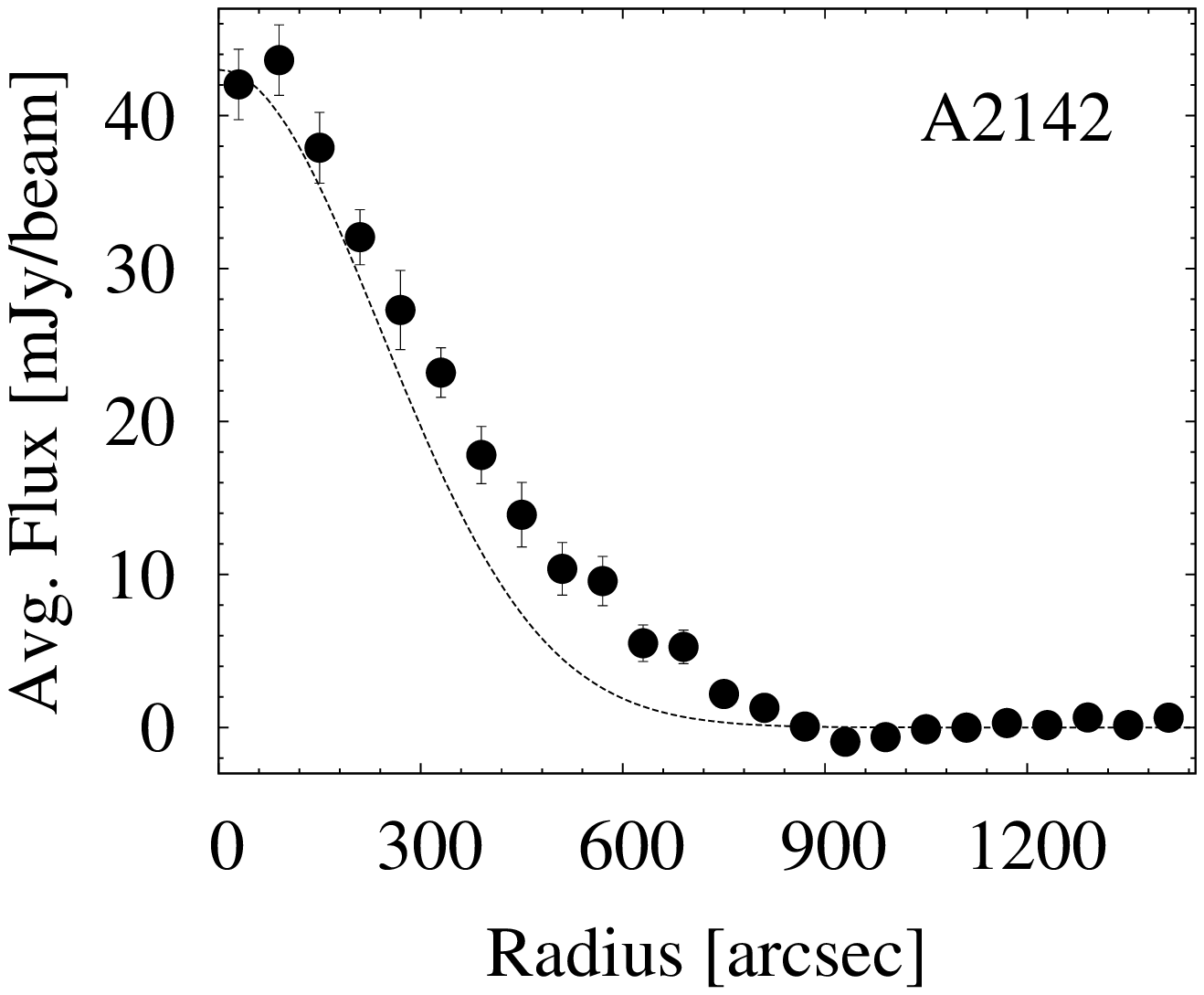,angle=0,width=0.32\textwidth} &
		\epsfig{file=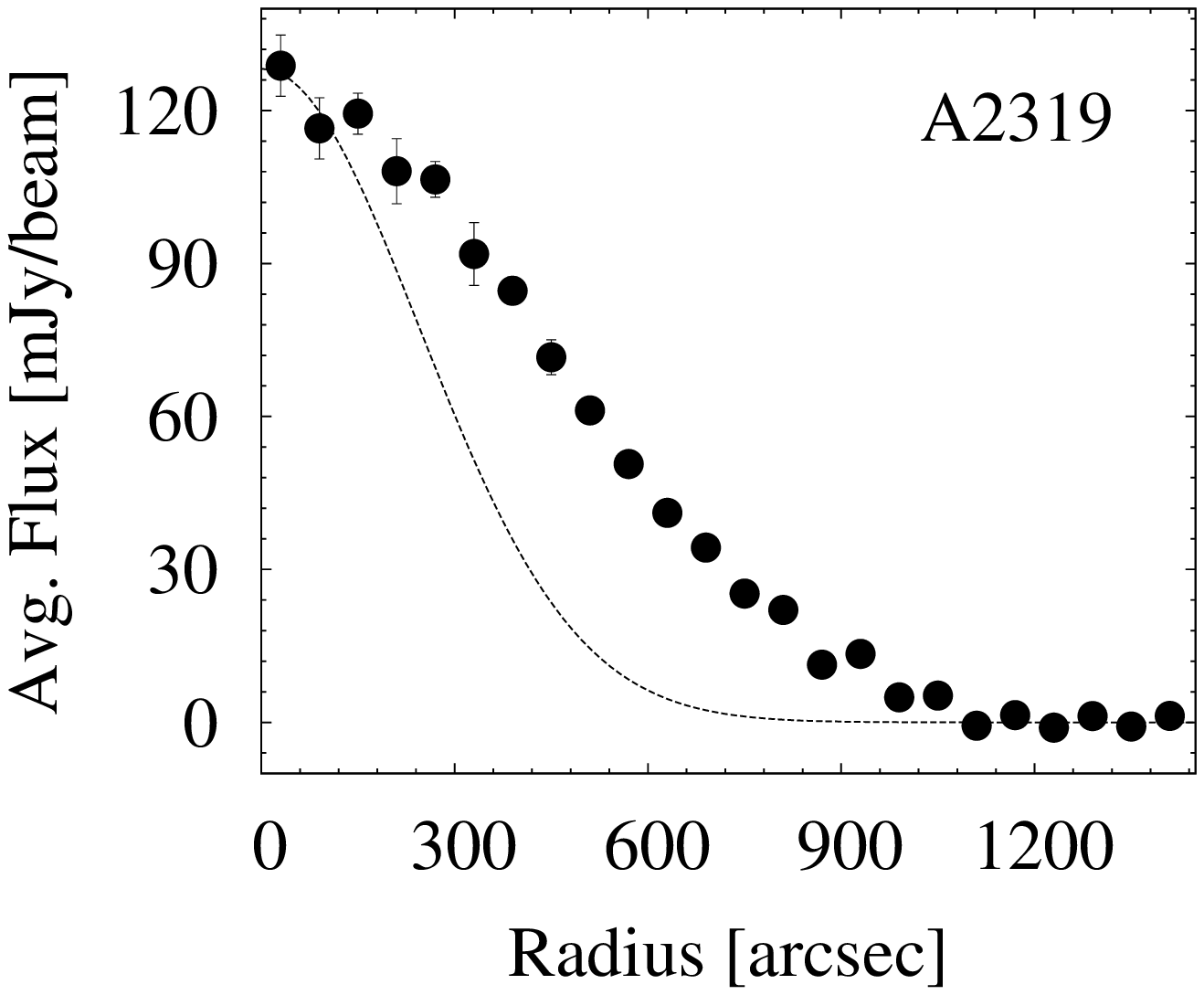,angle=0,width=0.32\textwidth} & \\
  \end{tabular}
	\caption{Azimuthally averaged radial flux profiles of the radio halo detections, as described in the text.  Error bars represent the standard deviation of pixel fluxes within the radial bin.  Note that the assumption of azimuthal symmetry allows the radial sampling to exceed the image pixel scale.  The effective circular Gaussian beam profile is shown as a dashed line.}
	\label{fig:radPros}
\end{figure*}

For the five tentative halo detections, we complemented our image analysis by extracting azimuthally averaged radio flux profiles to measure characteristic sizes and explore the physical mechanisms responsible for cosmic ray production.
The radial profiles were extracted from the GBT residual images using concentric annuli of width 60$''$ centered on the radio centroid, yielding an average flux and standard deviation within the annulus as a function of radius.
Because the point sources were already absent in our GBT residual images, only minimal masking was needed, e.g., in the presence of significant subtraction artefacts (e.g., from a nearby strong point source) or closely separated diffuse structures (e.g., the halo and relic in A2061).
The average radio flux as a function of radius for each of the halo detections is shown in Figure~\ref{fig:radPros}.

Finally, for each of the halo detections we measured the X-ray concentration, a good indicator of merger activity level (e.g., \citealt{cassano2010a}).

\subsection{X-ray/Radio Luminosity Correlation}
\label{sec:p14lx}
\begin{figure}[t]
  \centering
  \epsfig{file=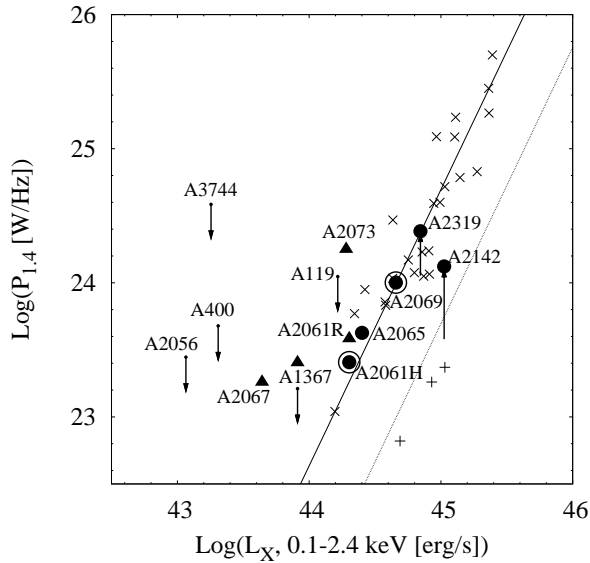,angle=0,width=0.47\textwidth}
  \caption{Plot of $P_{1.4}$ vs. $L_X$ for halo detections.  Xs are halos from the literature.  Crosses are statistical detections of off-state halos from \cite{brown2011b}.  From this work are halos (filled circles) and relics (filled, upward triangles); the (possibly) multi-structure halos A2061H and A2069 are shown as filled circles surrounded by open circles.  Upward arrows connecting previous (literature) $P_{1.4}$ measurements to those of this work are shown for A2142 and A2319.  Values of $P_{1.4}$ for detections are from the integrated 3$\sigma$ contours as described in the text.  Downward solid arrows show upper limits of $P_{1.4}$ for  A119, A400, A1367H, A2056, and A3744.  All of our halo detections are well above the ``off-state'' halo detections of \cite{brown2011b} (approximate upper limit marked by the dashed line).}
  \label{fig:p14lx}
\end{figure}

There is a known correlation between radio and X-ray luminosity for clusters hosting radio halos (e.g., \citealt{cassano2006}; \citealt{brunetti2009}).
For each of our detections and upper limits, we plot in Figure~\ref{fig:p14lx} the 1.4~GHz radio power measured from the 3$\sigma$ contours, $P_{1.4}$, vs. the literature 0.1-2.4~keV X-ray luminosity, $L_X$, along with the clusters in \cite{brunetti2009}.  
To supplement the sample of low $L_X$ clusters where some of our clusters reside we have included clusters from \cite{giovannini2009} and \cite{giacintucci2011}.
We note that the classification of the diffuse radio structure in A1213 as a halo (\citealt{giovannini2009}) is suspect $-$ primarily due to its small size and unusual morphology $-$ so we have omitted it from the plot.  
Our halo detections agree well with the observed correlation; in fact, our revised halo powers for A2142 and A2319 improve their agreement with the correlation predictions.  
It is apparent that the contribution of large scale, low surface brightness emission can be a significant fraction of the total radio halo power for some clusters.  
Our upper limits to $P_{1.4}$ for the three non-detections are well above the observed correlation and provide little insight to the low $L_X$ region they inhabit.

\subsection{Radio Halo Sizes and Emissivities}
\label{sec:SizesAndEmissivities}
The sizes of halos show a correlation with radio luminosity (e.g., \citealt{cassano2007}).
We now explore various measures of halo size which appear in the literature in order to compare our results with previous ones.
A brief description of each characteristic halo size follows, but we refer the reader to Appendix~\ref{sec:appendixA} for further details on the methodologies employed.
Note that for the following measures of size (i.e., LLS, $R_H$, and $R_{85}$), we assume a Gaussian halo profile unless otherwise stated (i.e., $R_e$ for an exponential profile).

\begin{figure}[t]
  \centering
  \epsfig{file=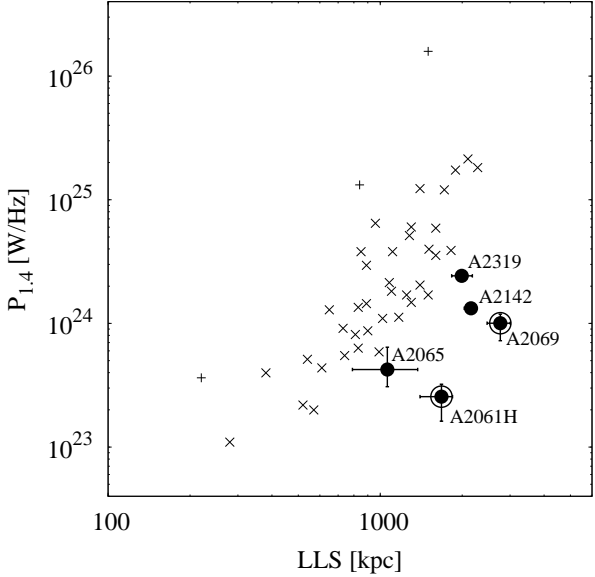,angle=0,width=0.47\textwidth}
  \caption{Plot of $P_{1.4}$ vs. LLS for radio halos.  Our halo detections (filled circles) are shown, along with 42 literature halos (shown as $\times$s, except for three ``peculiar'' objects shown as +s) compiled in \cite{feretti2012}.  The (possibly) multi-structure halos A2061H and A2069 are shown as filled circles surrounded by open circles.  Error bars represent sizes and luminosities determined from 2$\sigma$ and 4$\sigma$ contours.}
  \label{fig:p14lls}
\end{figure}

\begin{figure}[t]
  \centering
  \epsfig{file=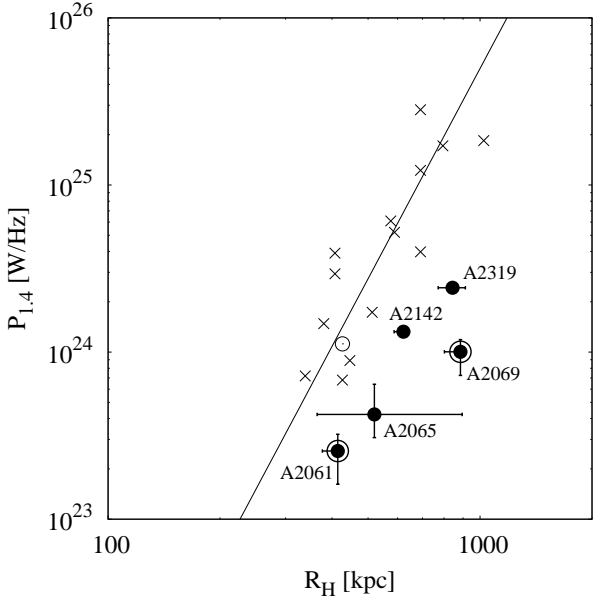,angle=0,width=0.47\textwidth}
  \caption{Plot of $P_{1.4}$ vs. $R_H$ for radio halos, estimated from the 3$\sigma$ contours.  Xs $-$ and an open circle for Abell 2319 $-$ are halos from \cite{cassano2007}.  The halo detections from this work are shown as filled circles.  The (possibly) multi-structure halos A2061H and A2069 are shown as filled circles surrounded by open circles.  Error bars represent sizes and luminosities determined from 2$\sigma$ and 4$\sigma$ contours.  The correlation from \cite{cassano2007} is drawn as a solid line.}
  \label{fig:p14rh}
\end{figure}

\begin{figure}[t]
  \centering
  \epsfig{file=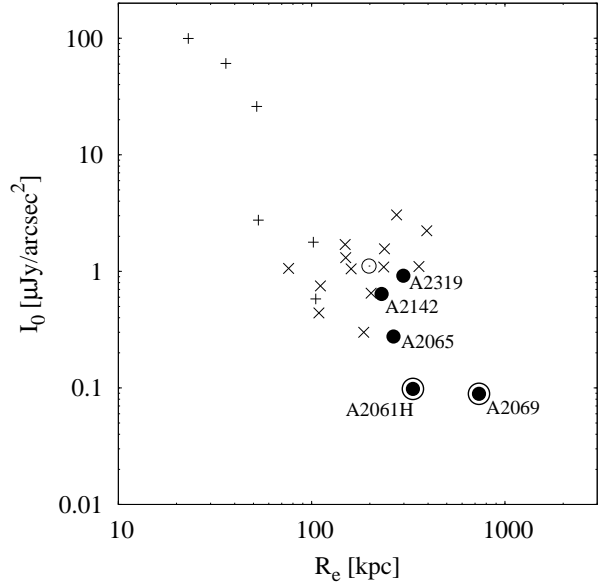,angle=0,width=0.47\textwidth}
  \caption{Plot of central surface brightness vs. e-folding radius (deconvolved quantities; see Table~\ref{tab:surfBrightRadial}) for GBT halo detections at 1.4 GHz, calculated from fits of synthetic radial flux profiles to the observed radial profiles extracted from each radio image (filled circles).  The (possibly) multi-structure halos A2061H and A2069 are shown as filled circles surrounded by open circles.  Also plotted, from \cite{murgia2009}, \cite{murgia2010}, and \cite{vacca2011}, are halos ($\times$s) and  mini-halos (+s) with exponential radial form; their A2319 datum is marked with an open circle.  In general, the volume averaged emissivity will be lower for objects with larger radius and lower central surface brightness.}
  \label{fig:surfBright_radFit}
\end{figure}

\begin{itemize}

  \item The largest linear scale (\textbf{LLS}) $-$ typically measured directly from an image's 3$\sigma$ isophotes $-$ is common in the literature.
  In Figure~\ref{fig:p14lls} we plot $P_{1.4}$ vs. LLS for our halo detections with 42 radio halos compiled from the literature by \cite{feretti2012}.
	We note that LLS can be heavily dependent upon the sensitivity of the observations.

 \item Another measure of halo size can be determined from the radio isophotes, following the technique of \cite{cassano2007}.
 Given the deconvolved major and minor widths of the 3$\sigma$ contours, $L_{maj}$ and $L_{min}$, we can calculate an effective halo radius ($\boldsymbol{R_{H}}$) by  $R_H = \frac{1}{2} \sqrt{L_{maj} \times L_{min}}$.
 We plot our detections with the sample and derived correlation for $P_{1.4}$ vs. $R_H$ from \cite{cassano2007} in Figure~\ref{fig:p14rh}.
 
 \item Another characteristic size of the halo can be estimated by the radius enclosing 85\% of the total integrated flux ($\boldsymbol{R_{85}}$) as in \cite{cassano2007}.
 This measure of the halo extent is less sensitive to map noise than $R_H$.
 For each of the halo detections, we measured the observed $R_{85}$ from the azimuthally averaged radial profiles and deconvolved it from the GBT beam using Equation \ref{eqn:general_deconv}.
 Because $R_{85}$ is measured from the azimuthally averaged radial profile rather than directly from the image, this size measure is less sensitive to image noise than either LLS or $R_H$.
 This can result in a measurement of $R_{85} > \textrm{LLS}/2$ as we see in A2065.
 \cite{cassano2007} found that $R_{85} \approx R_H$ for their halo sample.

 \item The e-folding radius ($\boldsymbol{R_e}$) of a model halo with an assumed exponential radial flux profile can be fit to the observed radial flux profile (see Appendix~\ref{sec:appendixA}) as done by \cite{orru2007}, \cite{murgia2009}, \cite{murgia2010}, and \cite{vacca2011}.
 We plot the model central surface brightness, $I_0$, vs. $R_e$ for each halo in Figure~\ref{fig:surfBright_radFit}, along with the exponential halo and mini-halo results of \cite{murgia2009}, \cite{murgia2010}, and \cite{vacca2011}.
 
\end{itemize}

When considering LLS and $R_H$, each of our halo detections is overly large for its radio luminosity when compared to halos in the literature.
As these measures of size are somewhat dependent upon sensitivity, it is not surprising that the GBT is able to detect low surface brightness emission to large cluster radii.
If we consider $R_e$, however, we see that the likely single-structure halos A2065, A2142, and A2319 are at the high end for halo size when compared to the sample of \cite{murgia2009}, but not anomalously so.
The likely multi-structure halos in A2061 and A2069, however, have very large $R_e$ for their respective central surface brightnesses, hinting at a complex nature of the diffuse emission.
A summary of the various sizes measured for our halo detections is given in Table~\ref{tab:haloRadii}.

\begin{figure}[t]
   \centering
   \epsfig{file=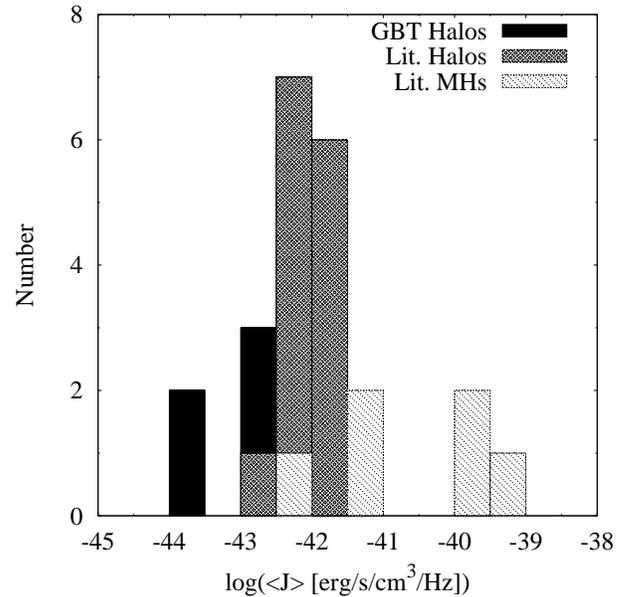,angle=0,width=0.47\textwidth}
   \caption{Histogram of volume averaged synchrotron emissivities for the halo detections of this work, from the results of the exponential flux profile fitting (see Appendix~\ref{sec:appendixA} and Table~\ref{tab:surfBrightRadial}).  Also shown are literature halo and mini-halo emissivities from \cite{murgia2009}, \cite{murgia2010}, and \cite{vacca2011}.  The lowest emissivity bin is populated exclusively by A2061 and A2069.}
   \label{fig:emissHist}
 \end{figure}

\cite{murgia2009} estimate the volume-averaged synchrotron emissivity for the exponential flux profile by assuming all the flux comes from a sphere of radius 3$R_e$:
\begin{equation}
 \langle J_{1.4} \rangle_{e} \approx 7.7 \times 10^{-41} (1+z)^{3+\alpha} \frac{I_0}{R_e}
	\label{eqn:Jmurg}
\end{equation}
where $\langle J_{1.4} \rangle_{e}$ is in erg~s$^{-1}$~cm$^{-3}$~Hz$^{-1}$, $I_0$ is the central surface brightness in units of $\mu$Jy arcsec$^{-2}$, $R_e$ is in kpc, (1+$z$)$^{3+\alpha}$ is a factor which accounts for the $k$-correction and cosmological dimming of surface brightness with redshift, $z$, and $\alpha$ is the spectral index.
We calculate the volume-averaged emissivities for each of our halo detections using the corresponding exponential halo model, adopting $\alpha = 1$ for comparison with the values of \cite{murgia2009}.
The calculated emissivities of our sample, listed in Table~\ref{tab:surfBrightRadial},  are one to two orders of magnitude smaller for most of our halo detections than the bulk of the sample in \cite{murgia2009}.
Figure~\ref{fig:emissHist} displays a histogram of emissivities for our sample, along with halos and mini-halos from \cite{murgia2009} for comparison, to illustrate this result.

\begin{table}[t]
	\centering
	\small
  \caption{Halo Sizes}
  \begin{tabular}{ l c c c c }
  \hline\hline
 Source & LLS & $R_H$ & $R_{85}$ & $R_e$ \\
				& (kpc) & (kpc) &  (kpc) & (kpc) \\
  \hline
A2061H & 1700 & 410 & 680 & 330 \\
A2065  & 1100 & 520 & 870 & 270 \\
A2069  & 2800 & 890 & 1120 & 730 \\
A2142  & 2200 & 620 & 630 & 230 \\
A2319  & 2000 & 840 & 650 & 300 \\
  \hline
  \end{tabular}
  \label{tab:haloRadii}
\end{table}

\subsection{X-ray Concentration}
We have calculated the X-ray concentration parameter (e.g., \citealt{santos2008}; \citealt{cassano2010a}; \citealt{brown2011b}) for each cluster with a tentative halo detection.
Following \cite{cassano2010a}, we adopt the definition
\begin{equation}
 c_X = \frac{S_X(r<100~\textrm{kpc})}{S_X(r<500~\textrm{kpc})},
\end{equation}
where $S_X$ is the integrated X-ray flux within an aperture of specified radius.
The concentration parameter is a measure of the dynamical disturbance of a cluster, and is a useful diagnostic of merger status.
\cite{cassano2010a} set $c_X = 0.2$ as the line between merging (low $c_X$) and non-merging (high $c_X$) clusters, while \cite{brown2011b} define $c_X = 0.156$ as that value.

Where available, we used Rosat PSPC images from \textit{Skyview}\footnote{http://skyview.gsfc.nasa.gov/} because of the higher resolution relative to the RASS images ($15''$/pixel for PSPC vs. $45''$/pixel for RASS).
For A2065, the only halo detection without an available PSPC image on \textit{Skyview}, the RASS image was used.
Background subtraction was performed using statistics from a circular annulus centered on the cluster with inner and outer radii of one and two Mpc, respectively, after masking of X-ray point sources.
A2061 required an irregular annulus in the 1-2~Mpc radial region due to nonuniform exposure in the region.
This issue was only present for $R>1$~Mpc, and so did not affect the statistics of the 100 and 500~kpc apertures.
A polygonal region including only areas of uniform exposure and absent of point sources was used, employing roughly 50\% of the total annulus.
Table~\ref{tab:XrayConcentration} lists the concentration parameters calculated for the five halo detections.

Of the five halo detections, A2061, A2065, and A2069 have $c_X < 0.156$, qualifying them as merging in the scenarios of \cite{cassano2010a} and \cite{brown2011b}.
Note that these are the three clusters with the lowest central radio surface brightnesses (each with $I_0 \lesssim 0.17$~$\mu$Jy~arcsec$^{-2}$) and, hence, the lowest volume averaged synchrotron emissivities among our detections.
By this classification, these are the clusters with the highest level of dynamical disturbance, and thus are most likely to display a radio halo.

The other two clusters, A2142 ($c_X = 0.25$) and A2319 ($c_X = 0.18$) $-$ the only two prior 1.4~GHz halo detections $-$ would be classified in the definition of \cite{brown2011b} as non-merging clusters.
Under the classification of \cite{cassano2010a}, however, A2319 would belong to the merging class.
A2142 was once called a cool core cluster, contributing to the early classification of the diffuse emission found in the NVSS as a mini-halo.
However, although it has a concentration index of 0.253, A2142 is now known to be dynamically active, with multiple cold fronts (e.g., \citealt{rossetti2013}) and minor merging activity at various cluster radii from multiple optical subclusters (\citealt{owers2011}).
A2319 and A2142 are also the two halo detections with the highest central radio surface brightnesses (each with $I_0 \gtrsim 0.38 ~\mu$Jy~arcsec$^{-2}$) and, hence, the highest volume averaged synchrotron emissivities among our detections.
In our small sample, low X-ray concentration correlates with low central radio surface brightness.

\begin{table}[t]
	\centering
	\small
	\caption{X-ray Concentration For Halo Detections}
		\begin{tabular}{ l c }
		\hline\hline
 Source & $c_X$ \\
		\hline
 A2061 & 0.071 \\
 A2065 & 0.11 \\
 A2069 & 0.071 \\
 A2142 & 0.25 \\
 A2319 & 0.18 \\
		\hline
		\end{tabular}
	\label{tab:XrayConcentration}
\end{table}

\section{Discussion}
\label{sec:discussion}
The observed correlation between $P_{1.4}$ and $L_X$ for clusters hosting a GRH is believed to reflect the role of the cluster merger history in the production of cosmic rays, whether by turbulent acceleration or secondary production (or both).
\cite{brunetti2009} used a sample of radio halos from the literature and the GMRT radio halo survey (\citealt{venturi2007}; \citealt{venturi2008}), which contained both radio halo detections and upper limits, to argue in favor of the primary mechanism of CRe generation.
They describe a scenario of evolution to and from an ``on-state'' where the halo radio luminosity obeys the observed correlation, dictated by the recent merger history of the cluster.
For clusters with no recent or major merging activity, an ``off-state'' is expected at which the halo radio power is an order of magnitude or more below the ``on-state'' correlation value.
This notion was supported by the radio power upper limits for non-detections in the GMRT radio halo survey.
\cite{brown2011b} performed a stacking analysis on SUMSS\footnote{The 843~MHz Sydney University Molonglo Sky Survey; http://www.physics.usyd.edu.au/sifa/Main/SUMSS} data, finding statistical detections of off-state halos in X-ray bright ($L_X>5\times10^{44}$~erg~s$^{-1}$) clusters, as predicted by \cite{brunetti2011b}.

Although our five halo detections agree well with the observed radio/X-ray luminosity correlation, they are all larger and fainter than typical radio halos.
As a result, the inferred volume-averaged synchrotron emissivities and corresponding equipartition magnetic field strengths are much lower than values estimated by \cite{murgia2009} for their sample of twelve radio halos.
We note that if the tentative halos in A2061 and A2069 are composed of multiple structures (e.g., halo+relic) blended within the GBT beam, then the emissivities calculated here would be underestimated due to the overestimated filling factor.
However, if they were confirmed by interferometric observations to be continuous radio structures, then they would be \textit{extreme} cases of low emissivity halos.

\subsection{Sensitivity Considerations And Halo Detectability}
Radio halo statistics such as frequency of occurrence are important to address questions about halo bimodality and the physical mechanism(s) of halo generation (e.g., CR acceleration) and evolution (e.g., timescales).

The overall fraction of X-ray clusters hosting radio halos is small, $<$10\% ($\sim$40\%) below (above) $L_X = 10^{44.9}$~erg~s$^{-1}$ (\citealt{cassano2008}).  
However, the detection rate increases significantly if the sample is restricted to dynamically disturbed systems.   
In a statistical study of 32 disturbed clusters with $L_X \gtrsim 10^{44.7}$~erg~s$^{-1}$ and $0.2 \leq z \leq 0.32$, \cite{cassano2010a} find a radio halo in 11 of 15 ($\sim$73\%).  
In this work, we find that this high fraction continues to lower luminosities; for the seven clusters with $L_X < 10^{44.9}$~erg~s$^{-1}$ which do not suffer confusion from bright radio galaxies, we detect four likely halos.
Two of these  are in A2061 and A2069, which are possibly multi-structure (e.g., halo+relic) sources or contaminated by faint radio galaxies not accounted for in our subtraction procedure.
We note that while our sample is neither unbiased nor large, our results are indeed consistent with a large occurrence of halos in merging systems.

Our success at halo detection at low X-ray luminosities (and hence, redshifts) is related to the superb surface brightness sensitivity of the GBT.
For comparison, we can estimate the surface brightness capabilities of the GMRT using sensitivities and beam sizes from \cite{venturi2008}.
The median (best) 3$\sigma$ surface brightness sensitivity for their 23 GMRT observations at 610~MHz is 0.70 (0.095)~$\mu$Jy~arcsec$^{-2}$ at 610~MHz.
With our 1.4~GHz GBT observations, we have achieved a median (best) 3$\sigma$ sensitivity of 0.03 (0.018)~$\mu$Jy~arcsec$^{-2}$.
Assuming a synchrotron spectral index of $\alpha = 1.3$, the equivalent median (best) 3$\sigma$ sensitivity for the GBT at 610~MHz would be 0.088 (0.053)~$\mu$Jy~arcsec$^{-2}$,  approximately eight (two) times better than the surface brightness  sensitivity of the GMRT.
We note that these sensitivity arguments are valid for sources that are well resolved.

In order to have unbiased statistics on clusters at low and high redshifts, it is thus critical to combine interferometric measurements, such as from the GMRT, complemented with single dish studies for low surface brightness and extended halos and the peripheries of brighter systems.

\subsection{Nature of the Diffuse Emission}
How can the radio/X-ray luminosities be in such good agreement with the classical picture of GRHs, while attributes such as size (alternatively, volume averaged emissivity) and surface brightness are not?

We argue here that the scaling of radio to X-ray luminosity for diffuse emission in clusters, tied to the thermal energy budget of the merger, is largely independent of the actual particle acceleration mechanism(s) in effect, e.g., shocks, turbulence, and hadronic collisions.
While the majority of the LSS formation energy budget goes into shock heating and virialisation of the thermal gas, a portion of this energy is extracted into nonthermal components and then transferred into radiation.
As we demonstrate in Appendix~\ref{sec:appendixB}, this cycle leads to a natural correlation between nonthermal and thermal luminosities, where different mechanisms of particle acceleration resulting from mergers lead to similar scalings, e.g., between $P_{1.4}$ and $L_X$, but not necessarily to size.

For relics, the synchrotron emission scales with the shock acceleration (reacceleration) of CRe, which depends on the kinetic energy flux at shocks; this is intimately tied to the cluster dark matter (DM) potential and, hence, the thermal gas energy density and X-ray luminosity.
For turbulent acceleration of CRe, the energy density in turbulence is again tied to that of the thermal gas due to the fact that merger shocks and oscillation of dark matter cores are the most important progenitor of large scale turbulence (not considering AGN jets which may stir the gas).
In the case of secondary CRe resulting from hadronic collisions, the injection rate of CRe depends on the number density of the thermal and relativistic protons; observationally this is tied to the thermal gas density and X-ray luminosity.

We address a few possible scenarios which could account for the large sizes and low surface brightnesses of the halo-like emission regions (with a more detailed discussion of radio $-$ X-ray scaling relationships given in Appendix~\ref{sec:appendixB}):
\begin{itemize}
	\item \textit{Shocks and turbulence} $-$ 
	Simulations suggest that merger shocks and minor merging activity at cluster outskirts are likely to generate considerable peripheral turbulence, thus contributing significant nonthermal pressure and relatively efficient particle acceleration (e.g., \citealt{burns2010}; \citealt{cavaliere2011}; \citealt{vazza2011}).
	For clusters such as A2319 and A2142, where halo emission has been previously seen but not to such large radii, we may be now picking up this extended, low surface brightness halo component previously unobserved in clusters.

	\item \textit{Particle acceleration near X-ray cold fronts} $-$
	Five clusters in our sample contain one or more X-ray cold fronts, four of which host tentative large-scale, halo-like emission.
	The remaining cluster, A1367, exhibits excess large-scale emission in the vicinity of the cold front, but subtraction artefacts from the tailed radio galaxy 3C264 prevent us from distinguishing between halo-like emission and diffuse radio tails. 
	These findings suggest a possible relationship between the physics of cold fronts and particle acceleration on large scales.
	Recent simulations by \cite{zuhone2013} have shown that diffuse, halo-like radio emission may be generated by turbulence at X-ray cold fronts.
	Additionally, simulations (e.g., \citealt{iapichino2008}; \citealt{dursi2008}; \citealt{vazza2011}) have shown significant turbulence to develop in the wake of a moving subcluster during a merger event (potentially the same event responsible for the sloshing of the cluster core and corresponding cold front), possibly giving rise to turbulent particle acceleration.

	\item \textit{Clumpy halos from multiple acceleration regions} $-$
	For less relaxed merging systems such as A2061 and A2069, evidenced by highly elongated radio and X-ray morphology and optical substructure, large scale turbulence believed to power GRHs may not have yet developed. 
	Rather than a single, large-scale GRH structure, in dynamically young systems CR acceleration may occur in multiple regions associated with internal shocks and inhomogeneous turbulent regions producing a ``clumpy'' halo (e.g., \citealt{venturi2013}); this morphology could be blended within the large GBT beam to give the appearance of a smooth Mpc-scale halo.
	As an example, the highly elongated, 2~Mpc radio structure in A2142 may arise from some combination of the multiple cold fronts present and the minor merging activity.

	\item \textit{CR propagation} $-$
	Secondary models of CR acceleration have long suffered from an inability to explain the extent of very large radio halos (e.g., \citealt{brunetti2004}; \citealt{planck2013a}; \citealt{brunetti2012}).
	This is due to the high content of CR protons required by such models at cluster peripheries (where the number density of thermal protons is very small), provided the magnetic field in these regions is not strong.
	However, low surface brightness halos require a smaller CRp energy density and may be powered by secondary CRe if the CR spatial distribution is very broad (\citealt{keshet2010}).
	
\end{itemize}

\subsubsection{The Ultra Steep Spectrum Halo in A2061}
\label{sec:USS}
One prediction of the turbulent reacceleration model for halo generation, which includes turbulent reacceleration of CRe, is that the integrated radio spectrum
of the halo should have a spectral index $\alpha > 1.5$ near GHz frequencies for a subset of clusters.
While they are predicted to be numerous (e.g., \citealt{cassano2010b}; \citealt{cassano2012}), only a handful of these so-called Ultra Steep Spectrum (USS) radio halos have been observed with present radio telescopes.
The prototype USS halo is A521, with $\alpha_{240 MHz}^{1.4 GHz} \sim 1.9$ (\citealt{brunetti2008}; \citealt{dallacasa2009}; \citealt{macario2013}).
The existence of USS is difficult to explain with purely hadronic models.
USS halos are theorized to result from conditions where turbulence is not strong enough to accelerate and maintain CRe at the energies necessary to emit at GHz frequencies, giving rise to a steep synchrotron spectrum.
\cite{cassano2010b} suggest that these USS halos should be more common in clusters with smaller mass and higher redshift, as a result of the smaller energetics of mergers and larger CRe energy losses, respectively.
According to numerical simulations of turbulent acceleration in binary cluster-cluster mergers, USS radio halos are also naturally produced during the initial and final phases of the evolution of a radio halo during a merger event (\citealt{donnert2013}).
A USS halo in A2061 would provide more evidence for theories of turbulent halo generation.
We note that the location of the peripheral relic to the SW of the cluster, suggested by \cite{vanweeren2011} to be a tracer of a shock wave from a prior cluster merger event, suggests a mature state of evolution for the merger event which created the relic-halo configuration.
Sensitive interferometric observations over 100-1000~MHz frequencies are needed to resolve the synchrotron morphology and better measure the integrated radio spectrum.

\subsubsection{A Possible Inter-cluster Filament In A2061-A2067}
Simulations suggest that up to 50\% of the baryonic matter at low redshift may reside in filamentary structures between galaxy clusters in a diffuse 10$^5$-10$^7$~K gas (e.g., \citealt{cen1999}) called the warm-hot intergalactic medium (WHIM).
These filaments are presumed to funnel matter onto clusters through accretion.
So far, however, these inter-cluster filaments have gone largely undetected due to the low temperatures and densities of the WHIM.
\cite{markevitch1998} found evidence for an inter-cluster filament in the A399-A401 system in their ASCA X-ray temperature map, and recent thermal SZ observations with Planck have further strengthened the case by detecting a hot (7~keV), diffuse gas bridge (\citealt{planck2013b}).
That filament detection, however, is at 8$\times$10$^7$~K $-$ similar to the temperature of the gas in the clusters $-$ making this unlikely to be a WHIM detection.
Radio synchrotron emission holds promise for filament detection due to the relatively high efficiency of shocks in these low density regions (e.g., \citealt{brown2011c} for a concise review).
Our tentative detection of an inter-cluster filament linking A2061 and A2067 is tantalizing, and warrants sensitive interferometric observations at $\lesssim$1~GHz.

\section{Conclusions}
\label{sec:conclusions}
We present results of a 1.4~GHz GBT study of twelve $z<0.2$ Abell galaxy clusters with $L_X \sim 10^{43} - 10^{45}$~erg~s$^{-1}$, each exhibiting some evidence of merging activity.
After subtraction of point sources using images from the NVSS, we reach a median  (best) 1$\sigma$ rms sensitivity level of 0.01 (0.006) $\mu$Jy~arcsec$^{-2}$, and find a significant excess of diffuse emission in eleven clusters.
These include two new halos and two new relics, increased sizes and integrated fluxes for known structures, and the tentative detection of an inter-cluster filament and ultra-steep spectrum radio halo in the Abell 2061-2067 system.
Residual contamination from faint galaxies $-$ e.g., starburst galaxies $-$ is unknown, and could contribute significantly to the fainter detections.
Sensitive interferometric observations are necessary to resolve the diffuse radio emission and further address the issue of residual contamination.
We also present a determination of the sensitivity of the NVSS as a function of source size.

While all five of the halo-type detections agree with the observed $P_{1.4} - L_X$ correlation, their sizes are larger than typically observed for their radio luminosities, implying volume averaged synchrotron emissivities 1-2 orders of magnitude below the average of \cite{murgia2009}.
We note, however, that for A2061 and A2069, the emissivities may be underestimated if these are multi-structure sources blended by the GBT beam.

The three new radio halo structures all have in common the presence of optical substructure, either an X-ray cold front or internal shock, and very low X-ray core concentration, implying all are in a relatively early stage of merging.
Due to the poor resolution of the GBT images, it is not possible to distinguish between a low surface brightness GRH and a blend of multiple smaller-scale structures associated with the merging activity (e.g., internal shocks, turbulent patches), which can be blended within the $\sim$9.5$'$ beam.
If a blend of multiple synchrotron structures is present, then it is interesting that the integrated radio luminosity should agree with the observed correlation for GRHs, supporting the idea that the merger energy input to cosmic rays and magnetic fields may be relatively independent of the particular mechanisms of particle acceleration (see Appendix~\ref{sec:appendixB}).

\acknowledgements
We gratefully acknowledge assistance and comments from Rossella Cassano, Dominique Eckert, Gabriele Giovannini, Thomas Jones, Mariachiara Rossetti, Craig Sarazin, and Tiziana Venturi.
Partial support for this work at the University of Minnesota was provided by the U.S. National Science Foundation through grants AST-09008668 and AST-1211595, and through award GSSP 09-0007 from the NRAO.
The GBT and VLA are operated by the National Radio Astronomy Observatory, a facility of the National Science Foundation operated under cooperative agreement by Associated Universities, Inc.
We acknowledge the use of NASA's \textit{SkyView} facility (http://skyview.gsfc.nasa.gov) located at NASA Goddard Space Flight Center.
This research has made use of the NASA/IPAC Extragalactic Database (NED) which is operated by the Jet Propulsion Laboratory, California Institute of Technology, under contract with the National Aeronautics and Space Administration.
This research has made use of the VizieR catalogue access tool, CDS, Strasbourg, France.
This research has made use of the X-Rays Clusters Database (BAX) which is operated by the Laboratoire d'Astrophysique de Tarbes-Toulouse (LATT), under contract with the Centre National d'Etudes Spatiales (CNES).

\pagebreak

\begin{appendix}

\section{Model Fitting To Radial Flux Profiles}
\label{sec:appendixA}
Here we give details of the radial flux profile fitting.
We use a method similar to that of \cite{murgia2009}, \cite{murgia2010}, and \cite{vacca2011}, who investigated the diffuse emission in a sample of twenty halos and mini-halos by fitting model halos of exponential radial form to their radio observations.
They estimated the central (peak) surface brightness, $I_0$, and the e-folding radius (i.e., the radius at which $I(r)$ falls to $I_0/e$), $R_e$, by fitting the model flux profile, convolved with a Gaussian beam, to azimuthally averaged radial profiles of their observed halo emission.
This was done by first constructing a 2-D synthetic image of the exponential halo, convolving with their beam, extracting the synthetic radial profile, and then evaluating the model radial profile with respect to their observations.
By iterating over many ($R_e$,$I_0$) pairs they then found the optimal model which would best fit the observed data.

Similarly, we have modeled the intrinsic flux profiles using models of exponential form with the intent of evaluating the flux profiles at large radii (low surface brightness), where our GBT observations provide an advantage over existing interferometer studies.
We employ the same exponential model as \cite{murgia2009},
\begin{equation}
I(r) = I_{0} e^{-r/R_{e}},
\end{equation}
where $I_{0}$ and $R_{e}$ are the intrinsic (i.e., deconvolved) central surface brightness and e-folding radius, respectively.

For each model we constructed a synthetic ``infinite resolution'' image, convolved it with the GBT beam for that field, extracted the model radial profile, and compared it with the observed radial profile.
Iterating over many ($R_e$,$I_0$) pairs enabled us to find the optimal exponential halo model.

The integrated halo flux is then computed as
\begin{equation}
    S = \int_0^R 2\pi I(r) r dr,
\end{equation}
which is analytically integrable for the simple profile assumed.
For the exponential fit to the radial flux profile we analytically calculate the integrated model flux to a sufficiently large radius; as a check, we verify that this integrated model flux is consistent with the observed cumulative flux from the azimuthally averaged radial profile used for $R_{85}$ determination.
As discussed in Section~\ref{sec:SizesAndEmissivities}, we then use the formalism of \cite{murgia2009} to estimate the volume-averaged synchrotron emissivity for the exponential flux profile by assuming all the flux comes from a sphere of radius 3$R_e$ (see Equation~\ref{eqn:Jmurg}).

\begin{figure*}[t]
  \begin{tabular}{c c c}
		\epsfig{file=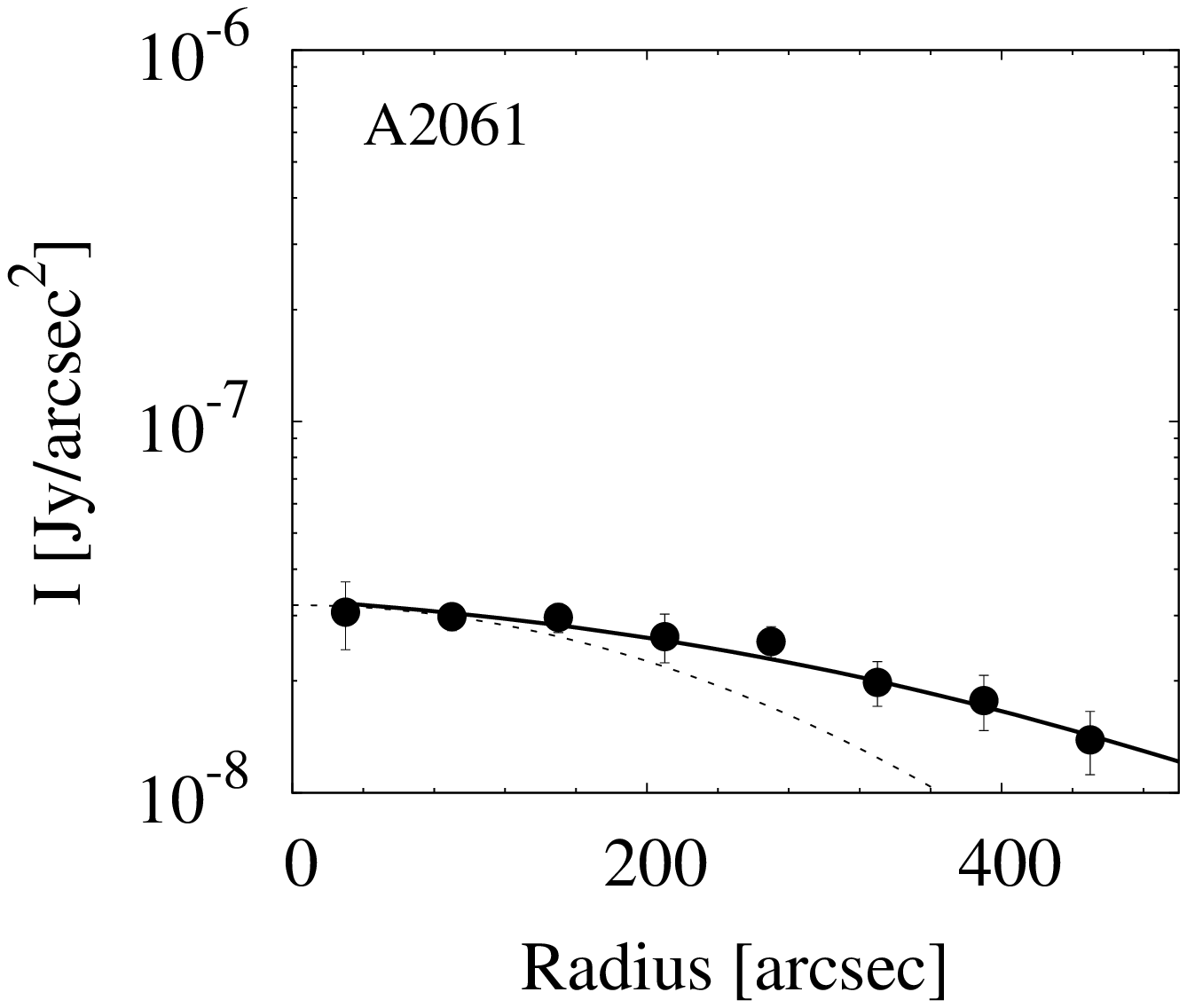,angle=0,width=0.32\textwidth} &
		\epsfig{file=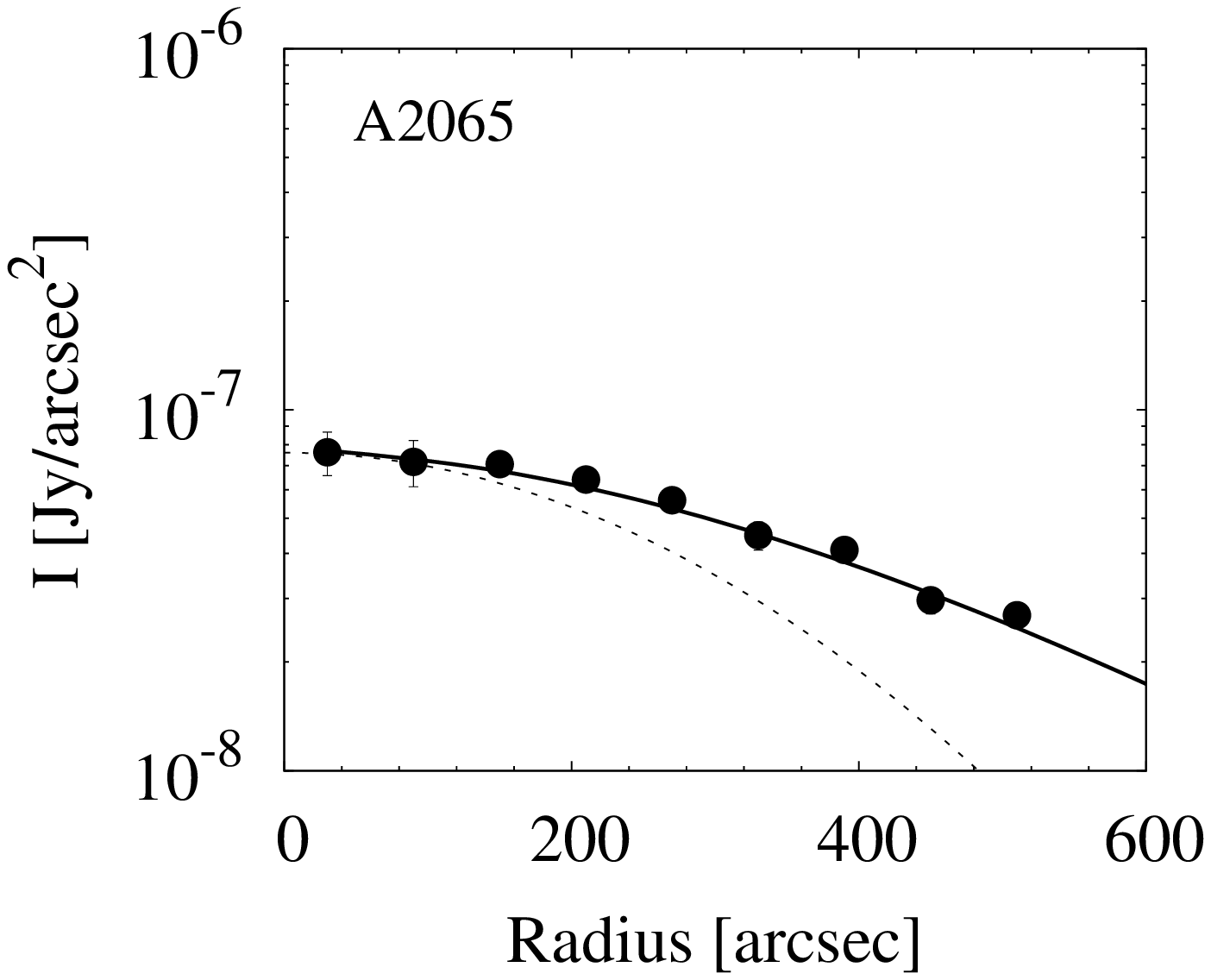,angle=0,width=0.32\textwidth} &
		\epsfig{file=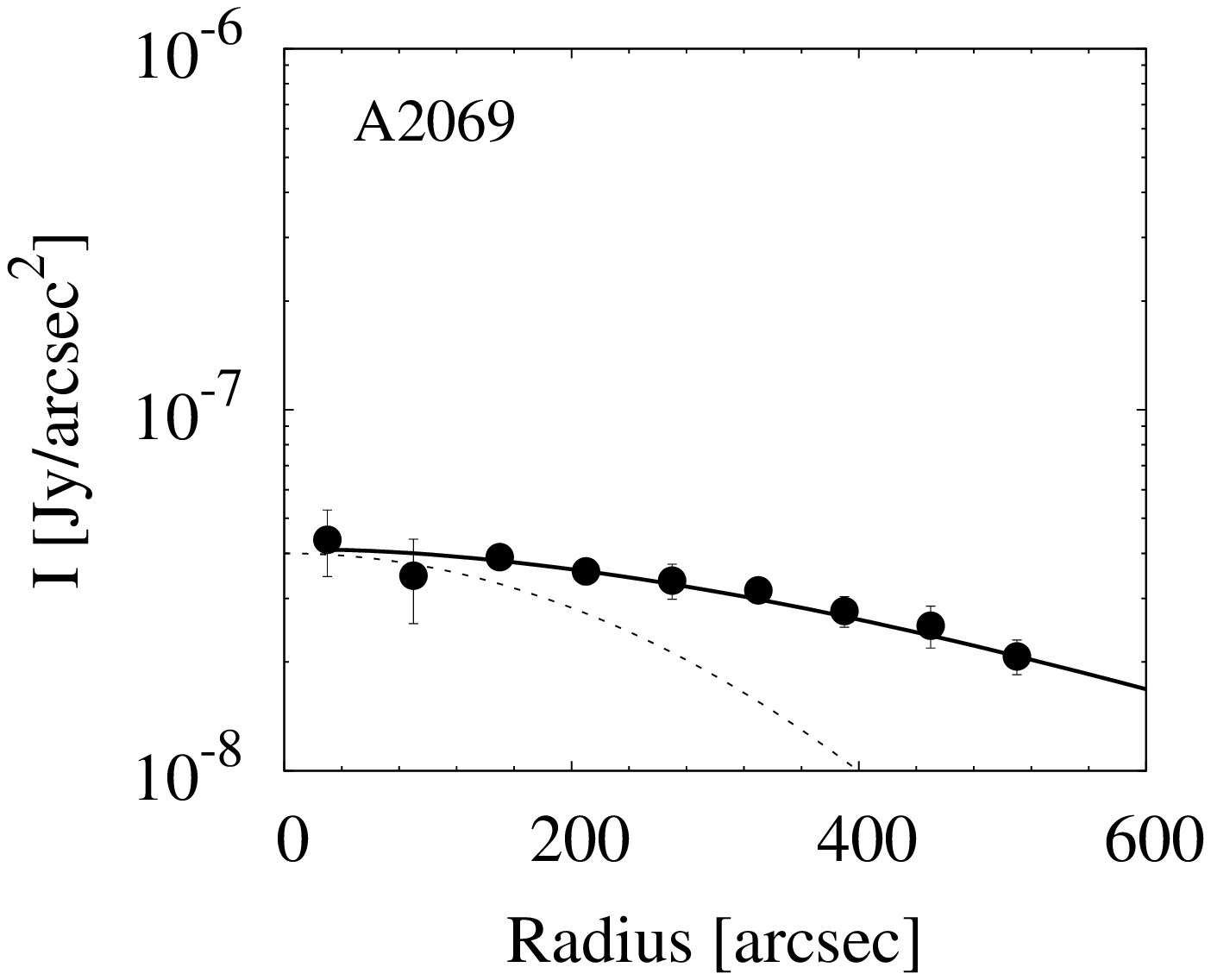,angle=0,width=0.32\textwidth} \\
		\epsfig{file=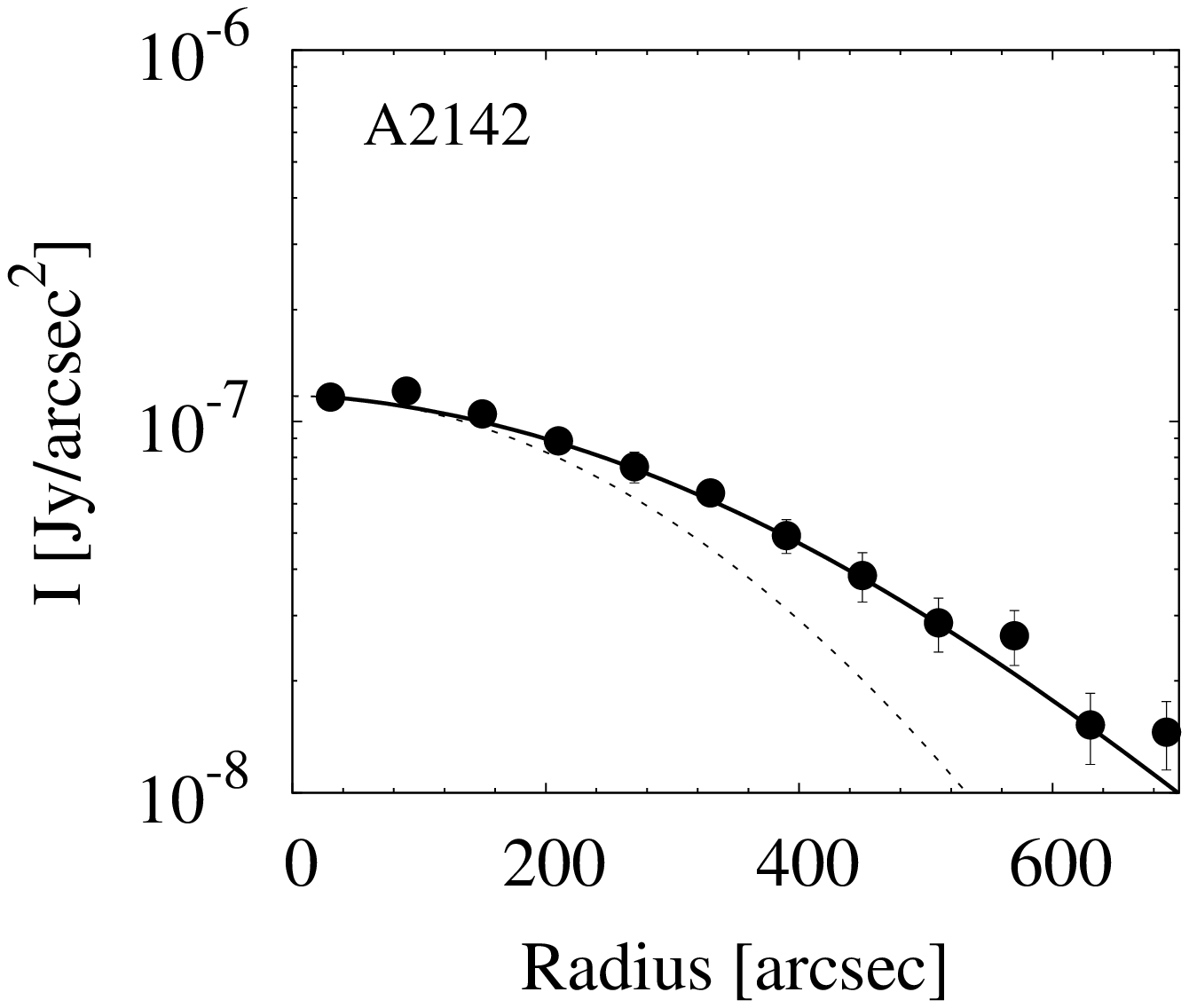,angle=0,width=0.32\textwidth} &
		\epsfig{file=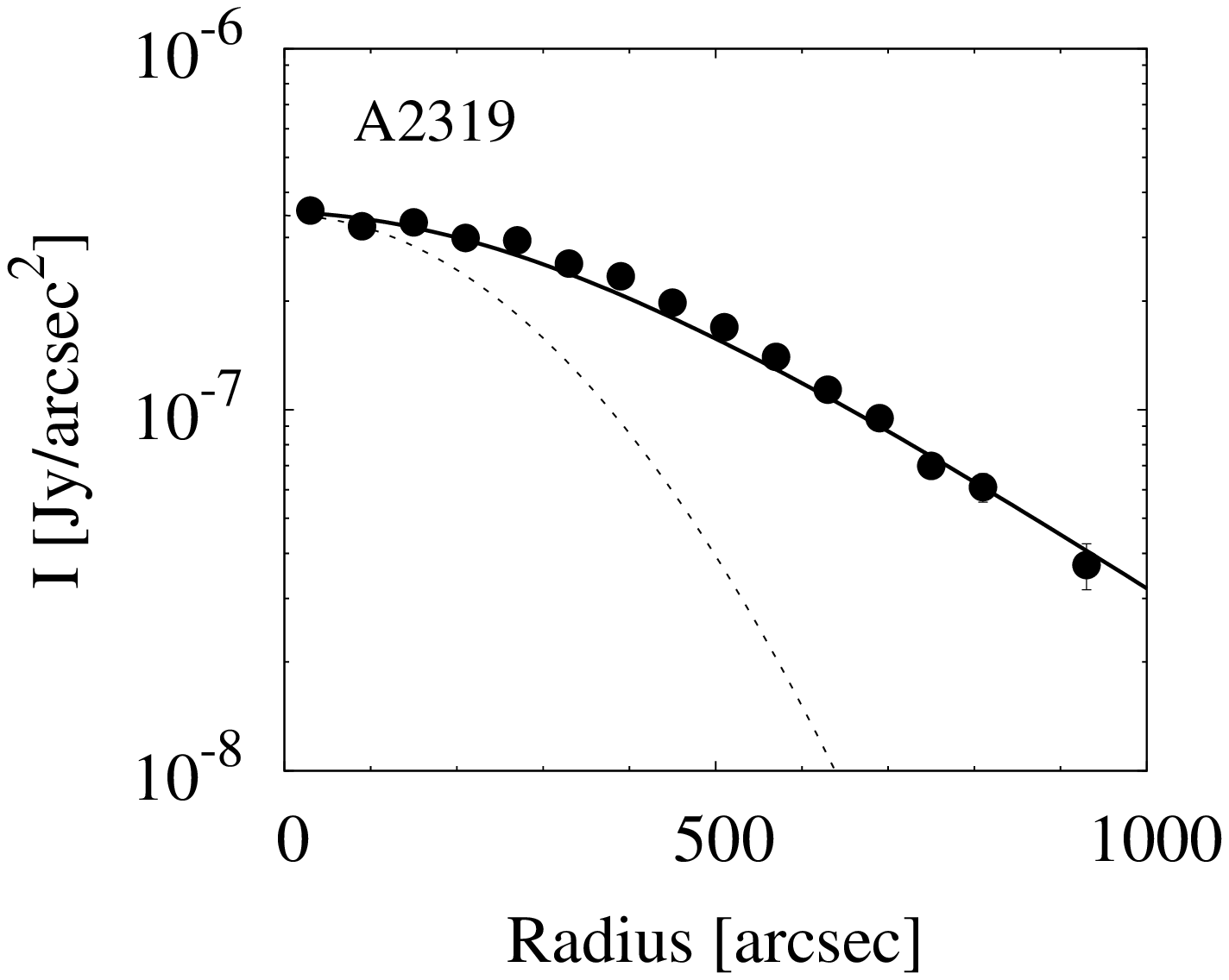,angle=0,width=0.32\textwidth} & \\
  \end{tabular}
	\caption{Fits to the azimuthally averaged brightness profiles of the radio halo detections.  The exponential halo model fits (solid black line), from which $I_0$ and $R_e$ are estimated (see text), are overlaid on the $\geq$2$\sigma$ data (filled circles) used for the profile fitting.  The effective circular Gaussian beam profile is shown (curved, dashed black line) to illustrate the extended nature of the halo detections.  Note that the assumption of azimuthal symmetry allows the radial sampling to exceed the image pixel scale.}
	\label{fig:radProFits}
\end{figure*}

\begin{figure}[t]
   \centering
   \epsfig{file=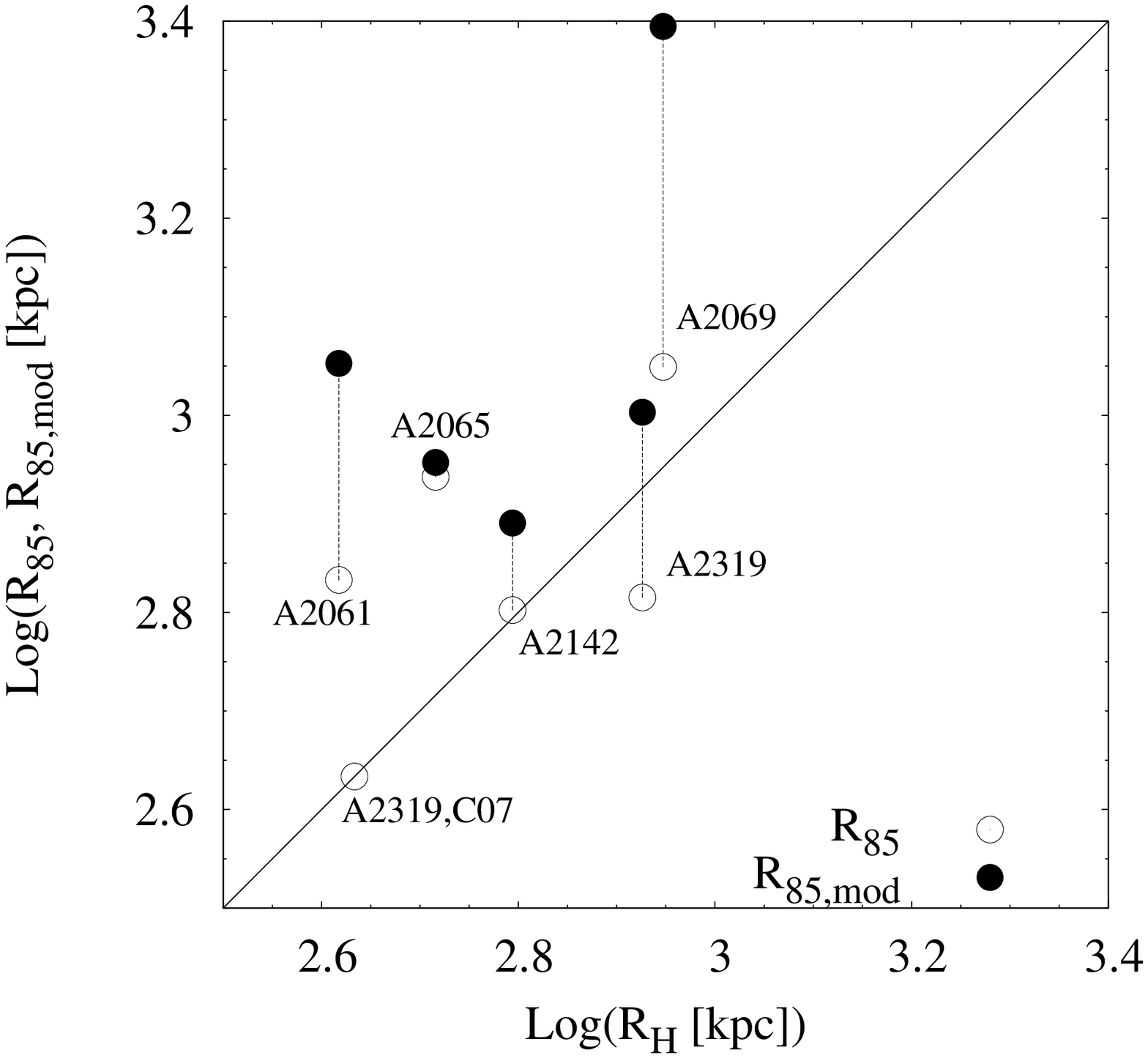,angle=0,width=0.47\textwidth}
   \caption{Plot of various measures of $R_{85}$ vs. $R_{H}$ for our radio halos.  $R_{85}$ (open circle) has been measured directly from the azimuthally averaged profiles, $R_{85,mod} \approx 3.38 R_e$ (filled circle) is from the exponential model fitting to azimuthally averaged radial profiles; the various values for each halo are connected with a dashed line.  The value of $R_H$ for A2319 derived by \cite{cassano2007} (from maps in \cite{feretti1997}) is shown as the solitary open circle, assuming $R_H = R_{85}$.}
   \label{fig:RHvR85s}
 \end{figure}

We illustrate the radial fits in Figure~\ref{fig:radProFits} and report the deconvolved fitting results and calculated emissivities in Table~\ref{tab:surfBrightRadial}; note that for the exponential model, $R_{85,mod} \approx 3.38$$R_e$.
Figure~\ref{fig:RHvR85s} shows $R_{85}$ and $R_{85,mod}$ vs. $R_H$ for the halo detections, illustrating that $R_H \approx R_{85}$ is not universal, contrary to the findings of \cite{cassano2007}.
Additionally, Figure~\ref{fig:RHvR85s} suggests that the assumption of an exponential radial profile tends to estimate a value of $R_{85}$ larger than that observed directly from the extracted radial flux profile.

\begin{table*}[t]
	\centering
	\small
	\caption{Surface brightness characteristics of halo detections from our radial profile fitting}
 \begin{tabular}{ l | *{3}{c} }
  \hline\hline
	Source & $R_{e}$ & $I_{0}$ & $\langle J_{1.4}\rangle_{e}$  \\
	 & (kpc) & ($\mu$Jy~arcsec$^{-2}$) & (erg~s$^{-1}$~cm$^{-3}$~Hz$^{-1}$) \\
	\hline
A2061H & 330 & 9.8$\times$10$^{-2}$ & 3.1$\times$10$^{-44}$ \\
A2065 & 270 & 2.8$\times$10$^{-1}$ & 1.1$\times$10$^{-43}$ \\
A2069 & 730 & 8.9$\times$10$^{-2}$ & 1.4$\times$10$^{-44}$ \\
A2142 & 230 & 6.4$\times$10$^{-1}$ & 3.0$\times$10$^{-43}$ \\
A2319 & 300 & 9.2$\times$10$^{-1}$ & 2.9$\times$10$^{-43}$ \\
	\hline
 \end{tabular}
 \label{tab:surfBrightRadial}
\end{table*}

\section{Radio - X-ray Luminosity Scaling For Particle Acceleration}
\label{sec:appendixB}
If gravity provides the energy budget for the nonthermal components in galaxy clusters, the luminosity of the diffuse emission should correlate with the thermal energy budget of the hosting clusters and this is expected independently of the precise mechanism that accelerates particles.
The main concept here is that the energy budget of large scale structure formation is mainly channeled into the heating/virialisation of the hot gas, but that a fraction of this energy is also extracted into nonthermal components and then transferred into radiation.
As we will show by considering the mechanisms of particle acceleration that are presently proposed for the origin of radio halos and relics, this cycle leads to a natural correlation between nonthermal and thermal luminosities (or cluster masses) that are not very different in slope and potentially also similar in normalization.
In this case, if unresolved, relics and halos in a cluster may mix and contribute with similar importance (within an order of magnitude) to the observed radio emission.
This is especially important for our paper because our observations do not have good spatial resolution. 
For example it should not be surprising to see that clusters like A2061 and A2069, whose diffuse radio emission is probably due to multiple components (e.g, halo+relic) not properly resolved by the GBT, follow the same correlation as classical radio halos (see Figure~\ref{fig:p14lx}).

In the following we will adopt a simplified approach to derive expected thermal $-$ nonthermal scaling relations resulting from different mechanisms that convert a fraction of the energy dissipated during mergers into particle acceleration.

\subsection{Merger Shocks}
The thermal energy budget in clusters is determined by the dark matter (DM) potential well.
The gas traces the DM potential and is heated up to the virial temperature which is proportional to $G M_v/R_v$, where $M_v$ and $R_v$ are the virial mass and radius of the cluster, respectively.
The gas reaches this temperature falling into the potential well where it is heated by shocks that form during the mass assembly of clusters.
The energy flux through each shock is
\begin{equation}
\frac{dE_{sh}}{dt} \sim \rho_{gas} v_{sh}^3 S ,
\end{equation}
where $\rho_{gas}$ is the gas density, $v_{sh}$ is the shock velocity, and $S$ is the shock surface area.
Essentially a large fraction of this energy flux times the number of shocks determine the cluster temperature and gas energy density, yet a sizable fraction of this energy can also be converted into the acceleration (or reacceleration) of CRp and CRe (e.g., \citealt{ryu2003}).

If we assume that radio relics are due to CRe accelerated at shock waves (the same shocks that heat up the gas), the energy flux generated in these CRe is
\begin{equation}
\frac{dE_{CRe}}{dt} \sim \eta_e \frac{dE_{sh}}{dt},
\end{equation}
where $\eta_e$ is the acceleration efficiency of CRe (much smaller, typically $\leq 1\%$, than that in CRp); the acceleration efficiency depends on the shock Mach number and on the presence of pre-existing CRs in the upstream region (e.g., \citealt{kang2009}), yet here we do not elaborate on this point.
Under stationary conditions all of the energy injected in ultra-relativistic CRe is converted into synchrotron ($B^2$) and inverse Compton ($B_{CMB}^2$) radiation. 
In this case the synchrotron luminosity emitted by the downstream region\footnote{If we are observing at a frequency $\nu_o$ this region will extend to a distance from the shock surface $\sim$$v_d \tau_{age}(\nu_o)$ where $v_d$ is the downstream velocity of the flow (in the reference frame of the shock) and $\tau_{age}$ is the lifetime of the CRe emitting at frequency $\nu_o$ in the downstream region.} will be 
\begin{equation}
L_{syn} \sim \frac{dE_{CRe}}{dt}  \frac{B^2}{B^2 + B_{CMB}^2}.
\end{equation}
In the case of strong $B$, essentially all of the energy will go into synchrotron radiation,
\begin{equation}
  L_{syn} \sim \frac{dE_{CRe}}{dt},
\end{equation}
while for weak $B$ only a small fraction of this energy will be in synchrotron,
\begin{equation}
L_{syn} \sim \frac{dE_{CRe}}{dt} \frac{B^2}{B_{CMB}^2}.
\end{equation}

The velocity of the shocks from mergers is essentially of the order of the freefall velocity (e.g., \citealt{sarazin2002}),
\begin{equation}
v_{sh} \sim \sqrt{G M_v/R_v},
\end{equation}
and as a first approximation we can assume that the shock surface area scales with $S \propto R_v^2$ (proportional to the cluster ``surface'') and $\rho_{gas} \propto f_b$ (for $f_b$, the baryon fraction, we use virial matter density that is constant, i.e., independent of cluster mass).

So one has:
\begin{equation}
\begin{split}
L_{syn} & \propto \eta_e f_b  \left(\frac{G M_v}{R_v}\right)^{3/2} R_v^2 \frac{B^2}{B^2+B_{CMB}^2} \\
				& \propto \eta_e f_b M_v^{5/3} \frac{B^2}{B^2+B_{CMB}^2}
\end{split}
\label{eq:shockshock}
\end{equation}
because $M_v \propto R_v^3$. 
Although here we have adopted a very basic approach, we note that Equation~\ref{eq:shockshock} is equivalent to Equation~32 in \cite{hoeft2007} in the case of strong shock approximation (and taking into account that Equation~\ref{eq:shockshock} refers to the bolometric synchrotron luminosity).

The magnetic field in galaxy clusters is likely connected with their thermal energy budget, $B^2 \propto \rho_{gas} T \propto f_b T \propto f_b M^{2/3}$, so in any case (i.e., weak or strong $B$) the synchrotron luminosity connected to CRe acceleration at shocks (e.g., relics) should depend on cluster $M_v$ or alternatively on other ``thermal'' observables.
One of the most common observables used as mass proxy is the X-ray luminosity of the cluster; note that throughout this appendix we will adopt the following notation for X-ray luminosities:  bolometric luminosity, $L_X$, or in the 0.1-2.4~keV band, $L_{[0.1-2.4]}$.

Because $L_X \propto T^{\alpha_T}$ (where $\alpha_T \sim 3$, e.g., \citealt{arnaud1999}; \citealt{pratt2009}), the expected scaling is
\begin{equation}
\begin{split}
L_{syn} & \propto \eta_e  T^{5/2} f_b \frac{B^2}{B^2+B_{CMB}^2} \\
				& \propto \eta_e f_b L_X^{5/2\alpha_T} \frac{B^2}{B^2+B_{CMB}^2}
\end{split}
\end{equation}
that in strong fields is
\begin{equation}
L_{syn} \propto \eta_e f_b L_X^{5/6}
\end{equation}
and in weak fields is
\begin{equation}
L_{syn} \propto \eta_e f_b^2 L_X^{7/6}.
\label{eq:shockweakB}
\end{equation}
Equation~\ref{eq:shockweakB} is equivalent to Equation~16 in \cite{kempner2001} if we assume a strong shock ($\alpha=-1$ in their equation).

In our paper we use the X-ray luminosity in the 0.1-2.4~keV band. 
For hot clusters this approximately scales with $L_X$ as $L_{[0.1-2.4]} \propto L_X^{1-0.6/\alpha_T}$ (e.g., \citealt{kushnir2009}) implying scalings in the form $L_{syn} \propto L_{[0.1-2.4]}^{ \frac{5/2}{\alpha_T -0.6} }$ and $L_{syn} \propto L_{[0.1-2.4]}^{ \frac{7/2}{\alpha_T -0.6} }$ in the case of strong and weak magnetic fields, respectively ($\alpha_T \sim 3$).
Note that slightly steeper scalings can be induced if we consider that the baryon fraction $f_b$ weakly depends on cluster temperature (e.g., \citealt{dai2010}).

\subsection{Turbulence}
One possibility for the origin of radio halos is that CRe are reaccelerated by turbulence in the ICM that can be generated during cluster mergers. 
In this case scaling relations between thermal and nonthermal properties depend on the scaling of turbulent energy vs. thermal energy.
Regardless of the details, as a first approximation it is natural to assume that the energy injection rate of turbulent motions scales with the thermal energy density divided by a reference timescale, such as the cluster-cluster crossing time (e.g., \citealt{cassano2005}). 
In the following we closely follow the derivation from \cite{cassano2007}.

Under the hypothesis given above, the total energy flux in turbulence is
\begin{equation}
\frac{dE_{tu}}{dt} \propto V_{tu} \frac{\rho f_b T}{\tau_{cross}},
\end{equation}
where $V_{tu}$ is the volume where turbulence ``illuminates'' the radio halo (the volume of the halo), $\rho f_b$ is the gas density within this volume, and $\tau_{cross} \sim R_v / v_i$ is a constant because the cluster-cluster impact velocity is $v_i \propto \sqrt{M_v/R_v}$.

A fraction of the turbulent energy flux is channeled into CRs (CRe and CRp); this fraction is the ratio of the damping rates of turbulence due to the interaction with a given species of particles divided by the total damping of the turbulence.
For electrons the fraction is 
\begin{equation}
f_{CRe} = \frac{\Gamma_{CRe}}{\Gamma_{th} + \Gamma_{CRp} + \Gamma_{CRe}} \propto X_e  \sqrt{T},
\end{equation}
where $X_e$ is the ratio of the CRe to thermal energy densities.
The energy flux that will be channeled into CRe is
\begin{equation}
\frac{dE_{CRe}}{dt} = f_{CRe} \frac{dE_{tu}}{dt}.
\end{equation}
Under stationary conditions this energy flux will be essentially converted into synchrotron and inverse Compton radiation:
\begin{equation}
L_{syn} \propto f_{CRe} M_{tu} f_b T^{3/2} \frac{B^2}{B^2 + B_{CMB}^2}
\end{equation}
where $M_{tu}=\rho V_{tu}$ is the cluster mass in the region of the halo. 
This can be calculated assuming hydrostatic equilibrium (isothermal gas; e.g., \citealt{sarazin1986}):
\begin{equation}
M_{tu} = \frac{3 k_B T R_{tu}^3 \beta}{\mu m_p G}
(R_{tu}^2 +r_c^2)^{-1},
\end{equation}
where $\beta$ and $r_c$ are the beta-model exponent and the core radius of the cluster, respectively.
Based on the analysis of \cite{cassano2007} for 14 GRHs, $M_{tu} \propto T^a$, where $a \simeq 2-3$, implying a scaling between synchrotron and X-ray luminosities in the form (using $L_X \propto T^{\alpha_T}$, with $\alpha_T \sim 3$, as in the previous subsection):
\begin{equation}
L_{syn} \propto X_e f_b L_X^{ \frac{ 3/2 + a }{\alpha_T} } \frac{B^2}{B^2 + B_{CMB}^2}.
\label{eq:turboturbo}
\end{equation}
Under the assumption (as in the previous section) that the magnetic field energy density scales with the thermal energy density, Equation~\ref{eq:turboturbo} implies scalings in the form $L_{syn} \propto X_e f_b L_X^{(3/2+a)/\alpha_T}$ and $L_{syn} \propto X_e f_b^2 L_X^{(5/2+a)/\alpha_T}$ for strong and weak fields, respectively.
Similarly (converting $L_X$ into $L_{[0.1-2.4]}$ as in the previous subsection), the model predicts $L_{syn} \propto X_e f_b L_X^{(3/2+a)/(\alpha_T-0.6)}$ and $L_{syn} \propto X_e f_b^2 L_X^{(5/2+a)/(\alpha_T-0.6)}$ for strong and weak fields, respectively ($\alpha_T \sim 3$).

\subsection{Secondary CRe}
An additional source of CRe in radio halos is provided by hadronic (CRp-p) collisions in the ICM.
The energy injection rate of secondaries is
\begin{equation}
\frac{dE_{CRe}}{d t} \propto n_{th} \epsilon_{CRp} R_v^3 \sigma_{pp},
\end{equation}
where $\sigma_{pp}$ is the CRp-p cross-section, $n_{th}$ is the thermal proton number density and $\epsilon_{CRp}$ is the energy density of CRp.
Under stationary conditions, the total synchrotron luminosity emitted by this process is
\begin{equation}
L_{syn} \propto n_{th} \epsilon_{CRp} R_v^3 \frac{B^2}{B^2+B_{CMB}^2}
\end{equation}
and the bolometric X-ray luminosity is
\begin{equation}
L_X \propto n_{th}^2 R_v^3 T^{1/2}.
\end{equation}
Consequently, defining $X_p$ as the ratio of the CRp to thermal energy densities, we would expect (using $L_X \propto T^{\alpha_T}$ and $\alpha_T \sim 3$ as in the previous subsections) a simple scaling between synchrotron and bolometric X-ray luminosity of the hosting cluster in the form
\begin{equation}
L_{syn} \propto X_p L_X^{1 + 1/2\alpha_T} \frac{B^2}{B^2+B_{CMB}^2}.
\end{equation}
In the case of a strong B field, this is equivalent to the scaling derived by \cite{kushnir2009}, i.e., $L_{syn} \propto X_p L_X^{1 + 1/2\alpha_T}$ or $L_{syn} \propto X_p L_{[0.1-2.4]}^{(\alpha_T + 0.5)/(\alpha_T -0.6)}$.
The weak magnetic fields case in these models is ruled out by the $\gamma$-ray upper limits obtained for nearby galaxy clusters (e.g., \citealt{ackermann2010}; \citealt{jeltema2011}; \citealt{brunetti2012}).

Finally, we note that, although not very different, hadronic and turbulent models predict different slopes for the nonthermal vs. thermal correlations in the case of radio halos. 
For example, if we focus on the $L_{syn}$ vs. $L_{[0.1-2.4]}$ correlation using the same assumptions in the two models, in the case of strong magnetic fields the hadronic models predict a slope $\sim$1.45 while turbulent models predict a slope in the range $\sim$1.45-1.9.
The slope becomes even steeper for weak fields in the turbulent model, $\sim$1.9--2.3; for reference, present observations give a slope of the correlation $=2.10 \pm 0.17$ (\citealt{cassano2013}).

\section{Supplemental Figures}
\label{sec:appendixC}

\begin{figure*}[t]
\begin{center}
  \begin{tabular}{c}
		\epsfig{file=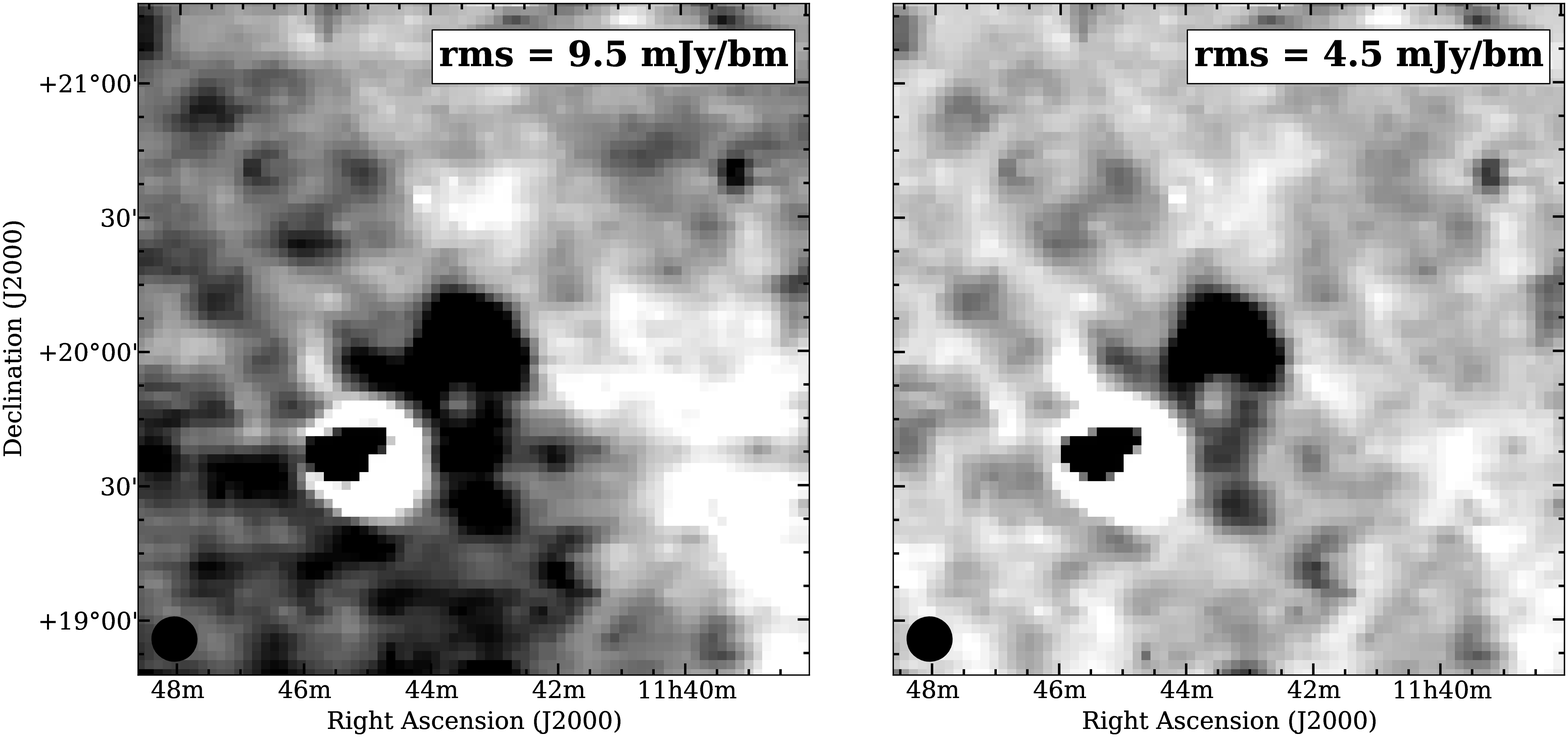,angle=0,width=0.75\textwidth} \\
		\epsfig{file=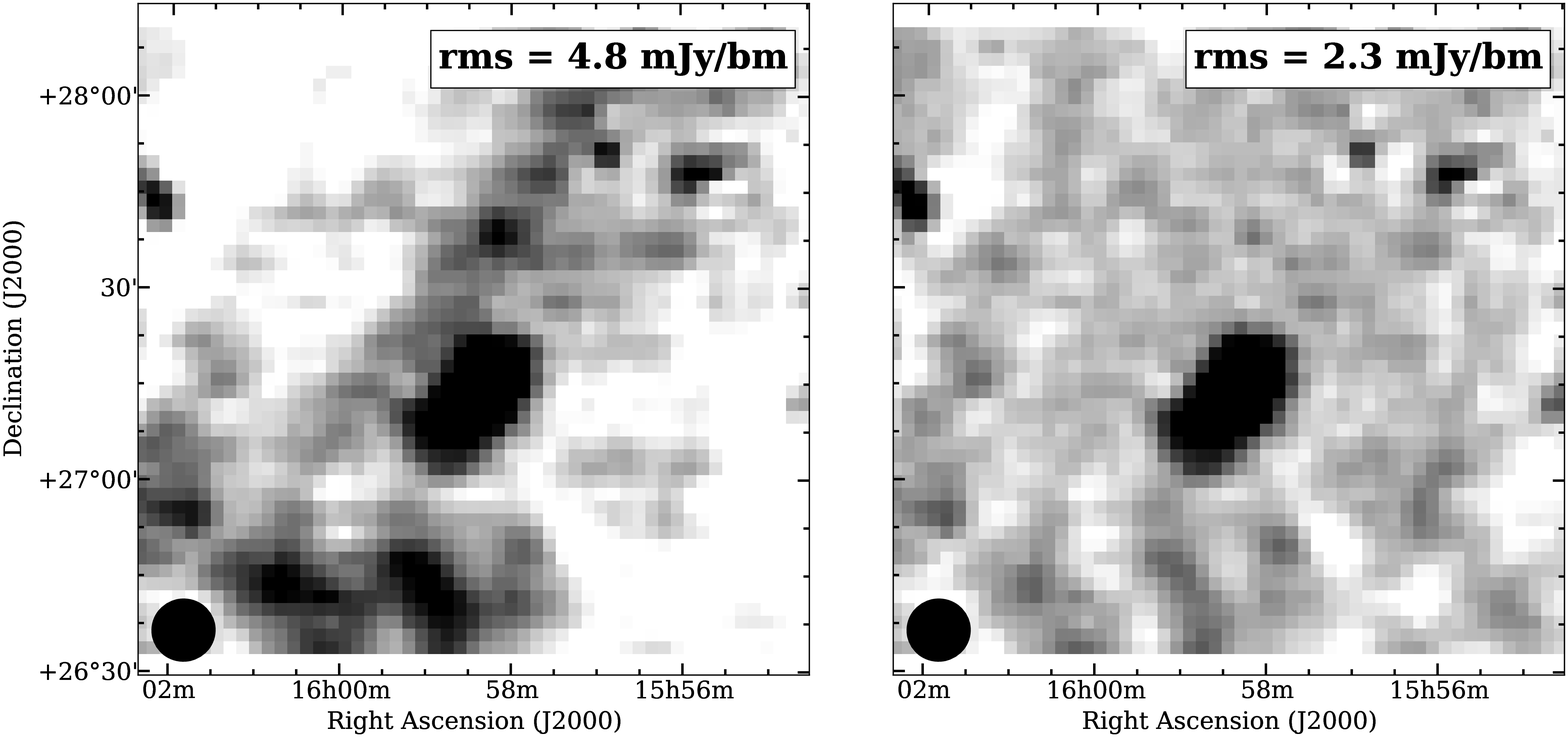,angle=0,width=0.75\textwidth} \\
		\epsfig{file=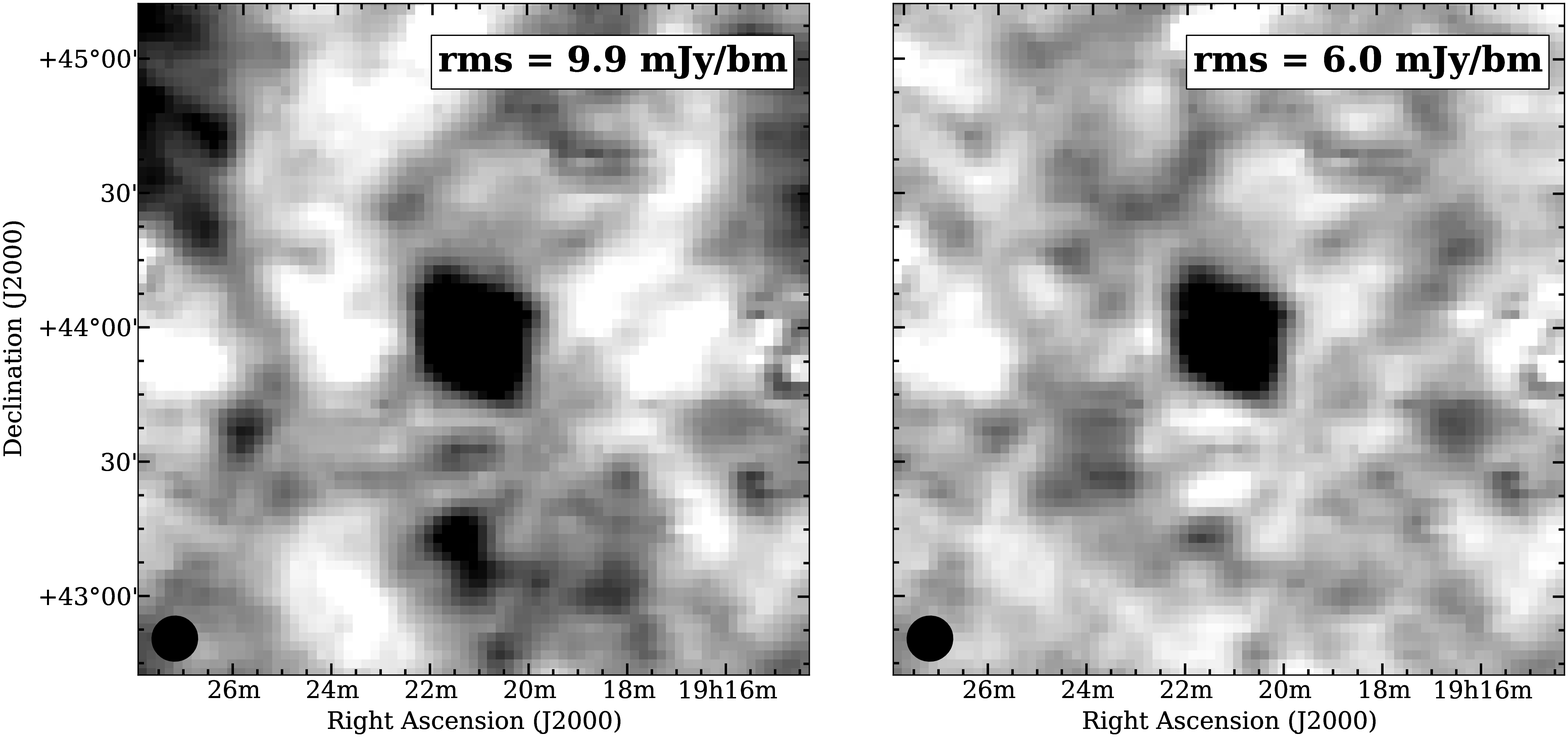,angle=0,width=0.75\textwidth}
  \end{tabular}
	\caption{A1367 (top), A2142 (center), A2319 (bottom) before (left) and after (right) the nonlinear Galactic foreground subtraction procedure described in Section~\ref{sec:results}.  For each image, a constant offset level has been added/subtracted to force the local background level to a mean of zero about the diffuse detection; the greyscale ranges from $-2\sigma_{map}$ to $6\sigma_{map}$, where $\sigma_{map}$ for each field is the post-subtraction background rms (listed in Table~\ref{tab:tableOfObservations}).  The GBT beam is shown in the lower left of each image.}
	\label{fig:galSubCompar}
\end{center}
\end{figure*}

Figure~\ref{fig:galSubCompar} illustrates the results of using elongated Gaussians to remove foreground Galactic ridges for three clusters, as described in the text.

\end{appendix}

\pagebreak

\end{document}